\documentclass[english,aps,preprintnumbers,superscriptaddress]{revtex4-2}
\usepackage[T1]{fontenc}
\usepackage[latin9]{inputenc}
\usepackage{verbatim}
\usepackage{mathrsfs}
\usepackage{amsmath}
\usepackage{ulem}
\usepackage{amssymb}
\usepackage{dsfont}
\usepackage{mathtools}
\usepackage{ifthen}
\usepackage{graphicx}
\usepackage[colorlinks=true,linktocpage=true,linkcolor=blue,citecolor=blue]{hyperref}
\usepackage{babel,oldgerm,combelow,slashed}
\usepackage{xcolor}
\def\dblone{\hbox{$1\hskip -1.2pt\vrule depth 0pt height 1.6ex width 0.7pt
\vrule depth 0pt height 0.3pt width 0.12em$}}
\definecolor{green}{rgb}{0,0.7,0}
\newcommand{\ms}{\mathfrak{s}}
\newcommand{\Dslash}[1]{\slash \hspace*{-0.18cm}{#1} }

\newcommand{\eq}{Eq.~}

\renewcommand{\d}{\mathrm{d}}
\newcommand{\s}{\mathfrak{s}}
\newcommand{\p}{\mathbf{p}}

\renewcommand{\u}{\mathbf{u}}
\newcommand{\ket}[1]{\left|#1\right\rangle}
\newcommand{\bra}[1]{\left\langle#1\right|}
\newcommand{\braket}[2]{\left\langle#1\right|\left.#2\right\rangle}
\renewcommand{\d}{\mathrm{d}}
\newcommand{\momint}[3]{\int \d {\MakeUppercase{#1}}^{#2}_{#3}}

\begin{document}

\title{Lorentz-covariant nonlocal collision term
for spin-1/2 particles}

\author{David Wagner}

\affiliation{Institute for Theoretical Physics, Goethe University, Max-von-Laue-Str.\ 1,
D-60438 Frankfurt am Main, Germany}

\author{Nora Weickgenannt}

\affiliation{Institute for Theoretical Physics, Goethe University, Max-von-Laue-Str.\ 1,
D-60438 Frankfurt am Main, Germany}

\author{Dirk H.\ Rischke}

\affiliation{Institute for Theoretical Physics, Goethe University, Max-von-Laue-Str.\ 1,
D-60438 Frankfurt am Main, Germany}

\affiliation{Helmholtz Research Academy Hesse for FAIR, Campus Riedberg, Max-von-Laue-Str.\ 12,
 D-60438 Frankfurt am Main, Germany}

\begin{abstract}
We revisit the derivation of the nonlocal collision term in
the Boltzmann equation for spin-1/2 particles,
using both the Wigner-function approach by de Groot, van Leeuwen, and van Weert, and the Kadanoff-Baym equation in $T$-matrix approximation.
Contrary to previous calculations, our results maintain full Lorentz covariance of the nonlocal collision term.
\end{abstract}

\maketitle

\section{Introduction and Summary}

The observation of polarization phenomena in relativistic heavy-ion collisions~\cite{Liang:2004ph,Voloshin:2004ha,Betz:2007kg,Becattini:2007sr,STAR:2017ckg,Adam:2018ivw,ALICE:2019aid,Mohanty:2021vbt} motivated a plethora of theoretical developments~\cite{Becattini:2013vja,Becattini:2013fla,Becattini:2015ska,Becattini:2016gvu,Karpenko:2016jyx,Pang:2016igs,Xie:2017upb,Becattini:2017gcx,Becattini:2020ngo,Florkowski:2019qdp,Florkowski:2019voj,Zhang:2019xya,Becattini:2019ntv,Xia:2019fjf,Wu:2019eyi,Sun:2018bjl,Liu:2019krs,Florkowski:2021wvk,Liu:2021uhn,Fu:2021pok,Becattini:2021suc,Becattini:2021iol,Yi:2021ryh,Wu:2022mkr,Li:2022pyw} in recent years, which aim at describing the dynamics of spin polarization in relativistic fluids.  Analogous to the nonrelativistic Barnett effect~\cite{Barnett:1935}, a rotating relativistic fluid with spin degrees of freedom can be polarized along the vorticity direction through the mutual conversion of orbital and spin angular momentum. In order to understand this process within a theoretical framework and to calculate the resulting spin polarization in a quantitatively reliable manner, considerable efforts have been made to derive a theory of relativistic spin hydrodynamics~\cite{Florkowski:2017ruc,Florkowski:2017dyn,Hidaka:2017auj,Florkowski:2018myy,Weickgenannt:2019dks,Bhadury:2020puc,Weickgenannt:2020aaf,Shi:2020htn,Speranza:2020ilk,Bhadury:2020cop,Singh:2020rht,Bhadury:2021oat,Peng:2021ago,Sheng:2022ssd,Hu:2021pwh,Hu:2022lpi,Fang:2022ttm,Wang:2022yli,Montenegro:2018bcf,Montenegro:2020paq,Gallegos:2021bzp,Hattori:2019lfp,Fukushima:2020ucl,Li:2020eon,She:2021lhe,Wang:2021ngp,Wang:2021wqq,Hongo:2021ona,Hongo:2022izs,Singh:2022ltu,Daher:2022xon,Weickgenannt:2022zxs,Weickgenannt:2022jes,Bhadury:2022qxd,Torrieri:2022xil,Wagner:2022gza,Weickgenannt:2022qvh,Daher:2022wzf}. 

It is convenient to first derive a kinetic theory  from quantum field theory, which in turn serves as the starting point for the derivation of spin hydrodynamics, e.g., by a Chapman-Enskog expansion~\cite{Bhadury:2020puc,Bhadury:2020cop} or by the method of moments~\cite{Weickgenannt:2022zxs,Weickgenannt:2022qvh}. It was found in previous works~\cite{Weickgenannt:2020aaf,Weickgenannt:2021cuo} (see also Refs.~\cite{Gao:2019znl,Hattori:2019ahi,Wang:2019moi,Liu:2020flb,Huang:2020kik,Wang:2020pej,Yang:2020hri,Wang:2020pej,Sheng:2021kfc,Das:2022azr} for related studies) that, in spin kinetic theory, a nonlocal collision term is responsible for the mutual conversion of orbital and spin angular momentum, and therefore polarizes the fluid along the vorticity when equilibrium is approached.

In Refs.~\cite{Weickgenannt:2020aaf,Weickgenannt:2021cuo}, we derived the
Boltzmann equation to order $\mathcal{O}(\hbar)$ for massive spin-1/2 particles
in the approach of Ref.\ \citep{DeGroot:1980dk}, termed
``GLW approach'' in this paper.
For Dirac particles with spin 1/2 and for binary elastic collisions, it assumes the form
\begin{align}
p \cdot \partial_x f(x,p,\ms)  = 
\mathfrak{C}[f] & = \frac{1}{4} \int \d\Gamma_1 \d\Gamma_2 \d\Gamma^\prime 
\d \bar{S}(p)\, 
(2\pi\hbar)^4\delta^{(4)}\left( p  +p' -p_1-p_2  \right) \nonumber \\
& \hspace*{0.5cm} \times \mathcal{W}[f(x+\Delta_1-\Delta,p_1,\ms_1)f(x+\Delta_2-\Delta,p_2,\ms_2)
-f(x,p,\bar{\ms})f(x+\Delta^\prime-\Delta,p^\prime,\ms^\prime)]\; .
\label{eq:finalcollisionterm}
\end{align}
Here, $\mathcal{W}$ is the
transition probability for the
scattering process and $f(x,p,\ms)$ is the spin-dependent single-particle distribution function in extended phase space, 
i.e., ordinary phase space extended by spin degrees of freedom. We also defined $\d\Gamma\coloneqq 
\d P \d S(p)$ as the integration measure over on-shell momentum space 
$\d P$ and spin space $\d S(p)$, see 
Eqs.\ (\ref{eq:measure_mom}) and (\ref{eq:measure_spin}). Note that in Eq.\ (\ref{eq:finalcollisionterm}) there is
in principle also a contribution from collisions
which exchange only spin but not momentum,
cf., e.g., Eq.\ (24) in Ref.\ \citep{Weickgenannt:2020aaf}. 
It can be shown that such a contribution 
corresponds to corrections
to the drift term, as well as
an additional Vlasov term, on the
left-hand side of the Boltzmann
equation. We defer a detailed
discussion to a subsequent work
\cite{Wagnertoappear} and, for the sake of simplicity, omit such a contribution in this paper.
We also note that the
collision term (\ref{eq:finalcollisionterm}) 
agrees with the one given in 
Refs.\ \cite{Weickgenannt:2020aaf,Weickgenannt:2021cuo} only when
replacing $x -\Delta$ by $x$.
The form given in Eq.\ (\ref{eq:finalcollisionterm}) is
actually the more accurate one.
The overall shift of all positions
by $-\Delta$ was neglected
in Refs.\ \cite{Weickgenannt:2020aaf,Weickgenannt:2021cuo} by an argument assuming an expansion around local thermodynamical equilibrium, for details see App.\ D of Ref.\ \cite{Weickgenannt:2021cuo}.

The nonlocality of the collision term (\ref{eq:finalcollisionterm}) manifests itself in the
fact that the distribution functions of the collision partners are taken at  space-time points shifted from position $x^\mu$ by
$\Delta'^\mu-\Delta^\mu $, $\Delta_1^\mu-\Delta^\mu $, and $\Delta_2^\mu-\Delta^\mu $,
respectively.
These shifts were calculated in Ref.\ \cite{Weickgenannt:2021cuo}
and are of the order of the
Compton wavelength of the particles, e.g.,\begin{equation}
\label{eq:deltanon}
\Delta_\mu\coloneqq \frac{\hbar}{2m(p\cdot\hat{t}+m)}\, \epsilon_{\mu\nu\alpha\beta}\, \hat{t}^\nu p^\alpha  \ms^{\beta}\;,
\end{equation}
and similarly for $\Delta^{\prime \mu}$,
$\Delta^\mu_1$, and $\Delta^\mu_2$, replacing
$(p^\alpha,\ms^\beta)$ by $(p^{\prime \alpha},\ms^{\prime \beta})$, $(p^\alpha_1,\ms^\beta_1)$, and $(p^\alpha_2,\ms^\beta_2)$,
respectively.
In Eq.\ (\ref{eq:deltanon}), $\hat{t}^\nu \coloneqq (1,\boldsymbol{0})$ is a time-like unit vector 
defining the frame where $p^\alpha$ is measured. 
The nonlocality of the collision term allows for 
the conversion of orbital into spin angular momentum. It was shown that the collision term vanishes
in global equilibrium and that the spin potential is then equal to a constant value of the
thermal vorticity. We remark that
the expression (\ref{eq:deltanon}) is
identical with the so-called Berry curvature, cf.\ e.g.\ Ref.\ \citep{Stone:2014fja}.

However, a serious shortcoming of the previous work \citep{Weickgenannt:2020aaf,Weickgenannt:2021cuo}
is that the Berry curvature violates Lorentz covariance, since it explicitly depends on the frame vector $\hat{t}^\mu$, which does not transform under Lorentz transformations. 
The technical reason for the occurrence of the Berry curvature is that at some point of the calculation one has to take 
momentum derivatives of the Dirac spinors appearing in the Wigner functions. At the same time, momentum derivatives of 
the scattering matrix elements are neglected, arguing that these derivatives should be small if the interaction is 
sufficiently localized. However, in doing so one neglects momentum derivatives of the Dirac spinors appearing in the 
matrix elements, which otherwise would lead to similar Berry-curvature terms,  which, if kept, restore Lorentz covariance. 

In this 
paper, we improve on the previously made approximation and derive a nonlocal collision term 
which manifestly respects Lorentz covariance. The form of this term is similar to 
Eq.~(\ref{eq:finalcollisionterm}), but the space-time shift (\ref{eq:deltanon}) is replaced by
a (much more complicated, but) Lorentz-covariant expression. For
a current-current interaction
as in the time-honored
Nambu--Jona-Lasinio (NJL) model~\cite{Nambu:1961tp,Nambu:1961fr}, i.e.,
an interaction which couples the fermion 
current $\overline{\psi} \Gamma^{(c)}
\psi$ with itself with coupling
strength $G_c$, where $\Gamma^{(c)}$ is a Dirac matrix and the index $c$
characterizes the particular
interaction channel (for details
see Sec.\ \ref{sec:Lagrangian}), we obtain
\begin{subequations}\label{eq:def_Deltas_NJL}
    \begin{eqnarray}
        \Delta_{1}^{\mu} &\coloneqq& \frac{\hbar}{m} \frac{4 G_c G_d}{\hbar^2}\frac{m^4}{\mathcal{W}} \, 
        \mathrm{Im}\left\{        \mathrm{Tr}\left[h\Gamma^{(d)} h_2  \Gamma^{(c)} \right] \mathrm{Tr}\left[\Gamma^{(d)} h_1 \gamma^\mu \Gamma^{(c)} h'\right]-\mathrm{Tr}\left[h \Gamma^{(d)}h_1 \gamma^\mu \Gamma^{(c)} h'\Gamma^{(d)} h_2 \Gamma^{(c)} \right] \right\}\;,\label{eq:delta_1_NJL}\\
        \Delta_{2}^{\mu} &\coloneqq& \frac{\hbar}{m}\frac{4 G_c G_d}{\hbar^2} \frac{m^4}{\mathcal{W}}\, \mathrm{Im}\left\{ \mathrm{Tr}\left[h\Gamma^{(d)}h_2\gamma^\mu  \Gamma^{(c)} \right]
        \mathrm{Tr}\left[\Gamma^{(d)}h_1\Gamma^{(c)}h'   \right]-\mathrm{Tr}\left[ h \Gamma^{(d)} h_1  \Gamma^{(c)} h'\Gamma^{(d)} h_2\gamma^\mu \Gamma^{(c)}\right] \right\}\;,\label{eq:delta_2_NJL}\\
        \Delta'^{\mu} &\coloneqq& \frac{\hbar}{m} \frac{4 G_c G_d}{\hbar^2}\frac{m^4}{\mathcal{W}} \, \mathrm{Im}\left\{\mathrm{Tr}\left[h\Gamma^{(d)} h_2  \Gamma^{(c)} \right]\mathrm{Tr}\left[\Gamma^{(d)}h_1  \Gamma^{(c)} h'\gamma^\mu\right]-\mathrm{Tr}\left[h \Gamma^{(d)}h_1  \Gamma^{(c)} h'\gamma^\mu\Gamma^{(d)} h_2 \Gamma^{(c)} \right] \right\}\;,\label{eq:delta_prime_NJL}\\
        \Delta^{\mu} &\coloneqq& \frac{\hbar}{m} \frac{4 G_c G_d}{\hbar^2}\frac{m^4}{\mathcal{W}} \, \mathrm{Im}\left\{\mathrm{Tr}\left[h\gamma^\mu\Gamma^{(d)}h_2  \Gamma^{(c)} \right]\mathrm{Tr}\left[\Gamma^{(d)}h_1  \Gamma^{(c)} h'\right]-\mathrm{Tr}\left[h\gamma^\mu\Gamma^{(d)}h_1  \Gamma^{(c)} h'\Gamma^{(d)} h_2 \Gamma^{(c)} \right] \right\}\;,\label{eq:delta_NJL}
    \end{eqnarray}
\end{subequations}
where
\begin{equation}
    \mathcal{W} \coloneqq m^4\, \frac{G_c G_d}{\hbar^2}\,  16 \, \mathrm{Re}\left\{\mathrm{Tr}\left[h  \Gamma^{(d)} h_2 \Gamma^{(c)} \bar{h}\right] \mathrm{Tr}\left[ \Gamma^{(d)} h_1\Gamma^{(c)}h' \right]-\mathrm{Tr}\left[h  \Gamma^{(d)} h_1\Gamma^{(c)} h' \Gamma^{(d)} h_2 \Gamma^{(c)}\bar{h}\right] \right\}\label{eq:W_NJL} \;.
\end{equation}
Here,
\begin{equation}
h \equiv h(p,\s) \coloneqq \frac{1}{4m}(\dblone +\gamma_5 \slashed{\ms})(\slashed{p} +m) \;.\label{eq:def_h}
\end{equation}
and similarly for $h'$, $h_1$, $h_2$,
and $\bar{h}$, with $(p,\ms)$ replaced
by $(p',\ms'), (p_1, \ms_1)$, $(p_2, \ms_2)$, and
$(p, \bar{\ms})$, respectively.
Equations (\ref{eq:def_Deltas_NJL}) are the main result of this work.

The paper is organized as follows. In Sec.\ \ref{sec:Wigner} we recall some facts about the Wigner function.
In Sec.\ \ref{sec:Lagrangian} we define
the Lagrangian underlying our
investigations.
In Sec.\ \ref{sec:de_groot} we  carefully repeat the calculation of the collision term via the GLW approach \cite{DeGroot:1980dk} performed in Refs.\ \citep{Weickgenannt:2020aaf,Weickgenannt:2021cuo}, but now paying attention to maintaining full 
Lorentz covariance throughout the calculation. For the purpose of making our calculations more concise, various definitions and conventions differ from those of 
Ref.\ \cite{DeGroot:1980dk} and used previously in Refs.\ \citep{Weickgenannt:2020aaf,Weickgenannt:2021cuo}. 
This also facilitates comparison with other works, which adhere to more commonly used notation. Our result is the nonlocal Boltzmann equation 
(\ref{eq:finalcollisionterm}) with the space-time shifts (\ref{eq:deltanon}) replaced by the expressions 
(\ref{eq:def_Deltas_NJL}). In Sec.\ \ref{sec:KB} we then confirm our results by repeating the calculation in the 
Kadanoff-Baym (KB) approach
(see, e.g., Refs.\ \cite{Landsman:1986uw,Mrowczynski:1992hq}), showing that the results coincide for an interaction of NJL-type.
The KB approach was previously used by some of us (N.W., D.H.R.) in Ref.\ \citep{Sheng:2021kfc}, 
where the nonlocal collision term was derived in $T$-matrix approximation. However, the use of 
the matrix-valued spin distribution functions in that work prevented a direct comparison with that of the 
GLW approach of Refs.\ \citep{Weickgenannt:2020aaf,Weickgenannt:2021cuo}.
In this paper, we complete the derivation of the nonlocal collision term in the KB approach,
using the scalar distribution function $f(x,p,\ms)$ in extended phase space 
that was used in the GLW approach. We note that
a similar study in 
the nonrelativistic limit has been performed in a recent paper~\cite{Dong:2022yzt}.
Finally, in Sec.\ \ref{sec:summary}, we conclude our work with an outlook for future studies.

We define the Lorentz-invariant measure in momentum space as
\begin{equation} \label{eq:measure_mom}
\d P \coloneqq \frac{\d^3 \mathbf{p}}{(2\pi \hbar)^3 p^0}\;.
\end{equation}
The Lorentz-invariant measure in spin space for particles with spin 1/2 is defined as \cite{Weickgenannt:2020aaf,Weickgenannt:2021cuo}
\begin{equation} \label{eq:measure_spin}
\d S(p)\coloneqq  \sqrt{\frac{p^2}{3 \pi^2}} \,  \d^4\ms\, \delta\left(\ms\cdot\ms+3\right)\,\delta(p\cdot \ms)\;.
\end{equation} 
This measure implies the following relations for integration over the spin 4-vector $\ms^\mu$:\begin{equation} \label{eq:intspin}
\int \d S(p) = 2\;, \quad \int \d S(p)\, \ms^\mu = 0\;, \quad \int \d S(p) \, \ms^\mu \ms^\nu =
- 2 \left( g^{\mu \nu}  - \frac{p^\mu p^\nu}{p^2} \right)\;.
\end{equation}
The measure (\ref{eq:measure_spin}) is chosen in such a way that integrating over the whole spin space gives
the number of spin degrees of freedom,
see the first relation in Eq.\ (\ref{eq:intspin}).

We adopt the following conventions:
the metric tensor is $g_{\mu\nu}=\mathrm{diag}(+,-,-,-)$ and 
the four-dimensional
unit matrix in Dirac space 
is denoted as $\dblone$,
while the Dirac matrices are denoted
as $\gamma^\mu$. The four-dimensional Levi-Civit\'{a} symbol is
$\epsilon^{0123}=-\epsilon_{0123}=1$,
and summation over repeated
indices is implied if not stated explicitly. 
The scalar product of two
four-vectors $a^\mu$ and $b^\mu$ is $a\cdot b \coloneqq a^{\mu}b_{\mu}$.
Furthermore, we define
$\slashed{a} \coloneqq \gamma^\mu a_\mu$. (Anti-)symmetrization in Lorentz indices is denoted as  $a_{[\mu}b_{\nu]}\coloneqq a_{\mu}b_{\nu}-a_{\nu}b_{\mu}$ and
$a_{(\mu}b_{\nu)}\coloneqq a_{\mu}b_{\nu}+a_{\nu}b_{\mu}$.
We choose natural Heaviside-Lorentz units, $c=\varepsilon_0 = \mu_0 = k_{B}=1$, but the reduced Planck constant $\hbar$
is kept explicitly in order to perform the power counting.
Lorentz indices are denoted by
Greek indices, except for
$\alpha,\,\beta,\,\gamma$ and $\delta$, which are
used for Dirac indices (if necessary with
appropriate sub-indices, e.g.,
$\alpha', \alpha_1, \alpha_2, \ldots$). 
Spin indices are denoted
by the letters $r,s,\ldots$.

\section{Wigner function}
\label{sec:Wigner}

In this section, we collect some well-known facts about the Wigner function, 
which will be used in the calculation of the collision term in the GLW as well as the KB approach. 
We start with a discussion of the two-particle correlation function in the closed-time path formalism 
and then focus on the definition of the Wigner function, 
its Clifford decomposition, as well as its equation of motion. We then establish a relation between the Wigner function 
and the single-particle distribution function in extended phase space.

\subsection{Two-particle correlation function in closed-time path formalism}

On the closed-time path
(see, e.g., Ref.\ \cite{Landsman:1986uw}),
the two-particle correlation function assumes the following matrix form,
\begin{equation}
G(x_{1},x_{2})=\left(\begin{array}{cc}
G^{++}(x_{1},x_{2}) & G^{+-}(x_{1},x_{2})\\
G^{-+}(x_{1},x_{2}) & G^{--}(x_{1},x_{2})
\end{array}\right)\equiv \left(\begin{array}{cc}
G^{F}(x_{1},x_{2}) & G^{<}(x_{1},x_{2})\\
G^{>}(x_{1},x_{2}) & G^{\bar{F}}(x_{1},x_{2})
\end{array}\right)\; .\label{eq:green-ctp}
\end{equation}
where $G^{ij}(x_{1},x_{2})$ (with $i,j=+,-$) means that the first time argument
$t_{1}=x_{1}^{0}$ lives on the time branch $i$ and the second time argument
$t_{2}=x_{2}^{0}$ lives on the time branch $j$. 
The definitions of the various Green's functions are
\begin{subequations}
\begin{eqnarray}
G_{\alpha\beta}^{F}(x_{1},x_{2}) & \coloneqq&
\left\langle T\psi_{\alpha}(x_{1})\overline{\psi}_{\beta}(x_{2})\right\rangle
\;, \label{eq:def-green_F}\\
G_{\alpha\beta}^{\bar{F}}(x_{1},x_{2}) & \coloneqq &
\left\langle T_{A}\psi_{\alpha}(x_{1})\overline{\psi}_{\beta}(x_{2})
\right\rangle \; ,\label{eq:def-green_barF} \\
G_{\alpha\beta}^{<}(x_{1},x_{2}) & \coloneqq &
\left\langle \overline{\psi}_{\beta}(x_{2})\psi_{\alpha}(x_{1})\right\rangle
\; ,\label{eq:def-green-lesser} \\
G_{\alpha\beta}^{>}(x_{1},x_{2}) & \coloneqq &
\left\langle \psi_{\alpha}(x_{1})\overline{\psi}_{\beta}(x_{2})\right\rangle
\; ,\label{eq:def-green-larger}
\end{eqnarray}
\end{subequations}
where $T$ and $T_{A}$ denote the time-ordering and anti-time-ordering
operators, respectively, and angular brackets denote averages computed
with respect to some initial state. 
Note that we define $G^<$ with opposite sign as in Ref.\ \citep{Sheng:2021kfc}, but with the same sign as
in Ref.~\citep{Yang:2020hri}.

Not all components of the correlation function (\ref{eq:green-ctp}) are independent. In fact,
one may express $G^F$ and $G^{\bar{F}}$ by $G^<$ and $G^>$ using the
definition of the time-ordering and anti-time-ordering operators, respectively,
\begin{subequations}
\begin{align}
G_{\alpha\beta}^{F}(x_{1},x_{2}) & =  \theta(t_1- t_2)  G^>_{\alpha\beta}(x_1,x_2) - \theta(t_2- t_1) G^<_{\alpha\beta}(x_1,x_2)\;, \\
 G_{\alpha\beta}^{\bar{F}}(x_{1},x_{2}) & =  - \theta(t_1- t_2)  G^<_{\alpha\beta}(x_1,x_2) + \theta(t_2- t_1) G^>_{\alpha\beta}(x_1,x_2)\;.
\end{align}
\end{subequations}

\subsection{Wigner function}

The Wigner function $G^<(x,p)$
in the KB approach is defined as
the Fourier transform of the 
two-point function (\ref{eq:def-green-lesser})
with respect to the difference $y\coloneqq x_{1}-x_{2}$ of the two space-time points
$x_1$ and $x_2$,
\begin{equation}
G_{\alpha\beta}^{<}(x,p)\coloneqq \int \d^{4}y\,e^{ip\cdot y/\hbar} \,
G_{\alpha \beta}^<(x_1,x_2)
\equiv \int \d^{4}y\,e^{ip\cdot y/\hbar}
\left\langle \overline{\psi}_{\beta}\left(x-\frac{y}{2}\right)\psi_{\alpha}\left(x+\frac{y}{2}\right)
\right\rangle\; ,
\label{eq:wigner-def}
\end{equation}
where $x\coloneqq(x_{1}+x_{2})/2$ is the arithmetic mean (or center) of the
two space-time points $x_1$ and $x_2$. Similarly,
\begin{equation}
G_{\alpha\beta}^{>}(x,p)\coloneqq 
\int \d^{4}y\,e^{ip\cdot y/\hbar}\, 
G_{\alpha \beta}^>(x_1,x_2) \equiv \int \d^{4}y\,e^{ip\cdot y/\hbar}
\left\langle \psi_{\alpha}\left(x+\frac{y}{2}\right)\overline{\psi}_{\beta}\left(x-\frac{y}{2}\right)
\right\rangle\; .
\label{eq:wigner-def_2}
\end{equation}

On the other hand, in the GLW approach \cite{DeGroot:1980dk},
the Wigner function is defined as 
\begin{equation}
\label{eq:def_Wigner}
    W_{\alpha \beta}(x,p)
     \coloneqq \int \d^4 y \, e^{-\frac{i}{\hbar}p\cdot y}\left\langle : \overline{\psi}_\beta \left(x + \frac{y}{2} \right)\psi_\alpha \left(x- \frac{y}{2}\right):\right\rangle\;,
\end{equation}
i.e., similar as in Eq.\ (\ref{eq:wigner-def}) when substituting
$y \rightarrow -y$ (in order to facilitate the comparison
with previous work, we do not 
perform this substitution explicitly), 
but with an additional normal-ordering
operation on the field operators. 
This bears no further
consequences, since we will neglect
antiparticles anyway. 

Note that compared to Refs.\ \cite{Weickgenannt:2019dks,Weickgenannt:2020aaf,Weickgenannt:2021cuo,DeGroot:1980dk} the factor $(2\pi\hbar)^{-4}$ in the
integration measure in Eq.\ (\ref{eq:def_Wigner}) is absent, because
we choose to absorb such factor into the four-dimensional momentum-space measure
$\d^4 p/(2 \pi \hbar)^4$. A further
consequence is that the single-particle
distribution function does not have
a prefactor $(2 \pi \hbar)^{-3}$
as, e.g., in Eq.\ (85) of Ref.\ \citep{Weickgenannt:2020aaf}.
This facilitates the expression for
Pauli-blocking factors, which
simply read $1-f$, instead of
$1-(2 \pi \hbar)^3 f$ in the notation
of Ref.\ \cite{DeGroot:1980dk}.

\subsection{Clifford decomposition}

We can expand $G^{<}(x,p)$
(or $W(x,p)$) in terms of the 16 independent generators
of the Clifford algebra, $\Gamma_a$, $a = 1, \ldots,16$, with
\begin{equation}
\Gamma_a \in \left\{ \dblone,\, \gamma^{\mu}, \, -i\gamma^{5}, \,
\gamma^{5}\gamma^{\mu},\, \sigma^{\mu\nu}
\right\}\;, \label{eq:c-generator}
\end{equation}
where $\sigma^{\mu \nu} \coloneqq \frac{i}{2} [ \gamma^\mu, \gamma^\nu]$,
such that
\begin{equation}
G^{<}(x,p) \equiv W(x,p) =\frac{1}{4}\left(\mathcal{F}+i\gamma^{5}\mathcal{P}
+ \gamma^\mu\mathcal{V}_{\mu}
+\gamma^{5}\gamma^\mu \mathcal{A}_{\mu}
+\frac{1}{2}\sigma^{\mu\nu}\mathcal{S}_{\mu\nu}\right)\;.
\label{eq:wigner-decomp}
\end{equation}
Similarly,
\begin{equation}
G^{>}(x,p)=\frac{1}{4}\left(\bar{\mathcal{F}}+i\gamma^{5}\bar{\mathcal{P}}
+ \gamma^\mu \bar{\mathcal{V}}_{\mu}
+\gamma^{5} \gamma^\mu\bar{\mathcal{A}}_{\mu}
+\frac{1}{2}\sigma^{\mu\nu} \bar{\mathcal{S}}_{\mu\nu}\right)\;.
\label{eq:wigner-decomp_2}
\end{equation}
The real-valued coefficient functions $\mathcal{F}$, $\bar{\mathcal{F}}$, $\mathcal{P}$, $\bar{\mathcal{P}}$, 
$\mathcal{V}_{\mu}$, $\bar{\mathcal{V}}_{\mu}$, $\mathcal{A}_{\mu}$, $\bar{\mathcal{A}}_{\mu}$, 
$\mathcal{S}_{\mu\nu}$, and $\bar{\mathcal{S}}_{\mu\nu}$ are the scalar, pseudo-scalar,
vector, axial-vector, and tensor components of $G^{\lessgtr}(x,p)$
(or $W(x,p)$), respectively, which can
be obtained by taking the trace of $G^{\lessgtr}(x,p)$ (or $W(x,p)$), multiplied with the appropriate
generator $\Gamma_a$ of the Clifford algebra. 

\subsection{Equation of motion}

The equation of motion for the
Wigner function $G^<(x,p)$ 
can be derived
from the Dyson-Schwinger equation
for the two-particle correlation
function. In the quasi-particle
approximation one obtains
\cite{Sheng:2021kfc}
\begin{equation}
\left(\slashed{K}-m\right)G^{<}(x,p) = I_{\mathrm{coll}}\;,
\label{eq:kb-main}
\end{equation}
where $m$ is the mass of the particles
and
\begin{equation}
K^\mu \coloneqq p^\mu + \frac{i\hbar}{2} \partial_{x}^\mu \;.\label{eq:k-op}
\end{equation}
The collision term $I_{\mathrm{coll}}$ is given by
\begin{eqnarray}
I_{\mathrm{coll}}& \coloneqq & \frac{i \hbar}{2}\left[\Sigma^{<}(x,p)G^{>}(x,p)
-\Sigma^{>}(x,p)G^{<}(x,p)\right]
\nonumber \\
 & + & \frac{\hbar^{2}}{4}\left[\left\{ \Sigma^{<}(x,p),G^{>}(x,p)\right\} _{\mathrm{PB}}
 -\left\{ \Sigma^{>}(x,p),G^{<}(x,p)\right\} _{\mathrm{PB}}\right]\;.
\label{eq:collision_term}
\end{eqnarray}
Note the change of sign in the collision term as compared to Ref.\ \citep{Sheng:2021kfc}, which is due to the
opposite sign in the definition of $G^<$.
In Eq.\ (\ref{eq:collision_term}), $\Sigma^{\gtrless}(x,p)$ are the 
Wigner transforms of the self-energies 
$\Sigma^\gtrless(x_1,x_2)$ on
the closed-time path and
we introduced the Poisson bracket
\begin{equation}
\left\{ A, B \right\}_\textrm{PB} \coloneqq(\partial_x A) \cdot (\partial_p B) - (\partial_p A) \cdot (\partial_x B)\;.
\end{equation}

By multiplying Eq.\ (\ref{eq:kb-main}) with the generators of the Clifford algebra and taking the trace, we can derive a system of equations of motion for the Clifford components of the Wigner function. The real parts of these equations read
\begin{subequations}
\begin{eqnarray}
p^\mu \mathcal{V}_\mu - m \mathcal{F}
& = & \mathrm{Re\, Tr} \left( I_{\mathrm{coll}}\right)\;,
\label{eq:scalar_real} \\
m \mathcal{P} + \frac{\hbar}{2}\, \partial_x^\mu \mathcal{A}_\mu
& = &  \mathrm{Re\, Tr} \left(i \gamma^5 I_{\mathrm{coll}}\right)\;,
\label{eq:pseudoscalar_real}  \\
p_\mu \mathcal{F} - m \mathcal{V}_\mu
+ \frac{\hbar}{2} \, \partial_x^\nu \mathcal{S}_{\mu \nu}
& = & \mathrm{Re\, Tr} \left( \gamma_\mu I_{\mathrm{coll}}\right)\;,
\label{eq:vector_real}  \\
\frac{1}{2} \, \epsilon_{\mu \nu \rho\sigma} p^\nu \mathcal{S}^{\rho \sigma}
+ m \mathcal{A}_\mu
- \frac{\hbar}{2} \, \partial_{x,\mu} \mathcal{P} & = &
\mathrm{Re\, Tr} \left( \gamma^5 \gamma_\mu I_{\mathrm{coll}}\right)\;,
\label{eq:axialvector_real}  \\
\epsilon_{\mu \nu \rho\sigma} p^\rho \mathcal{A}^\sigma + m \mathcal{S}_{\mu \nu}
- \frac{\hbar}{2} \partial_{x[\mu} \mathcal{V}_{\nu]}
& = & - \mathrm{Re\, Tr} \left( \sigma_{\mu \nu} I_{\mathrm{coll}}\right)\;,
\label{eq:tensor_real}
\end{eqnarray}
while the imaginary parts are
\begin{eqnarray}
\frac{\hbar}{2}\, \partial_x^\mu \mathcal{V}_\mu & = &
\mathrm{Im\, Tr} \left( I_{\mathrm{coll}}\right)\;,
\label{eq:scalar_im} \\
p^\mu \mathcal{A}_\mu & = & \mathrm{Im\, Tr} \left(- i \gamma^5 I_{\mathrm{coll}}\right)\;,
\label{eq:pseudoscalar_im} \\
p^\nu \mathcal{S}_{\nu \mu} + \frac{\hbar}{2} \, \partial_{x,\mu} \mathcal{F}
& = & \mathrm{Im\, Tr} \left( \gamma_\mu I_{\mathrm{coll}}\right)\;, \label{eq:vector_im} \\
p_\mu \mathcal{P} + \frac{\hbar}{4} \, \epsilon_{\mu \nu \rho\sigma}
\partial_x^\nu \mathcal{S}^{\rho\sigma}
& = & \mathrm{Im\, Tr} \left( \gamma^5 \gamma_\mu I_{\mathrm{coll}}\right)\;,
\label{eq:axialvector_im} \\
p_{[\mu} \mathcal{V}_{\nu]} + \frac{\hbar}{2}\, \epsilon_{\mu \nu \rho\sigma}
\partial_x^\rho
\mathcal{A}^\sigma & = & - \mathrm{Im\, Tr} \left( \sigma_{\mu \nu} I_{\mathrm{coll}}\right)\;.
\label{eq:tensor_im}
\end{eqnarray}
\end{subequations}

Acting with the operator $\slashed{K} + m$ onto Eq.\ (\ref{eq:kb-main}) and combining
the resulting equation with its Hermitian conjugate, multiplied from
the left- and the right-hand side with $\gamma^0$, we can
derive a mass-shell constraint and a Boltzmann equation for the Wigner function $G^<(x,p)$,
\begin{subequations}
\label{eq:mass_boltz_KB_0}
\begin{eqnarray}
\left(p^{2}-m^{2}-\frac{\hbar^{2}}{4}\partial_x^{2}\right)G^{<}(x,p)
& = & \frac{1}{2}\left\{(\slashed{K}+m)I_{\mathrm{coll}}+\gamma^{0}
\left[(\slashed{K}+m)I_{\mathrm{coll}}\right]^{\dagger}\gamma^{0}\right\}\;,
\label{eq:massshell-wig}\\
\hbar\, p\cdot\partial_x G^{<}(x,p) & = &
-\frac{i}{2}\left\{ (\slashed{K}+m)I_{\mathrm{coll}}-\gamma^{0}
\left[(\slashed{K}+m)I_{\mathrm{coll}}\right]^{\dagger}\gamma^{0}\right\} \;,
\label{eq:boltzmann-wig}
\end{eqnarray}
\end{subequations}
where we have used
$\gamma^0 (G^{<})^\dagger \gamma^0 \equiv G^<$, which follows from Eq.\ (\ref{eq:wigner-decomp}).
Taking the trace with the appropriate basis elements of
the Clifford algebra, we obtain for the components of the Wigner function
\begin{subequations}
\begin{eqnarray}
\left(p^{2} -m^{2}-\frac{\hbar^{2}}{4}\partial_x^{2}\right)\,
\mathrm{Tr}\left(\Gamma_{a}G^{<}\right) & = &
\mathrm{Re}\mathrm{Tr}\left[\Gamma_{a}(\slashed{K}+m)I_{\mathrm{coll}}\right]\;,
\label{eq:massshell-wigner} \\
\hbar\,  p\cdot\partial_x\, \mathrm{Tr}\left(\Gamma_{a}G^{<}\right) & = &
\mathrm{Im}\mathrm{Tr}\left[\Gamma_{a}(\slashed{K}+m)I_{\mathrm{coll}}\right] \;,
\label{eq:boltzmann-wigner}
\end{eqnarray}
\end{subequations}
where we have used $\gamma^0 \Gamma_a^\dagger \gamma^0 = \Gamma_a$.
It was shown in Ref.~\citep{Sheng:2021kfc} that off-shell contributions in Eq.~(\ref{eq:boltzmann-wigner})
are of higher order, $\mathcal{O}(G_c^4)$, in the coupling constant. They are
therefore neglected in the following. In this approximation, we will also show by an explicit computation that, at least 
to order $\mathcal{O}(\hbar)$,
all propagators are on the mass shell.

Similarly, in the GLW approach the equations of motion for the Wigner function (\ref{eq:def_Wigner}) read \cite{DeGroot:1980dk, Weickgenannt:2021cuo}
\begin{subequations}
\label{eq:mass_boltz_GLW}
\begin{eqnarray}
    \left(p^2-m^2-\frac{\hbar^2}{4}\partial_x^2  \right)W(x,p) &=& \hbar \delta M(x,p)\;,\label{eq:mass_shell}\\
    p \cdot \partial_x W(x,p)&=&\mathcal{C}(x,p)\;.\label{eq:coll}
\end{eqnarray}
\end{subequations}
Here we defined
\begin{subequations}
\begin{eqnarray}
   \delta M_{\alpha\beta}(x,p)&\coloneqq&  \frac12 \int \d^4 y e^{-\frac{i}{\hbar}y\cdot p} \left\langle : \left[\overline{\rho}(x_1)\left(i\hbar \overleftarrow{\slashed{\partial}}_x+m\right)\right]_\beta \psi_\alpha(x_2)+\overline{\psi}_\beta(x_1) \Big[\left(-i\hbar \slashed{\partial}_x+m\right)\rho(x_2)\Big]_\alpha :\right\rangle\;,\label{eq:def_deltaM}\\
   \mathcal{C}_{\alpha\beta}(x,p)&\coloneqq&  \frac{i}{2} \int \d^4 y e^{-\frac{i}{\hbar}y\cdot p} \left\langle : \left[\overline{\rho}(x_1)\left(-i\hbar \overleftarrow{\slashed{\partial}}_x+m\right)\right]_\beta\psi_\alpha (x_2)-\overline{\psi}_\beta(x_1) \Big[\left(i\hbar \slashed{\partial}_x+m\right)\rho(x_2)\Big]_\alpha :\right\rangle\;,\label{eq:def_C}
\end{eqnarray}
\end{subequations}
where
\begin{equation}
\rho(x) \coloneqq - \frac{1}{\hbar} \, \frac{\partial \mathcal{L}_{\text{int}}}{\partial \overline{\psi}(x)}\;,
\end{equation}
with the interaction Lagrangian $\mathcal{L}_{\text{int}}$ of the
theory under consideration,
and where $\overline{\psi}(x) \coloneqq
\psi^\dagger(x) \gamma^0$, $\overline{\rho}(x) \coloneqq \rho^\dagger (x) \gamma^0$
are the Dirac adjoints of
the fermion field $\psi(x)$ and
the source term $\rho(x)$. Note the
formal similarity between Eqs.\ (\ref{eq:mass_boltz_KB_0}) and
(\ref{eq:mass_boltz_GLW}).

Similar to
Eqs.\ (\ref{eq:massshell-wig}) and (\ref{eq:boltzmann-wig}), Eq.\ \eqref{eq:mass_shell} is a mass-shell equation for the Wigner function, while Eq.\ \eqref{eq:coll} represents a Boltzmann-type equation.
Again, off-shell terms in Eq.\ \eqref{eq:coll} can be shown to
cancel~\cite{Weickgenannt:2021cuo}, such that we have
\begin{equation}
    p \cdot \partial_x W_{\text{on-shell}}(x,p)=\mathcal{C}_{\text{on-shell}}(x,p)\;,\label{eq:coll_onshell}
\end{equation}
where the Wigner function and collision terms are decomposed as
\begin{subequations}\label{eq:def_onshell_terms}
\begin{eqnarray}
\label{eq:def_W_onshell}
    W(x,p)&=& 4\pi m\hbar\,\delta(p^2-m^2)\, W_{\text{on-shell}}(x,p)+W_{\text{off-shell}}(x,p)\;,\\
    \mathcal{C}(x,p)&=& 4\pi m \hbar\,\delta(p^2-m^2)\, \mathcal{C}_{\text{on-shell}}(x,p)+\mathcal{C}_{\text{off-shell}}(x,p)\;.\label{eq:def_C_onshell}
\end{eqnarray}
\end{subequations}
Note that the prefactor is chosen such that the usual momentum-space measure is recovered for the on-shell terms, as is discussed in App.\ \ref{EMTensor}.
This also implies that the $\hbar$ in the prefactor
of Eqs.\ (\ref{eq:def_W_onshell}), (\ref{eq:def_C_onshell}) does not participate in the $\hbar-$power counting.

\subsection{Single-particle distribution function in extended phase space}

Our goal is to derive
a Boltzmann equation for
the single-particle distribution
function $f(x,p,\ms)$ in
extended phase space
from the Boltzmann-type equations
(\ref{eq:boltzmann-wig}) and
(\ref{eq:coll}), respectively.
To this end, we need to establish
a relation between
$f(x,p,\ms)$ and
$G^<(x,p)$ or $W(x,p)$, respectively.
Since we work in the semi-classical expansion, we can do this order by order in $\hbar$. We work up
to order $\mathcal{O}(\hbar^2)$ in
the equation of motion (\ref{eq:kb-main}) for
the Wigner function. Since the
collision term $I_{\text{coll}}$ is
already of order $\mathcal{O}(\hbar)$,
cf.\ Eq.\ (\ref{eq:collision_term}),
for the computation of the latter
we only need to determine
$f(x,p,\ms)$ and
$G^<(x,p)$ or $W(x,p)$ up to first order
in $\hbar$.

First of all, following Refs.\ \citep{Weickgenannt:2020aaf,Weickgenannt:2021cuo} we introduce a scalar
single-particle distribution function in extended phase space,
\begin{equation} \label{eq:f}
\mathfrak{f}(x,p,\ms) \coloneqq \frac{1}{2} \left[ \mathcal{F}(x,p) - \ms \cdot \mathcal{A}(x,p) \right]\;,
\end{equation}
and analogously
\begin{equation} \label{eq:fbar}
\bar{\mathfrak{f}}(x,p,\ms) \coloneqq \frac{1}{2} \left[ \bar{\mathcal{F}} (x,p) - \ms \cdot \bar{\mathcal{A}}(x,p) \right]\;.
\end{equation}
Note that 
$\mathfrak{f}(x,p,\ms)$ and
$\bar{\mathfrak{f}}(x,p,\ms)$
are distribution-valued and thus not identical, but proportional, to $f(x,p,\ms)$
and $\bar{f}(x,p,\ms)$, respectively.

We now expand all
quantities in powers of $\hbar$,
e.g.,
\begin{equation} \label{eq:expansion_f}
\mathfrak{f}(x,p,\ms) =
\mathfrak{f}^{(0)}(x,p) + \hbar\,
\mathfrak{f}^{(1)}(x,p,\ms) + \hbar^2
\mathfrak{f}^{(2)}(x,p,\ms) + 
\mathcal{O}(\hbar^3)\ldots\;,
\end{equation}
and similarly for all other quantities.
Here we have assumed that
spin effects enter
$\mathfrak{f}$ first at order 
$\mathcal{O}(\hbar)$,
cf.\ the discussion in
Refs.\ \cite{Weickgenannt:2020aaf,Weickgenannt:2021cuo}.

\subsubsection{Zeroth order in $\hbar$}

Since the collision term 
$I_{\text{coll}}$ is
already of first order in $\hbar$,
cf.\ Eq.\ (\ref{eq:collision_term}),
at zeroth order in $\hbar$ we obtain
from Eqs.\ (\ref{eq:scalar_real}) -- (\ref{eq:axialvector_real})
\begin{subequations}
\begin{align}
p^{\mu}\mathcal{V}_{\mu}^{(0)}-m\mathcal{F}^{(0)} & =  0\;,\label{eq:scalar_real_0} \\
\mathcal{P}^{(0)} & =  0\;,\label{eq:pseudoscalar_real_0} \\
p_{\mu}\mathcal{F}^{(0)}-m\mathcal{V}_{\mu}^{(0)}& =  0\;,\label{eq:vector_real_0}\\
\frac{1}{2}\, \epsilon_{\mu\nu \rho\sigma}p^\nu\mathcal{S}^{(0)\rho\sigma}
+m\mathcal{A}_{\mu}^{(0)} & =  0\;.\label{eq:axialvector_real_0}
\end{align}
\end{subequations}
In order to proceed, we make the additional assumption
that $\mathcal{A}_\mu = \hbar \, \mathcal{A}_\mu^{(1)} + \mathcal{O}(\hbar^2)$, i.e., that $\mathcal{A}^{(0)}_\mu =0$,
see Refs.\ \citep{Weickgenannt:2020aaf,Weickgenannt:2021cuo}. This can be justified by assuming that
polarization effects are at most generated dynamically within the system, but are not induced already from the 
outset by, e.g., external fields. In this case, Eqs.~(\ref{eq:pseudoscalar_real_0}) -- (\ref{eq:axialvector_real_0}) imply that
\begin{equation} \label{eq:orderzero}
\mathcal{P}^{(0)} = 0\;, \quad \mathcal{V}_\mu^{(0)} = \frac{p_\mu}{m} \, \mathcal{F}^{(0)}\;, \quad
\mathcal{A}_\mu^{(0)} = 0\;, \quad \mathcal{S}_{\mu \nu}^{(0)} =0\;,
\end{equation}
and the only independent Lorentz component of the Wigner function at order $\mathcal{O}(\hbar^0)$ 
is $\mathcal{F}^{(0)}$. Analogous relations hold for 
the Clifford components $\bar{\mathcal{P}}^{(0)}$, $\bar{\mathcal{V}}^{(0)}_\mu$,
$\bar{\mathcal{A}}^{(0)}_\mu$, and $\bar{\mathcal{S}}^{(0)}_{\mu \nu}$ of the Wigner function $G^>$.

Furthermore, we conclude by combining Eqs.~(\ref{eq:scalar_real_0}) and (\ref{eq:vector_real_0}) that
\begin{equation}
p^2 \mathcal{F}^{(0)} = m p^\mu \mathcal{V}_\mu^{(0)} = m^2 \mathcal{F}^{(0)} \;,
\end{equation}
i.e., $\mathcal{F}^{(0)}$ is on-shell, and thus also $\mathcal{V}^{(0)}_\mu$ is on-shell. Similar arguments
apply to $\bar{\mathcal{F}}^{(0)}$ and $\bar{\mathcal{V}}^{(0)}_\mu$.

Inserting Eq.~(\ref{eq:orderzero}) into Eq.\ (\ref{eq:wigner-decomp}), we immediately derive
\begin{equation}
G^{<(0)}(x,p) \equiv W^{(0)}(x,p) =  \frac{\slashed{p}+m}{4m}\, \mathcal{F}^{(0)}(x,p)
\equiv \frac{1}{2}\, \Lambda^{+}(p) 
\mathcal{F}^{(0)}(x,p)\;,
\end{equation}
and similarly 
\begin{equation}
G^{>(0)}(x,p) = \frac{1}{2}\, \Lambda^{+}(p)\, \bar{\mathcal{F}}^{(0)}(x,p)\;,
\end{equation}
where we used the projector 
\begin{equation}
\label{eq:projector}
\Lambda^+(p) \coloneqq \frac{\slashed{p}+m}{2m}
\end{equation}
onto positive-energy states. 
On the other hand, because $\mathcal{A}^{(0)}_\mu =0$, to order $\mathcal{O}(\hbar^0)$ Eq.\ (\ref{eq:f}) reads
\begin{equation} \label{eq:f_frakf}
\mathfrak{f}^{(0)}(x,p) = \frac{1}{2} \mathcal{F}^{(0)}(x,p)\;,
\end{equation}
and similarly Eq.~(\ref{eq:fbar}) reads
\begin{equation}
\bar{\mathfrak{f}}^{(0)}(x,p) = \frac{1}{2} \bar{\mathcal{F}}^{(0)}(x,p)\;.
\end{equation}
Since $\mathcal{F}^{(0)}$ and $\bar{\mathcal{F}}^{(0)}$ are on-shell, we can factor out a 
mass-shell delta function from $\mathfrak{f}^{(0)}$ and $\bar{\mathfrak{f}}^{(0)}$,
\begin{equation} \label{eq:frakf_onshell}
\mathfrak{f}^{(0)}(x,p) \coloneqq 4 \pi m \hbar \,\delta(p^2-m^2) f^{(0)}(x,p)\;, \quad 
\bar{\mathfrak{f}}^{(0)}(x,p) \coloneqq 4 \pi m\hbar\, \delta(p^2-m^2) \bar{f}^{(0)}(x,p)\;,
\end{equation}
where the prefactor $4 \pi m \hbar$ is introduced to make $f^{(0)}$ and $\bar{f}^{(0)}$ dimensionless and to ensure
that $f^{(0)}$ converges
to the Fermi distribution function
in the thermodynamical limit (see App.\ \ref{EMTensor}),
while $\bar{f}^{(0)}$
becomes a Pauli-blocking
factor $1 - f^{(0)}$. Note
that the factor $2 \pi \hbar$ in the prefactor 
in Eq.\ (\ref{eq:frakf_onshell}) does
not contribute to the power-counting, since it does not appear with either
a gradient or a spin-related quantity. It merely serves to cancel a $(2 \pi \hbar)^{-1}$ from the four-dimensional
momentum-space measure, cf.\ App.\ \ref{EMTensor}.
We thus obtain the final expressions for $G^{\lessgtr(0)}$ and $W^{(0)}$ in terms of $f^{(0)}$ and $\bar{f}^{(0)}$,
\begin{subequations} \label{eq:Gsmallerlarger_orderzero}
\begin{align}
G^{<(0)}(x,p) \equiv W^{(0)}(x,p) & = 4\pi m \hbar\, \delta(p^2-m^2) \Lambda^+(p) f^{(0)}(x,p)\;, \\
G^{>(0)}(x,p) & = 4 \pi m \hbar\, \delta(p^2-m^2) \Lambda^+(p) \bar{f}^{(0)}(x,p)\;.
\end{align}
\end{subequations}
Following the discussion in Sec.\ V of Ref.~\citep{Sheng:2021kfc}, we
expect $\bar{f}^{(0)} \equiv 1- f^{(0)} $ (note the sign difference in our definition of $G^<$ with respect to that
reference), i.e., it is sufficient to know $f^{(0)}$ to reconstruct both
$G^{<(0)}$ and $G^{>(0)}$. 

\subsubsection{First order in $\hbar$}

To first order in $\hbar$, we
employ Eqs.\ (\ref{eq:scalar_real}) -- (\ref{eq:tensor_real}), as
well as Eq.\ (\ref{eq:pseudoscalar_im}), which read
\begin{subequations}
\begin{align}
p^{\mu}\mathcal{V}_{\mu}^{(1)}-m\mathcal{F}^{(1)}
& =  \mathrm{Re\, Tr}\left(I_{\mathrm{coll}}^{(1)}\right)\;,\label{eq:scalar_real_1} \\
\frac{1}{2}\, \partial^{\mu}_{x}\mathcal{A}^{(0)}_{\mu}+m\mathcal{P}^{(1)}
& =  \mathrm{Re\, Tr}\left(i\gamma_{5}I_{\mathrm{coll}}^{(1)}\right)\;,
\label{eq:pseudoscalar_real_1} \\
\frac{1}{2}\, \partial_{x}^{\nu}\mathcal{S}_{\nu\mu}^{(0)}- p_{\mu}\mathcal{F}^{(1)}
+ m\mathcal{V}_{\mu}^{(1)}
& =  - \mathrm{Re\, Tr}\left(\gamma_{\mu}I_{\mathrm{coll}}^{(1)}\right)\;,
\label{eq:vector_real_1} \\
\frac{1}{2}\, \epsilon_{\mu\nu\rho\sigma}p^{\nu}
\mathcal{S}^{(1)\rho\sigma} +m\mathcal{A}_{\mu}^{(1)}
& =  \mathrm{Re\,Tr}\left(\gamma^{5}\gamma_{\mu}I_{\mathrm{coll}}^{(1)}\right)\;,
\label{eq:axialvector_real_1} \\
\frac{1}{2}\,\partial_{x[\mu}\mathcal{V}_{\nu]}^{(0)}
-\epsilon_{\mu\nu\rho\sigma}p^{\rho}\mathcal{A}^{(1)\sigma}
-m\mathcal{S}_{\mu\nu}^{(1)}
& =  \mathrm{Re\, Tr}\left(\sigma_{\mu\nu}I_{\mathrm{coll}}^{(1)}\right)\;,
\label{eq:tensor_real_1} \\
p^\mu \mathcal{A}^{(1)}_\mu
& =  \mathrm{Im\, Tr}\left(-i \gamma^{5}I_{\mathrm{coll}}^{(1)}\right)\;.
\label{eq:pseudoscalar_im_1}
\end{align}
\end{subequations}
Because $G^{>(1)} \equiv - G^{<(1)}$
\citep{Sheng:2021kfc}, we have $\bar{\mathcal{F}}^{(1)} \equiv -\mathcal{F}^{(1)}$, 
$\bar{\mathcal{P}}^{(1)} \equiv - \mathcal{P}^{(1)}$, $\bar{\mathcal{V}}^{(1)}_\mu \equiv - \mathcal{V}^{(1)}_\mu$,
 $\bar{\mathcal{A}}^{(1)}_\mu \equiv - \mathcal{A}^{(1)}_\mu$, and
 $\bar{\mathcal{S}}^{(1)}_{\mu \nu}\equiv - \mathcal{S}^{(1)}_{\mu \nu}$, respectively.
We also have $\bar{f}^{(1)} = - f^{(1)}$.

With Eq.~(\ref{eq:orderzero}) we conclude from Eq.~(\ref{eq:pseudoscalar_real_1}) that
\begin{equation} \label{eq:P_orderone}
\mathcal{P}^{(1)} = \frac{1}{m} \, \mathrm{Re}\mathrm{Tr} \left(i \gamma_5 I_{\textrm{coll}}^{(1)} \right)
= \mathcal{O}(G_c^2)\;.
\end{equation}
Since we are ultimately interested in computing the collision term, we may make further approximations.
Namely, when using the Clifford decomposition for $G^{<(1)}$ in
the collision term, a term
such as in Eq.\ (\ref{eq:P_orderone})
gives an overall contribution of order
$\mathcal{O}(G_c^4)$, which can be neglected. We can therefore safely assume that $\mathcal{P}^{(1)} \simeq 0$.

Equation (\ref{eq:vector_real_1}) together with Eq.~(\ref{eq:orderzero}) yields
\begin{equation} \label{eq:V_orderone}
\mathcal{V}^{(1)}_\mu = \frac{p_\mu}{m}\, \mathcal{F}^{(1)} - \frac{1}{m} \, \mathrm{Re}\mathrm{Tr} \left( \gamma_\mu
 I_{\textrm{coll}}^{(1)} \right) \simeq \frac{p_\mu}{m}\, \mathcal{F}^{(1)}  \;.
\end{equation}
In the last step, we have again neglected the contribution of order $\mathcal{O}(G_c^2)$ 
from the collision term, since this gives rise to an overall contribution of order
$\mathcal{O}(G_c^4)$ when inserting the Clifford decomposition for $G^{<(1)}$ into
the collision term.

Finally, Eq.~(\ref{eq:tensor_real_1}) together with Eq.~(\ref{eq:orderzero}) gives
\begin{equation} \label{eq:S_orderone}
\mathcal{S}^{(1)}_{\mu \nu} = -\frac{1}{m} \epsilon_{\mu\nu\rho\sigma} p^\rho \mathcal{A}^{(1) \sigma}
+ \frac{1}{2m} \partial_{x, [\mu} \mathcal{V}^{(0)}_{\nu]} - \frac{1}{m} \, \mathrm{Re}\mathrm{Tr} \left( \sigma_{\mu \nu}
 I_{\textrm{coll}}^{(1)} \right) \simeq  -\frac{1}{m} \epsilon_{\mu\nu\rho\sigma} p^\rho \mathcal{A}^{(1) \sigma}
- \frac{1}{2m^2} p_{[\mu} \partial_{x, \nu]} \mathcal{F}^{(0)}\;.
\end{equation}
By the same arguments as above, the contribution from the collision term can be neglected. 

Furthermore, combining Eqs.~(\ref{eq:scalar_real_1}) and (\ref{eq:vector_real_1}) and using Eq.\ (\ref{eq:orderzero}),
we conclude that
\begin{equation}
p^2 \mathcal{F}^{(1)} = m^2 \mathcal{F}^{(1)} + \mathrm{Re}\mathrm{Tr}\left[ (\slashed{p}+m)
I_{\textrm{coll}}^{(1)}\right]  \simeq m^2 \mathcal{F}^{(1)}\;,
\end{equation}
i.e., $\mathcal{F}^{(1)}$ is on-shell up to collisional contributions, which can be neglected when inserting
the Clifford decomposition for $G^{<(1)}$ into the collision term. 

We can also derive a mass-shell condition for $\mathcal{A}^{(1)}_\mu$. To this end, multiply 
Eq.\ (\ref{eq:tensor_real_1}) by $p_\lambda \epsilon^{\mu \nu \lambda \tau}$ and insert Eqs.\ (\ref{eq:axialvector_real_1})
and (\ref{eq:pseudoscalar_im_1}). Using Eq.~(\ref{eq:orderzero}), this results in
\begin{equation}
p^2 \mathcal{A}^{(1)\tau} = m^2 \mathcal{A}^{(1)\tau} - \mathrm{Re}\mathrm{Tr}\left[ \gamma_5 \gamma^\tau
(\slashed{p}+m) I_\textrm{coll}^{(1)} \right] \simeq m^2 \mathcal{A}^{(1)\tau} \;,
\end{equation}
i.e., $\mathcal{A}^{(1)}_\mu$ is on-shell up to collisional contributions, which can be neglected when inserting
the Clifford decomposition for $G^{<(1)}$ into the collision term. 
On account of Eqs.\ (\ref{eq:f}) and (\ref{eq:fbar}), also
$\mathfrak{f}^{(1)}$ and $\bar{\mathfrak{f}}^{(1)}$ are on-shell up to collisional contributions.

We can now insert Eqs.~(\ref{eq:P_orderone}), (\ref{eq:V_orderone}), and (\ref{eq:S_orderone}) into the
Clifford decomposition (\ref{eq:wigner-decomp}) of the Wigner function $G^{<(1)}(x,p) \equiv W^{(1)}(x,p)$ and obtain
up to collisional contributions
\begin{equation} \label{eq:Gonesplit}
G^{<(1)} (x,p) \equiv W^{(1)} (x,p) \simeq G_{\textrm{qc}}^{<(1)} (x,p)+ G_\nabla^{<(1)}(x,p)\;,
\end{equation}
with
\begin{subequations}
\begin{align}
G_{\textrm{qc}}^{<(1)} (x,p) & = \frac{1}{2} \, \Lambda^+(p) \left[ \mathcal{F}^{(1)}(x,p) + \gamma_5 \gamma \cdot
\mathcal{A}^{(1)}(x,p) \right]\;, \label{eq:qp_one} \\
G_\nabla^{<(1)} (x,p) & =  \frac{1}{8m^2} \sigma^{\mu \nu} p_\nu \partial^x_\mu \mathcal{F}^{(0)}(x,p)\;,
\label{eq:gradient_one} \end{align} 
\end{subequations}
where in the terminology
of Ref.\ \cite{Sheng:2021kfc} the subscript ``qc'' 
denotes the so-called quasi-classical contribution,
while the subscript ``$\nabla$''
denotes the gradient contribution. 
The gradient contribution $G_\nabla^{<(1)}$ can be immediately expressed in terms of
$f^{(0)}(x,p)$ using Eqs.\ (\ref{eq:f_frakf}) and (\ref{eq:frakf_onshell}),
\begin{equation} \label{eq:Gsmallernabla1}
G_\nabla^{<(1)}(x,p) = \frac{\pi \hbar}{m}\, \delta(p^2-m^2) \, \sigma^{\mu \nu} p_\nu  \partial^x_\mu f^{(0)}(x,p)\;,
\end{equation} 
and similarly
\begin{equation} \label{eq:Glargernabla1}
G_\nabla^{>(1)}(x,p) = \frac{\pi \hbar}{m}\, \delta(p^2-m^2) \, \sigma^{\mu \nu} p_\nu  \partial^x_\mu \bar{f}^{(0)}(x,p)\;.
\end{equation} 
Both the
quasi-classical and the gradient parts are ``quasi-particle''
contributions in the sense that they are on the mass
shell.
Since $\partial^x_\mu \bar{f}^{(0)} \equiv - \partial^x_\mu f^{(0)}$, we confirm that
$G_\nabla^{>(1)} \equiv - G_\nabla^{<(1)}$.

In order to express the quasi-classical contribution
$G_{\textrm{qc}}^{<(1)}$ in terms of $f^{(1)}$, we need to 
invert Eq.\ (\ref{eq:f}). 
This can be done with the relations
\citep{Weickgenannt:2020aaf,Weickgenannt:2021cuo}
\begin{equation}
\mathcal{F}(x,p) = \int \d S(p)\, \mathfrak{f}(x,p,\ms)\;, \quad
\mathcal{A}_\mu(x,p) = \int \d S(p)\, \ms_\mu \, \mathfrak{f}(x,p,\ms)\;.
\end{equation}
Using the definition (\ref{eq:f}) of $\mathfrak{f}(x,p,\ms)$ and the identities (\ref{eq:intspin}),
one proves that the first relation is strictly valid 
to all orders in $\hbar$,
while the second relation is
strictly valid up to order 
$\mathcal{O}(\hbar^0)$ and
valid up to order $\mathcal{O}(\hbar)$ if collisional contributions are disregarded, such that
$p \cdot \mathcal{A}^{(1)} \simeq 0$,
cf.\ Eq.\ (\ref{eq:pseudoscalar_im_1}).

Inserting the first-order expressions $\mathcal{F}^{(1)}$ and $\mathcal{A}^{(1)}_\mu$ into Eq.\ (\ref{eq:qp_one}),
and factoring out a mass-shell delta function from $\mathfrak{f}^{(1)}(x,p,\ms)$,
\begin{equation} \label{eq:f1}
\mathfrak{f}^{(1)}(x,p,\ms) = 
4\pi m\hbar\, \delta(p^2-m^2) f^{(1)}(x,p,\ms)\;,
\end{equation}
we then obtain
\begin{equation} \label{eq:Gsmallerqp1}
G_{\textrm{qc}}^{<(1)}(x,p) = 4\pi m\hbar\, \delta(p^2-m^2)  \int \d S(p) \, h(p,\ms)\,
f^{(1)}(x,p,\ms)\;,
\end{equation}
where we have used Eq.\ (\ref{eq:def_h}) and the relation
\begin{equation}
\label{eq:useful_id}
(\dblone + \gamma_5 \slashed{\ms}) \Lambda^+(p) = \Lambda^+(p) (\dblone + \gamma_5 \slashed{\ms})\;,
\end{equation}
which holds since $p \cdot \ms = 0$.
Similarly,
\begin{equation} \label{eq:Glargerqp1}
G_{\textrm{qc}}^{>(1)} (x,p) = 4 \pi m\hbar \,\delta(p^2-m^2)  \int \d S(p) \, h(p,\ms) \,
\bar{f}^{(1)}(x,p,\ms)\;,
\end{equation}
Since $\bar{f}^{(1)} \equiv - f^{(1)}$, we confirm that also $G_{\textrm{qc}}^{>(1)} = - G_{\textrm{qc}}^{<(1)}$. 

To summarize the results of
this subsection, we
have expressed the Wigner functions $G^{\lessgtr}(x,p)$ and $W(x,p)$
up to first order in $\hbar$
in terms of $f^{(0)}$, $\bar{f}^{(0)}$, $f^{(1)}$, and $\bar{f}^{(1)}$,
cf.\ Eqs.\ (\ref{eq:Gsmallerlarger_orderzero})
and (\ref{eq:Gonesplit}) with Eqs.\ (\ref{eq:Gsmallernabla1}),
(\ref{eq:Glargernabla1}),
(\ref{eq:Gsmallerqp1}), and (\ref{eq:Glargerqp1}).
For further use, we note that, because
of Eq.\ (\ref{eq:intspin}), up
to first order in $\hbar$
we may write
\begin{equation}
\label{eq:Gidentity}
G^<(x,p) \equiv W(x,p)
= 4 \pi m \hbar \, \delta(p^2-m^2) 
\int \d S(p) \, h(p,\ms)\,
f(x,p,\ms) + G^<_\nabla(x,p)\;.
\end{equation}

\section{Interaction Lagrangian}
\label{sec:Lagrangian}

For the interaction
$\mathcal{L}_{\text{int}}$ between
fermions we consider one-boson exchange. Assuming the interaction range to be much
smaller than all other scales in the problem, we can integrate out
the boson fields and reduce
the interaction to a four-fermion vertex, similar to the NJL model \citep{Nambu:1961tp,Nambu:1961fr}.
Thus, the interaction Lagrangian reads
\begin{equation}
\mathcal{L}_{\text{int}}=
\sum_{c} G_{c}\,
\sum_{a,b} \left[\overline{\psi}(x) \Gamma^{(c)}_{a}\psi(x)\right]g^{ab}_{(c)}
\left[\overline{\psi}(x) \Gamma^{(c)}_{b}\psi(x)\right]\; .
\label{eq:int_Lagrangian}
\end{equation}
The sum over $c$ runs over all possible interaction channels, e.g., scalar ($c=S$), pseudo-scalar
$(c=P)$, vector ($c=V$), axial-vector $(c=A)$, and tensor $(c=T)$ channel. The matrices
$\Gamma^{(c)}_{a}, \Gamma^{(c)}_{b}$ represent the corresponding elements of the Clifford algebra:
$\Gamma^{(S)}_a = \dblone$, $\Gamma^{(P)}_a = -i \gamma_5$, $\Gamma^{(V)}_a = \gamma^\mu$,
$\Gamma^{(A)}_a = \gamma_5 \gamma^\mu$, and $\Gamma^{(T)}_a = \sigma^{\mu \nu}$. In the scalar and pseudo-scalar channels,
the sum over $a$ and $b$ only contains one element and $g^{ab}_{(S,P)} \equiv 1$.
In the vector and axial-vector channels,
$a$ and $b$ are Lorentz indices, which are summed over with $g^{ab}_{(V,A)} \equiv g^{\mu \nu}$.
In the tensor channel, $a$ and $b$ represent pairs of (unequal) Lorentz indices, say $a = (\mu \nu)$,
$b=(\rho\sigma)$ and $g^{ab}_{(T)} \equiv g^{\mu \rho} g^{\nu \sigma}$. Finally,
$G_c$ denotes the four-fermion coupling in channel $(c)$.

In order to simplify the notation, in the
remainder of this work we will omit the
indices $a,b$ and the metric $g^{ab}_{(c)}$, and just indicate
the particular interaction channel $c$ 
at the element $\Gamma^{(c)}$ of the
Clifford algebra, i.e., an
appropriate summation over $(a,b)$
is implied,
\begin{equation}
\sum_{a,b} \left[\overline{\psi}(x) \Gamma^{(c)}_{a}\psi(x)\right]g^{ab}_{(c)}
\left[\overline{\psi}(x) \Gamma^{(c)}_{b}\psi(x)\right]
\equiv 
\left[\overline{\psi}(x) \Gamma^{(c)}\psi(x)\right]
\left[\overline{\psi}(x) \Gamma^{(c)}\psi(x)\right]\;.
\end{equation}

\section{The nonlocal collision term in the GLW approach}\label{sec:de_groot}

In this section, we
rederive the collision term in the
GLW approach using slightly
different and more commonly used
conventions than in
Refs.\ \cite{Weickgenannt:2020aaf, Weickgenannt:2021cuo}, in order to
facilitate comparison with
results from the KB approach.
We first repeat parts of the 
derivation of the collision
term as presented in Ref.\ \cite{Weickgenannt:2021cuo} and then
focus separately on the local and the
nonlocal parts of the collision term.
Finally, we summarize our
results.

\subsection{The collision term revisited}

In the following we restrict ourselves to the particle sector, i.e., $p^0>0$ for all on-shell momenta $p^\mu$; the antiparticle sector can be derived analogously. The positive-frequency
``in''-fields are given by
\begin{equation}
    \psi_{\text{in}}(x)=\frac12 \sum_{s} \momint{p}{}{} u_s (p) \hat{a}_s (p) e^{-i p\cdot x/\hbar}\;,
\end{equation}
where $p^\mu$ is on-shell, $\d P$ was defined in Eq.\ (\ref{eq:measure_mom}), and $\hat{a}_s(p)$ annihilates a particle with momentum $p$ and spin $s$. 
In our convention (which differs
from that of Ref.\ \cite{DeGroot:1980dk}) the nonzero anticommutation relations for the creation and annihilation operators are
\begin{equation}
    \{\hat{a}_s(p),\hat{a}^\dagger_r(p')\}=(2\pi\hbar)^3 2p^0 \delta^{(3)}(\p-\p')\delta_{rs}\;,
\end{equation}
while the orthogonality and completeness relation of the basis spinors read
\begin{equation}
\bar{u}_{r,\alpha} (p) 
u_s^\alpha(p)=2m \delta_{rs}\;,\quad \sum_{r} u_r^\alpha (p) \bar{u}_{r,\beta} (p) = (\slashed{p}+m)^{\alpha}_{\;\;\beta}\;.
\end{equation}
Note that, from now on, we will not distinguish between upper and lower Dirac indices. When evaluating the expressions, it is implied that the Dirac indices are on their natural position, e.g., upper (lower) indices for spinors (adjoint spinors), and
repeated indices are simply summed over (without additional sign change).
Following Ref.\ \cite{DeGroot:1980dk}, we may cast the collision term \eqref{eq:def_C} in the following form, 
\begin{eqnarray}
    \mathcal{C}_{\alpha\beta}(x,p)&=&\frac12 \frac{1}{(2\pi\hbar)^{12}(2m)^4} \sum_{r^2,s^2} \int\d^4 x^2 \int \d^4 p^2 \int \d^4 u^2 \prescript{}{\text{in}}{\bra{p^2-\frac{u^2}{2};r^2}}\Phi_{\alpha\beta}(p)\ket{p^2+\frac{u^2}{2};s^2}_{\text{in}} \nonumber\\
    &&\hspace{3cm} \times \prod_{j=1}^2 e^{\frac{i}{\hbar}u_j\cdot x_j} \bar{u}_{s_j,\alpha_j}\left(p_j+\frac{u_j}{2}\right) W^{\alpha_j\beta_j}(x+x_j,p_j) u_{r_j,\beta_j}\left(p_j-\frac{u_j}{2}\right)\;.\label{eq:coll_1}
\end{eqnarray}
Note that the different prefactor as compared to Eq.\ (11) of Ref.\ \cite{Weickgenannt:2021cuo} is due to the different definition of the momentum-space measure. 
Here, for the sake of brevity
we used the notation of Ref.\ \cite{DeGroot:1980dk} for sums, integrals, and two-particle states,
\begin{equation}
    \sum_{r^2,s^2}\coloneqq \sum_{r_1=1}^2\sum_{r_2=1}^2\sum_{s_1=1}^2\sum_{s_2=1}^2\;,\quad \int \d^4x^2\coloneqq \int \d^4 x_1 \int \d^4 x_2\;,\quad \ket{p^2\pm \frac{u^2}{2};r^2}\coloneqq \ket{p_1\pm \frac{u_1}{2}, p_2\pm \frac{u_2}{2}; r_1,r_2}\;.
\end{equation}
The operator $\Phi$ in Eq.\ \eqref{eq:coll_1} reads
\begin{eqnarray}
    \Phi_{\alpha\beta}(p)&\coloneqq& \frac{i}{2} \frac{1}{(2\pi\hbar)^4} \int \d^4 y\, e^{-\frac{i}{\hbar} p\cdot y } :\bigg\{ \left[P_\mu,\bar{\rho}\left(\frac{y}{2}\right)\gamma^\mu\right]_\beta\psi_\alpha\left(-\frac{y}{2}\right)+m\bar{\rho}_\beta\left(\frac{y}{2}\right) \psi_\alpha\left(-\frac{y}{2}\right)\nonumber\\
    && \hspace*{3.8cm} -\bar{\psi}_\beta\left(\frac{y}{2}\right)\left[\gamma^\mu \rho\left(-\frac{y}{2}\right),P_\mu\right]_\alpha-m\bar{\psi}_\beta\left(\frac{y}{2}\right)\rho_\alpha\left(-\frac{y}{2}\right)\bigg\} :\;,
\end{eqnarray}
where $P_\mu$ denotes the momentum operator.
The calculation needed to compute the matrix element in Eq.\ (\ref{eq:coll_1}) has been discussed in Refs. \cite{DeGroot:1980dk, Weickgenannt:2021cuo}, but we will nonetheless show the main steps. After the insertion of a complete set of ``out''-states and performing the $y$-integration, one arrives at
\begin{eqnarray}
    &&\prescript{}{\text{in}}{\bra{p^2-\frac{u^2}{2};r^2}}\Phi_{\alpha\beta}(p)\ket{p^2+\frac{u^2}{2};s^2}_{\text{in}}=\frac{i}{4} \sum_{r'} \momint{p'}{}{} \delta^{(4)}(p+p'-p_1-p_2) \nonumber\\
    &&\times\Bigg[ \prescript{}{\text{out}}{\bigg\langle p';r'\bigg|}\psi_\alpha(0)\ket{p^2+\frac{u^2}{2};s^2}_{\text{in}} \prescript{}{\text{in}}{\bra{p^2-\frac{u^2}{2};r^2}}:\bar{\rho}_{\alpha'}(0):\bigg|p';r'\bigg\rangle_{\text{out}}
    \left(\slashed{p}-\frac{\slashed{u}_1+\slashed{u}_2}{2}+m\right)_{\alpha'\beta}\nonumber\\
    &&-\left( \slashed{p}+\frac{\slashed{u}_1+\slashed{u}_2}{2}+m\right)_{\alpha\alpha'} \prescript{}{\text{out}}{\bigg\langle p';r'\bigg|}:\rho_{\alpha'}(0):\ket{p^2+\frac{u^2}{2};s^2}_{\text{in}} \prescript{}{\text{in}}{\bra{p^2-\frac{u^2}{2};r^2}}\overline{\psi}_\beta(0)\bigg| p';r'\bigg\rangle_{\text{out}}\Bigg]\;.\label{eq:exp:Phi_1}
\end{eqnarray}
Next we evaluate the matrix elements of the fields $\psi,\overline{\psi}$. These fields can be written in terms of the ``in''-fields as
\begin{equation}
    \psi(0)=\psi_{\text{in}}(0)+\int \d^4 x\,S_R(-x)\rho(x)\;,
\end{equation}
where $S_R$ is the retarded fermion propagator, whose Fourier transform is given by $\tilde{S}_R(p)=-(1/\hbar)(\slashed{p}+m)\tilde{G}(p)$, with the scalar propagator $\tilde{G}(p)\coloneqq -\hbar^2/(p^2-m^2+i\epsilon p^0)$. Using the orthogonality relation of the momentum eigenstates,
\begin{equation}
    \prescript{}{\text{in}}{\braket{p^2;r^2}{p'^2;r'^2}}_{\text{in}}=[2(2\pi\hbar)^3]^2p_1^0 p_2^0\left[\delta^{(3)}(\p_1-\p_1')\delta_{r_1 r_1'}\delta^{(3)}(\p_2-\p_2')\delta_{r_2 r_2'}-\delta^{(3)}(\p_1-\p_2')\delta_{r_1 r_2'}\delta^{(3)}(\p_2-\p_1')\delta_{r_2 r_1'}\right]\;,\label{eq:orthogonality_states}
\end{equation}
in conjunction with the fact that for one-particle states ``in''- and ``out''-states are identical, we find
\begin{eqnarray}
    \prescript{}{\text{out}}{\bigg\langle p';r'\bigg|}\psi(0)\ket{p^2+\frac{u^2}{2};s^2}_{\text{in}}&=&2(2\pi\hbar)^3 p'^0\left[u_{s_1}\left(p_1+\frac{u_1}{2}\right)\delta^{(3)}\left(\p'-\p_2-\frac{\mathbf{u}_2}{2}\right)\delta_{r' s_2}-(1\leftrightarrow 2)\right]\nonumber\\
    &&+\tilde{S}_R\left(p_1+\frac{u_1}{2}+p_2+\frac{u_2}{2}-p'\right)\prescript{}{\text{out}}{\bigg\langle p';r'\bigg|}:\rho(0):\ket{p^2+\frac{u^2}{2};s^2}_{\text{in}}\;.\label{eq:psi_exp}
\end{eqnarray}
Employing this relation and using the projector $\Lambda^+(p)$
defined in Eq.\ (\ref{eq:projector}), we can rewrite Eq.\ \eqref{eq:exp:Phi_1} as
\begin{eqnarray}
    &&\prescript{}{\text{in}}{\bra{p^2-\frac{u^2}{2};r^2}}\Phi_{\alpha\beta}(p)\ket{p^2+\frac{u^2}{2};s^2}_{\text{in}}\nonumber\\
    &=&im  \bigg( \bigg\{u_{s_1,\alpha}\left(p_1+\frac{u_1}{2}\right)\delta^{(3)}\left(\p-\p_1+\frac{\u_2}{2}\right)\delta\left[p^0+\sqrt{\left(\p_2+\frac{\u_2}{2}\right)^2+m^2}-p_1^0-p_2^0\right]\nonumber\\
    &&\times\prescript{}{\text{in}}{\bra{p^2-\frac{u^2}{2};r^2}}:\bar{\rho}_{\alpha'}(0):\ket{p_2+\frac{u_2}{2};s_2}_{\text{out}}\Lambda^{+}_{\alpha'\beta}\left(p-\frac{u_1+u_2}{2}\right)+(1\leftrightarrow 2)\bigg\}\nonumber\\
    &&- \bigg\{\bar{u}_{r_1,\beta}\left(p_1-\frac{u_1}{2}\right)\delta^{(3)}\left(\p-\p_1-\frac{\u_2}{2}\right)\delta\left[p^0+\sqrt{\left(\p_2-\frac{\u_2}{2}\right)^2+m^2}-p_1^0-p_2^0\right]\nonumber\\
    &&\times \Lambda^{+}_{\alpha\alpha'}\left(p+\frac{u_1+u_2}{2}\right) \prescript{}{\text{out}}{\bra{p_2-\frac{u_2}{2};r_2}}:\rho_{\alpha'}(0):\ket{p^2+\frac{u^2}{2};s^2}_{\text{in}}+(1\leftrightarrow 2)\bigg\}\nonumber\\
    &&-\frac{m}{\hbar} \sum_{r'}\momint{p'}{}{} \delta^{(4)}(p+p'-p_1-p_2)\left[\tilde{G}\left(p+\frac{u_1+u_2}{2}\right)-\tilde{G}^*\left(p-\frac{u_1+u_2}{2}\right)\right]\Lambda^{+}_{\alpha\alpha'}\left(p+\frac{u_1+u_2}{2}\right) 
    \nonumber\\
    &&\times\prescript{}{\text{out}}{\bigg\langle p';r'\bigg|}:\rho_{\alpha'}(0):\ket{p^2+\frac{u^2}{2};s^2}_{\text{in}} \prescript{}{\text{in}}{\bra{p^2-\frac{u^2}{2};r^2}}:\bar{\rho}_{\beta'}(0):\bigg|p';r'\bigg\rangle_{\text{out}}\Lambda^{+}_{\beta'\beta}\left(p-\frac{u_1+u_2}{2}\right)\bigg)\;.\label{eq:exp:Phi_2}
\end{eqnarray}
Here we used that $\ket{p_1,p_2;s_1,s_2}=-\ket{p_2,p_1;s_2,s_1}$.
Next, we have to make use of the relation between the source terms and scattering-matrix elements, which is given by \cite{DeGroot:1980dk}
\begin{eqnarray}
   \hspace*{-15pt} (2\pi\hbar)^4 \delta^{(4)}(p+p'-p_1-p_2)\prescript{}{\text{in}}{\bra{p,p';r,r'}}\hat
{t}\ket{p^2;r^2}_{\text{in}}&\coloneqq&-i\prescript{}{\text{in}}{\bra{p,p';r,r'}}(\hat{S}-\mathds{1})\ket{p^2;r^2}_{\text{in}}\nonumber\\
    &=&-(2\pi\hbar)^4 \delta^{(4)}(p+p'-p_1-p_2) \bar{u}_r(p)\prescript{}{\text{out}}{\bra{p';r'}}\rho(0)\ket{p^2;r^2}_{\text{in}}\;,\label{eq:rel_t_rho}
\end{eqnarray}
where $\hat{S}$ is the scattering matrix. We now split the transfer matrix into real and imaginary parts and make use of the optical theorem \cite{DeGroot:1980dk},
\begin{eqnarray}
     &&\frac{i}{2} \prescript{}{\text{in}}{\bra{p,p';r,r'}}(\hat{t}-\hat{t}^\dagger)\ket{p^2;r^2}_{\text{in}}\nonumber\\
     &=&-\frac{(2\pi\hbar)^4}{16}\delta^{(4)}(p+p'-p_1-p_2)\sum_{s^2} \momint{q}{}{1} \d Q_2 \prescript{}{\text{in}}{\bra{p,p';r,r'}}\hat{t}\ket{q^2;s^2}_{\text{in}} \prescript{}{\text{in}}{\bra{q^2;s^2}}\hat{t}^\dagger \ket{p^2;r^2}_{\text{in}} \;.
\end{eqnarray}
In the remainder of this paper, we will neglect the real parts of the transfer matrix and defer a more detailed discussion of the latter to a subsequent work
\cite{Wagnertoappear}.

We now employ the optical theorem and the following expression of the transfer-matrix elements
\begin{subequations}
\label{matel}
\begin{eqnarray}
\bra{p,p';r,r'}\hat{t}\ket{p^2;r^2}&=&\frac{1}{\hbar}\bra{p,p';r,r'}:\mathcal{L}_{\text{int}}(0):\ket{p^2;r^2}\nonumber \\
& = &\bar{u}_{r,\alpha}(p) \bar{u}_{r',\alpha'}(p') u_{r_1,\alpha_1}(p_1) u_{r_2,\alpha_2}(p_2) M^{\alpha\alpha'\alpha_1\alpha_2}(p,p',p_1,p_2)\;,\\
\bra{p^2;r^2}\hat{t}^\dagger\ket{p,p';r,r'}&=&\frac{1}{\hbar}\bra{p^2;r^2}:\mathcal{L}_{\text{int}}^\dagger(0):\ket{p,p';r,r'}
\nonumber \\
&=& \bar{u}_{r_1,\alpha_1}(p_1) \bar{u}_{r_2,\alpha_2}(p_2) u_{r,\alpha}(p) u_{r',\alpha'}(p')  \overline{M}^{\alpha_1\alpha_2\alpha\alpha'}(p_1,p_2,p,p')
\;,
\end{eqnarray}
\end{subequations}
 where $M$ is the tree-level vertex function of the theory in momentum space, i.e, the Fourier transform of the fourth functional derivative of the classical action with respect to the fields, and $\overline{M}^{\alpha_1\alpha_2\alpha\alpha'}\coloneqq \gamma^0_{\alpha\beta}\gamma^0_{\alpha'\beta'}\gamma^0_{\alpha_1\beta_1}\gamma^0_{\alpha_2\beta_2} M^{*\beta\beta'\beta_1\beta_2}$.
 With these we
are able to rewrite the source terms in the second and fourth lines of Eq.\ \eqref{eq:exp:Phi_2} as
\begin{subequations}\label{eq:rels_rho}
\begin{eqnarray}    &&\Lambda^{+}_{\alpha\alpha'}\left(p+\frac{u_1+u_2}{2}\right)\prescript{}{\text{out}}{\bra{p_2-\frac{u_2}{2};r_2}}:\rho_{\alpha'}(0):\ket{p^2+\frac{u^2}{2};s^2}_{\text{in}}\nonumber\\
&=&-\frac{i}{4}m^2 (2\pi\hbar)^4 \int \d Q_1  \d Q_2 \, \delta^{(4)}\left(p+p_2+\frac{u_1}{2}-q_1-q_2\right)M^{\alpha_1 \alpha_2 \beta_1 \beta_2} \overline{M}^{ \gamma_1 \gamma_2\delta_1 \delta_2}\Lambda^{+}_{\alpha\alpha_1}\left(p+\frac{u_1+u_2}{2}\right) \nonumber\\
&&\hspace*{4cm} \times  \Lambda^{+}_{\beta_1\gamma_1}(q_1)\Lambda^{+}_{\beta_2\gamma_2}(q_2 )\bar{u}_{r_2,\alpha_2}\left(p_2-\frac{u_2}{2}\right)  u_{s_1,\delta_1}\left(p_1+\frac{u_1}{2}\right)u_{s_2,\delta_2}\left(p_2+\frac{u_2}{2}\right)\label{eq:rel_rho}
    \end{eqnarray}
    and
    \begin{eqnarray}
    &&\prescript{}{\text{out}}{\bra{p^2-\frac{u^2}{2};s^2}}:\bar{\rho}_{\alpha'}(0):\ket{p_2+\frac{u_2}{2};r_2}_{\text{in}}\Lambda^{+}_{\alpha'\beta}\left(p-\frac{u_1+u_2}{2}\right)\nonumber\\
    &=&\frac{i}{4}m^2 (2\pi\hbar)^4 \int \d Q_1  \d Q_2 \, \delta^{(4)}\left(p+p_2-\frac{u_1}{2}-q_1-q_2\right) \overline{M}^{\alpha_1 \alpha_2\beta_1 \beta_2} M^{\gamma_1 \gamma_2\delta_1 \delta_2}\Lambda^{+}_{\beta_1\beta} \left(p-\frac{u_1+u_2}{2}\right)\nonumber\\
    && \hspace*{4cm} \times  \Lambda^{+}_{\delta_1\alpha_1}(q_1)\Lambda^{+}_{\delta_2\alpha_2}(q_2) u_{r_2,\beta_2}\left(p_2+\frac{u_2}{2}\right)  \bar{u}_{s_1,\gamma_1}\left(p_1-\frac{u_1}{2}\right)\bar{u}_{s_2,\gamma_2}\left(p_2-\frac{u_2}{2}\right)\;,\quad\label{eq:rel_rho_bar}
\end{eqnarray}
respectively. On the other hand, the source terms in the last line can be written as
\begin{align}
& \sum_{r'}\Lambda^{+}_{\alpha\alpha'}\left(p+\frac{u_1+u_2}{2}\right) 
    \prescript{}{\text{out}}{\bigg\langle p';r'\bigg|}:\rho_{\alpha'}(0):\ket{p^2+\frac{u^2}{2};s^2}_{\text{in}} \prescript{}{\text{in}}{\bra{p^2-\frac{u^2}{2};r^2}}:\bar{\rho}_{\beta'}(0):\bigg| p';r'\bigg\rangle_{\text{out}}\Lambda^{+}_{\beta'\beta}\left(p-\frac{u_1+u_2}{2}\right)\nonumber\\
    &= 2m M^{\alpha_1 \alpha_2 \beta_1 \beta_2}\overline{M}^{ \gamma_1 \gamma_2\delta_1 \delta_2} \Lambda^{+}_{\alpha\alpha_1}\left(p+\frac{u_1+u_2}{2}\right)\Lambda^{+}_{\delta_1\beta}\left(p-\frac{u_1+u_2}{2}\right) \Lambda^{+}_{\delta_2\alpha_2}(p')\nonumber\\
    & \quad \times u_{s_1,\beta_1}\left(p_1+\frac{u_1}{2}\right)u_{s_2,\beta_2}\left(p_2+\frac{u_2}{2}\right)
    \bar{u}_{r_1,\gamma_1}\left(p_1-\frac{u_1}{2}\right)\bar{u}_{r_2,\gamma_2}\left(p_2-\frac{u_2}{2}\right)\;.\label{eq:rel_rho_rho_bar}
\end{align}
\end{subequations}
Furthermore, in the Boltzmann equation only on-shell terms contribute, i.e., it has to hold that $p^2=m^2$ \cite{Weickgenannt:2021cuo,Sheng:2021kfc}. For this reason we may use the relation
\begin{equation}
    \tilde{G}\left(p+\frac{u_1+u_2}{2}\right)-\tilde{G}^*\left(p-\frac{u_1+u_2}{2}\right)= 2\pi i\hbar^2 \delta(p^2-m^2)\label{eq:rel_prop}
\end{equation}
in Eq.\ \eqref{eq:exp:Phi_2}, since the neglected terms are all off-shell contributions.

Inserting Eqs.\ \eqref{eq:rels_rho} and \eqref{eq:rel_prop} into Eq.\ \eqref{eq:exp:Phi_2} and the result into Eq.\ \eqref{eq:coll_1}, we arrive with the definition of the on-shell quantities \eqref{eq:def_onshell_terms} at the following result for the collision term,
\begin{eqnarray}
    &&\mathcal{C}_{\text{on-shell},\alpha\beta}(x,p)\nonumber\\
    &=& (2\pi\hbar)^4 \frac{m^4}{2}  \momint{p}{}{1}\,\d P_2 \,\d P' \int \d^4 u^2  \,
    M^{\alpha_1 \alpha_2 \beta_1 \beta_2}M^{\gamma_1\gamma_2 \delta_1 \delta_2}\Lambda^{+}_{\alpha\alpha'}\left(p+\frac{u_1+u_2}{2}\right)\Lambda^{+}_{\beta'\beta}\left(p-\frac{u_1+u_2}{2}\right)\nonumber\\
    &&\times \bigg(  \delta^{(4)}(p+p'-p_1-p_2) \delta^{\alpha'}_{\alpha_1}\delta^{\beta'}_{\delta_1}
    \Lambda^{+}_{\delta_2\alpha_2}(p')\Lambda^{+}_{\beta_1\gamma_1'}\left(p_1+\frac{u_1}{2}\right)\Lambda^{+}_{\beta_2\gamma_2'}\left(p_2+\frac{u_2}{2}\right)\nonumber\\
    && \hspace*{0.5cm} \times \Lambda_{\delta_1'\gamma_1}^{+}\left(p_1-\frac{u_1}{2}\right)\Lambda^{+}_{\delta'_2 \gamma_2}\left(p_2-\frac{u_2}{2}\right) \prod_{j=1}^2 \left\{W_{\text{on-shell}}^{\gamma'_j\delta'_j} (x, p_j)\delta^{(4)}(u_j)- i\hbar\left[\partial^\rho_{u_j} \delta^{(4)}(u_j)\right]\partial_\rho^x W_{\text{on-shell}}^{\gamma'_j \delta'_j} (x, p_j)\right\}\nonumber\\
    && \hspace*{0.5cm}  -\frac12\delta^{\alpha'}_{\gamma'_1}\delta^{\beta'}_{\beta_1}\delta^{(4)}\left(p+p'-p_1-p_2-\frac{u_1}{2}\right)\Lambda^{+}_{\beta_2\gamma'_2}\left(p'+\frac{u_2}{2}\right)\Lambda^{+}_{\delta'_1 \gamma_1}\left(p+\frac{u_2-u_1}{2}\right)\Lambda^{+}_{\delta'_2 \gamma_2}\left(p'-\frac{u_2}{2}\right)\nonumber\\
    &&\hspace*{0.5cm} \times \Lambda^{+}_{\delta_1 \alpha_1}(p_1)\Lambda^{+}_{\delta_2 \alpha_2}(p_2) \left\{W_{\text{on-shell}}^{\gamma'_1\delta'_1} \left(x, p+\frac{u_2}{2}\right)\delta^{(4)}(u_1)- i\hbar\left[\partial^\mu_{u_1} \delta^{(4)}(u_1)\right]\partial_\mu^x W_{\text{on-shell}}^{\gamma'_1 \delta'_1} \left(x, p+\frac{u_2}{2}\right)\right\}\nonumber\\
    &&\hspace*{3.5cm} \times \left\{W_{\text{on-shell}}^{\gamma'_2\delta'_2} (x, p')\delta^{(4)}(u_2)- i\hbar\left[\partial^\mu_{u_2} \delta^{(4)}(u_2)\right]\partial_\mu^x W_{\text{on-shell}}^{\gamma'_2\delta'_2} (x, p')\right\}\nonumber\\
    &&\hspace*{0.5cm} -\frac12\delta^{\alpha'}_{\gamma_1}\delta^{\beta'}_{\delta'_1}\delta^{(4)}\left(p+p'-p_1-p_2+\frac{u_1}{2}\right)\Lambda^{+}_{\beta_2\gamma'_2}\left(p'-\frac{u_2}{2}\right)\Lambda^{+}_{\beta_1 \gamma'_1}\left(p+\frac{u_1-u_2}{2}\right)\Lambda^{+}_{\delta'_2 \gamma_2}\left(p'+\frac{u_2}{2}\right)\nonumber\\
    &&\hspace*{0.5cm} \times \Lambda^{+}_{\delta_1 \alpha_1}(p_1)\Lambda^{+}_{\delta_2 \alpha_2}(p_2) \left\{W_{\text{on-shell}}^{\gamma'_1\delta'_1} \left(x, p-\frac{u_2}{2}\right)\delta^{(4)}(u_1)- i\hbar\left[\partial^\mu_{u_1} \delta^{(4)}(u_1)\right]\partial_\mu^x W_{\text{on-shell}}^{\gamma'_1 \delta'_1} \left(x, p-\frac{u_2}{2}\right)\right\}\nonumber\\
    &&\hspace*{3.5cm} \times \left\{W_{\text{on-shell}}^{\gamma'_2\delta'_2} (x, p')\delta^{(4)}(u_2)- i\hbar\left[\partial^\mu_{u_2} \delta^{(4)}(u_2)\right]\partial_\mu^x W_{\text{on-shell}}^{\gamma'_2\delta'_2} (x, p')\right\} \bigg)\;.\label{eq:coll_master}
\end{eqnarray}
Here we used the completeness relation of the basis spinors multiple times, assumed that $\overline{M}=M$, and took $M$ to be independent of momentum. Furthermore, we expanded the Wigner function to first order around $x_j=0$ and considered only the on-shell part of the collision term, cf.\ Eq.\ \eqref{eq:coll_onshell}.
The first terms in curly brackets in Eq.\ \eqref{eq:coll_master} provide the local contributions, while the respective second terms, which are proportional to space-time derivatives of the Wigner functions, constitute the nonlocal parts of the collision term.

\subsection{Local collisions}
From Eq.\ \eqref{eq:coll_master} we can read off the local collision term as
\begin{eqnarray}
    &&\mathcal{C}^{\text{local}}_{\text{on-shell},\alpha\beta}(x,p)\nonumber\\
    &=&  \frac{m^4}{2} \momint{p}{}{1}\, \d P_2 \, \d P'
    \, (2\pi\hbar)^4\delta^{(4)}(p+p'-p_1-p_2)
    M^{\alpha_1 \alpha_2 \beta_1 \beta_2}M^{\gamma_1\gamma_2 \delta_1 \delta_2}\Lambda^{+}_{\alpha \alpha'}\left(p\right)\Lambda^{+}_{\beta' \beta}\left(p\right)\nonumber\\
    &&\times \bigg\{  \delta^{\alpha'}_{ \alpha_1}\delta^{\beta'}_{\delta_1}
    \Lambda^{+}_{\delta_2 \alpha_2}(p')\Lambda^{+}_{\beta_1 \gamma'_1}(p_1)\Lambda^{+}_{\beta_2 \gamma'_2}(p_2)\Lambda^{+}_{\delta'_1 \gamma_1}(p_1)\Lambda^{+}_{\delta'_2 \gamma_2}(p_2) \prod_{j=1}^2 W_{\text{on-shell}}^{\gamma'_j\delta'_j} (x, p_j)\nonumber\\
    &&-\frac12\Lambda^{+}_{\beta_2\gamma'_2}(p')\Lambda^{+}_{\delta'_2 \gamma_2}(p')\Lambda^{+}_{\delta_1 \alpha_1}(p_1)\Lambda^{+}_{\delta_2 \alpha_2}(p_2)\bigg[\delta^{\alpha'}_{\gamma'_1 }\delta^{\beta'}_{\beta_1}\Lambda^{+}_{\delta'_1 \gamma_1}(p) +\delta^{\alpha'}_{\gamma_1}\delta^{\beta'}_{\delta'_1}\Lambda^{+}_{\beta_1 \gamma'_1}(p)\bigg] W_{\text{on-shell}}^{\gamma'_1\delta'_1} (x, p)W_{\text{on-shell}}^{\gamma'_2\delta'_2} (x, p')\bigg\}\;. \label{eq:78}
\end{eqnarray}
Next we show how to translate this expression into extended phase space. 
To this end, we first note
that all Wigner functions
in Eq.\ (\ref{eq:78}) are
sandwiched between energy projectors. 
Because of the relation
\begin{equation}
\Lambda^+(p) \sigma^{\mu \nu}
p_\nu \Lambda^+(p) =0
\end{equation}
this has the consequence that
the gradient part of the
Wigner function vanishes,
cf.\ Eqs.\ (\ref{eq:gradient_one}) and (\ref{eq:Gidentity}), such that
\begin{equation}
\label{eq:expr_W_h}
\Lambda^+(p) W_{\text{on-shell}}(x,p)
\Lambda^+(p)
= \int \d S(p)\, h(p,\ms) f(x,p,\ms)\;.
\end{equation}
One may wonder whether this
introduces a discrepancy to the KB approach (where no such cancellation occurs). However, in the GLW approach another gradient contribution is generated at order $\mathcal{O}(\hbar)$ by an integration by parts, so that in the end both approaches yield the same result.

Inserting Eq.\ \eqref{eq:expr_W_h}
into Eq.\ \eqref{eq:78}, the local collision term becomes
\begin{eqnarray}
    &&\mathcal{C}^{\text{local}}_{\text{on-shell},\alpha\beta}(x,p)\nonumber\\
    &=&  \frac{m^4}{4}  \int \d \Gamma_1 \, \d\Gamma_2 \,\d\Gamma'  \, \d\bar{S}(p) \,(2\pi\hbar)^4\delta^{(4)}(p+p'-p_1-p_2)
    M^{\alpha_1 \alpha_2 \beta_1 \beta_2}M^{\gamma_1\gamma_2 \delta_1 \delta_2}\nonumber\\
    &&\times \bigg\{ \Lambda^{+}_{\alpha \alpha_1}\left(p\right)\Lambda^{+}_{\delta_1 \beta}\left(p\right)
    h_{\delta_2 \alpha_2}(p',\s') h_{\beta_1\gamma_1} (p_1,\s_1)h_{\beta_2\gamma_2} ( p_2,\s_2) f(x,p_1,\s_1)f(x,p_2,\s_2)\nonumber\\
    &&\hspace*{0.3cm} -f(x,p,\bar{\s})f(x,p',\s') h_{\delta_1 \alpha_1}(p_1,\s_1)h_{\delta_2 \alpha_2}(p_2,\s_2)\bigg[h_{\alpha\gamma_1} (p,\bar{\s})\Lambda^{+}_{\beta_1 \beta}\left(p\right) +\Lambda^{+}_{\alpha \gamma_1}\left(p\right)h_{\beta_1 \beta} (p,\bar{\s})\bigg] h_{\beta_2\gamma_2} (p',\s')\bigg\}\;,
\end{eqnarray}
where we used that 
\begin{equation}
\int \d S(p) h(p,\s)=\Lambda^{+}(p)\;.\label{eq:int_h}
\end{equation}
Lastly, we employ that $\mathfrak{C}_{\text{on-shell}}\coloneqq \frac12(1+\gamma_5 \slashed{\s})_{\beta\alpha} \mathcal{C}_{\text{on-shell}}^{\alpha\beta}$,
which then gives
\begin{eqnarray}
\mathfrak{C}^{\text{local}}_{ \text{on-shell}}(x,p,\s)
&=& \frac{m^4}{4} \int \d \Gamma_1  \, \d \Gamma_2 \, \d\Gamma'\,  \d \bar{S}(p)\, (2\pi\hbar)^4\delta^{(4)}(p+p'-p_1-p_2)M^{\alpha_1 \alpha_2 \beta_1 \beta_2}M^{\gamma_1\gamma_2 \delta_1 \delta_2}  \nonumber\\
&& \times h_{\beta_1 \gamma_1}(p_1,\s_1)h_{\beta_2 \gamma_2} (p_2,\s_2) h_{\delta_2 \alpha_2}(p',\s')\left[h_{\delta_1 \beta}(p,\s) h_{\beta \alpha_1}(p,\bar{\s})+ h_{\delta_1 \beta}(p,\bar{\s}) h_{\beta \alpha_1}(p,\s) \right]\nonumber\\
&&\times \left[f(x,p_1,\s_1)f(x,p_2,\s_2)-f(x,p,\bar{\s})f(x,p',\s')\right]\;, \label{eq:C_loc}
\end{eqnarray}
where we also made use of the relation
\begin{equation}
    \int \d \bar{S}(p) \,\left[h_{\delta_1 \beta}(p,\s) h_{\beta \alpha_1}(p,\bar{\s})+ h_{\delta_1 \beta}(p,\bar{\s}) h_{\beta \alpha_1}(p,\s) \right]= 2 h_{\delta_1\alpha_1}(p,\s)\;.
\end{equation}

Note that the dependence
on $\bar{\ms}$ in Eq.\ (\ref{eq:C_loc}) can be eliminated employing
a so-called
``weak equivalence principle''
\cite{Weickgenannt:2020aaf,Weickgenannt:2021cuo}. This then gives
a clearer interpretation of
the last term in \eq\eqref{eq:C_loc} as a loss term corresponding to particles with momentum $p$ and spin $\ms$. The new collision term then has the form
\begin{equation}
 \tilde{\mathfrak{C}}^{\text{local}}_{ \text{on-shell}}(x,p,\ms)
=  \int \d \Gamma_1 \, \d \Gamma_2 \, \d\Gamma'\, \widetilde{\mathcal{W}} \left[f(x,p_1,\s_1)f(x,p_2,\s_2)-f(x,p',\s')f(x,p,\s)\right]  \;, 
\end{equation}
with
\begin{equation}
    \widetilde{W}\coloneqq \frac{m^4}{2}(2\pi\hbar)^4\delta^{(4)}(p+p'-p_1-p_2)M^{\alpha_1 \alpha_2 \beta_1 \beta_2}M^{\gamma_1\gamma_2 \delta_1 \delta_2} h_{\beta_1 \gamma_1}(p_1,\s_1)h_{\beta_2 \gamma_2} (p_2,\s_2) h_{\delta_2 \alpha_2}(p',\s')h_{\delta_1 \alpha_1}(p,\s)\; . \end{equation}
    This agrees with the local collision term derived in Refs.~\cite{Weickgenannt:2020aaf,Weickgenannt:2021cuo} up to the part corresponding to collisions without momentum exchange.

If we employ an NJL-type interaction according to Eq.\ (\ref{eq:int_Lagrangian}), we
obtain
\begin{equation}
    M^{\alpha_1 \alpha_2 \beta_1 \beta_2}=\frac{2G_c}{\hbar} \left( \Gamma^{(c)\alpha_1 \beta_1}
    \Gamma^{(c)\alpha_2 \beta_2}- \Gamma^{(c)\alpha_1 \beta_2}
    \Gamma^{(c) \alpha_2 \beta_1}\right)\;,\label{eq:M_NJL}
\end{equation}
from which it follows that 
\begin{eqnarray}
    &&M^{\alpha_1 \alpha_2 \beta_1 \beta_2}M^{\gamma_1\gamma_2 \delta_1 \delta_2} h_{\beta_1 \gamma_1}(p_1,\s_1)h_{\beta_2 \gamma_2} (p_2,\s_2) h_{\delta_2 \alpha_2}(p',\s')h_{\delta_1 \beta}(p,\s) h_{\beta \alpha_1}(p,\bar{\s})\nonumber\\
    &\equiv&\frac{8G_cG_d}{\hbar^2}\left\{\mathrm{Tr}\left[h_2\Gamma^{(d)}h'\Gamma^{(c)}\right]\mathrm{Tr}\left[h_1\Gamma^{(d)}h\bar{h}\Gamma^{(c)}\right]
    -\mathrm{Tr}\left[h_2\Gamma^{(d)}h\bar{h}\Gamma^{(c)}h_1\Gamma^{(d)}h'\Gamma^{(c)}\right]\right\}\;,\label{eq:W_NJL_1}
\end{eqnarray}
where we abbreviated $h_1\coloneqq h(p_1,\s_1)$ and likewise for $h_2,h',\bar{h}$, and $h$.
Here, the symbol ``$\equiv$'' means that the expressions are equal under the respective integrals where they appear. This allowed us to use the symmetry under exchanging the integrations over $(p_1,\s_1)$ and $(p_2,\s_2)$, which is reflected in an additional factor of two in Eq.\ \eqref{eq:W_NJL_1}. Taking the complex conjugate of Eq.\ \eqref{eq:W_NJL_1} and using that $h^\dagger=\gamma^0 h \gamma^0$ as well as
$\gamma^0 \Gamma^{(c)\dagger} \gamma^0 \equiv \Gamma^{(c)}$, we find 
\begin{eqnarray}
    &&\left[M^{\alpha_1 \alpha_2 \beta_1 \beta_2}M^{\gamma_1\gamma_2 \delta_1 \delta_2} h_{\beta_1 \gamma_1}(p_1,\s_1)h_{\beta_2 \gamma_2} (p_2,\s_2) h_{\delta_2 \alpha_2}(p',\s')h_{\delta_1 \beta}(p,\s) h_{\beta \alpha_1}(p,\bar{\s})\right]^*\nonumber\\
    &\equiv&\frac{8G_cG_d}{\hbar^2}\left\{\mathrm{Tr}\left[h_2\Gamma^{(d)}h'\Gamma^{(c)}\right]\mathrm{Tr}\left[h_1\Gamma^{(d)}\bar{h}h\Gamma^{(c)}\right]
    -\mathrm{Tr}\left[h_2\Gamma^{(d)}\bar{h}h\Gamma^{(c)}h_1\Gamma^{(d)}h'\Gamma^{(c)}\right]\right\}\;.
\end{eqnarray}
This allows us to bring Eq.\ \eqref{eq:C_loc} into the following form, 
\begin{eqnarray}
\mathfrak{C}^{\text{local}}_{ \text{on-shell}}(x,p,\s)
&=&\frac{4G_cG_d}{\hbar^2} m^4 \int \d \Gamma_1 \, \d \Gamma_2 \, \d\Gamma' \, \d \bar{S}(p) \,  (2\pi\hbar)^4\delta^{(4)}(p+p'-p_1-p_2) \nonumber\\
&& \times \mathrm{Re}\left\{\mathrm{Tr}\left[h_2\Gamma^{(d)}h'\Gamma^{(c)}\right]\mathrm{Tr}\left[h_1\Gamma^{(d)}h\bar{h}\Gamma^{(c)}\right]
-\mathrm{Tr}\left[h_2\Gamma^{(d)}h\bar{h}\Gamma^{(c)}h_1\Gamma^{(d)}h'\Gamma^{(c)}\right]\right\}\nonumber\\
&&\times \left[f(x,p_1,\s_1)f(x,p_2,\s_2)-f(x,p',\bar{\s})f(x,p,\s')\right]\;. \label{eq:C_loc_NJL}
\end{eqnarray}

\subsection{Nonlocal collisions}

In order to obtain the nonlocal contribution to the collision term, we have to integrate by parts in the variables $u_1, u_2$ in Eq.\ \eqref{eq:coll_master}. Fortunately, the derivatives acting on the projectors are straightforwardly evaluated, e.g., 
\begin{equation}
\partial^\mu_{u} \Lambda^{+}\left(p+\frac{u}{2}\right)\Big|_{u=0}=\frac{1}{4m}\gamma^\mu\;.
\end{equation}
We split the effect of the $u_1,u_2$-derivatives into four contributions
(enumerated by capital Roman
numbers): First, the derivatives act on the projectors $\Lambda^{+}
(p\pm u_1/2 \pm u_2/2)$ in front of everything, giving 
\begin{align}
\mathcal{C}^{\text{nonlocal}}_{\text{on-shell,I,}\alpha\beta}(x,p)
    &= \frac{i\hbar}{4m}  \frac{m^4}{2}  \momint{p}{}{1}\, \d P_2 \, \d P' \, (2\pi\hbar)^4\delta^{(4)}(p+p'-p_1-p_2)
    M^{\alpha_1 \alpha_2 \beta_1 \beta_2}M^{\gamma_1\gamma_2 \delta_1 \delta_2}
    \nonumber\\
    &\times \left[\gamma^{\mu}_{\alpha\alpha'}\Lambda^{+}_{\beta'\beta}\left(p\right)-\Lambda^{+}_{\alpha\alpha'}\left(p\right)\gamma^\mu_{\beta'\beta}\right]\bigg\{  \delta_{\alpha' \alpha_1}\delta_{\beta' \delta_1}
    \Lambda^{+}_{\delta_2 \alpha_2}(p')  \partial_\mu^x\left[W_{\text{on-shell}}^{\beta_1\gamma_1} (x, p_1)W_{\text{on-shell}}^{\beta_2\gamma_2} (x, p_2)\right]\nonumber\\
    &-\frac12\Lambda^{+}_{\delta_1 \alpha_1}(p_1)\Lambda^{+}_{\delta_2 \alpha_2}(p_2)\bigg[\delta_{\alpha' \gamma'_1 }\delta_{\beta_1\beta'}\delta_{\delta'_1 \gamma_1} +\delta_{\alpha' \gamma_1}\delta_{\delta'_1 \beta'}\delta_{\beta_1 \gamma'_1}\bigg] \partial_\mu^x\left[W_{\text{on-shell}}^{\gamma'_1\delta'_1} (x, p)W_{\text{on-shell}}^{\beta_2\gamma_2} (x, p')\right]\bigg\}\;,
\end{align}
where we already simplified some contractions of energy projectors and Wigner functions.
Translating this expression into extended phase space, we find
\begin{align}   \mathfrak{C}^{\text{nonlocal}}_{\text{on-shell,I}}(x,p,\s)
    &= \frac{i\hbar}{4m} \frac{m^4}{4}  \int \d \Gamma_1\, \d \Gamma_2\, \d \Gamma' \, \d \bar{S}(p)\,  (2\pi\hbar)^4\delta^{(4)}(p+p'-p_1-p_2)
    M^{\alpha_1 \alpha_2 \beta_1 \beta_2}M^{\gamma_1\gamma_2 \delta_1 \delta_2}
    \nonumber\\
    &\times \left[h(p,\s),\gamma^\mu\right]_{\beta'\alpha'}\bigg\{  \delta_{\alpha' \alpha_1}\delta_{\beta' \delta_1}
    h_{\delta_2 \alpha_2}(p',\s') h_{\beta_1\gamma_1}(p_1,\s_1)h_{\beta_2\gamma_2}(p_2,\s_2) \partial_\mu^x\left[f(x, p_1,\s_1)f (x, p_2,\s_2)\right]\nonumber\\
    &- h_{\delta_1 \alpha_1}(p_1,\s_1)h_{\delta_2 \alpha_2}(p_2,\s_2)h_{\beta_2\gamma_2}(p',\s')\bigg[h_{\alpha'\gamma_1}(p,\bar{\s})\delta_{\beta' \beta_1}+h_{\beta_1\beta'}(p,\bar{\s})\delta_{\alpha' \gamma_1}\bigg] \partial_\mu^x\left[f (x, p,\bar{\s})f (x, p',\s')\right]\bigg\}\;.
\end{align}
Since the parts of $f(x,p,\s)$ that are proportional to $\s^\mu$ are at least of order $\mathcal{O}(\hbar)$, we may perform the $\d\bar{S}(p)$-integral trivially to obtain
\begin{align} \mathfrak{C}^{\text{nonlocal}}_{\text{on-shell,I}}(x,p,\s)
    &= \frac{i\hbar}{4m}  \frac{m^4}{2}  \int \d \Gamma_1\, \d \Gamma_2\, \d \Gamma'\, (2\pi\hbar)^4\delta^{(4)}(p+p'-p_1-p_2)
    M^{\alpha_1 \alpha_2 \beta_1 \beta_2}M^{\gamma_1\gamma_2 \delta_1 \delta_2}
    \nonumber\\
    &\times h_{\delta_2 \alpha_2}(p',\s') h_{\beta_1\gamma_1}(p_1,\s_1)h_{\beta_2\gamma_2}(p_2,\s_2)\left[h(p,\s),\gamma^\mu\right]_{\delta_1 \alpha_1}
     \partial_\mu^x\left[f(x, p_1)f (x, p_2)-\frac12 f (x, p)f (x, p')\right]\;.\label{eq:coll_nonloc_1}
\end{align}
Here we used that $\{[h(p,\s),\gamma^\mu],\Lambda^{+}(p)\}=[h(p,\s),\gamma^\mu]$ since $h(p,\s)(\slashed{p}-m)=0$.

As a second nonlocal contribution, after integration by parts the $u_1,u_2$-derivatives in Eq.\ \eqref{eq:coll_master} act on the remaining projectors. Performing the same steps that led to Eq.\ \eqref{eq:coll_nonloc_1}, we find
\begin{align}
\mathfrak{C}^{\text{nonlocal}}_{\text{on-shell, II}}(x,p,\s)
&=-\frac{i\hbar}{4m} \frac{m^4}{2} \int \d \Gamma_1 \, \d \Gamma_2 \, \d\Gamma' \, (2\pi\hbar)^4\delta^{(4)}(p+p'-p_1-p_2)M^{\alpha_1 \alpha_2 \beta_1 \beta_2}M^{\gamma_1\gamma_2 \delta_1 \delta_2} \nonumber\\
& \; \times \bigg\{ f(x,p_2)\Big[\partial_\mu^x f(x,p_1) \Big]h_{\delta_2 \alpha_2}(p',\s') h_{\beta_2 \gamma_2} (p_2,\s_2)  h_{\delta_1 \alpha_1}(p,\s) \left[h(p_1,\s_1),\gamma^\mu\right]_{\beta_1\gamma_1} \nonumber\\
&\quad +f(x,p_1)\Big[\partial_\mu^x f(x,p_2) \Big]h_{\delta_2 \alpha_2}(p',\s') h_{\beta_1 \gamma_1} (p_1,\s_1)  h_{\delta_1 \alpha_1}(p,\s) \left[h(p_2,\s_2),\gamma^\mu\right]_{\beta_2\gamma_2} \nonumber\\
& \quad -f(x,p)\left[\partial_\mu^x f(x,p')\right]h_{\beta_1 \gamma_1}(p_1,\s_1)  h_{\beta_2 \gamma_2} (p_2,\s_2) h_{\delta_1 \alpha_1}(p,\s)  \left[h(p',\s'),\gamma^\mu\right]_{\delta_2\alpha_2}\nonumber\\
&\quad -\frac12 \Big[f(x,p')\partial_\mu^x f(x,p)-f(x,p)\partial_\mu^x f(x,p')\Big]h_{\beta_1 \gamma_1}(p_1,\s_1)  h_{\beta_2 \gamma_2} (p_2,\s_2) h_{\delta_2 \alpha_2}(p',\s')  \left[h(p,\s),\gamma^\mu\right]_{\delta_1\alpha_1}\bigg\}\;.\label{eq:coll_nonloc_2}
\end{align}
Thirdly, a $u_2$-derivative acts on the Wigner functions $W^{\gamma_1' \delta_1'}_{\text{on-shell}}(x,p\pm u_2/2)$ in the loss term, yielding
\begin{align}
\mathfrak{C}^{\text{nonlocal}}_{\text{on-shell, III}}(x,p,\s)
&=-\frac{i\hbar}{4m}\frac{m^4}{4} \int \d \Gamma_1 \, \d \Gamma_2 \, \d\Gamma' \, \d \bar{S}(p) \,(2\pi\hbar)^4\delta^{(4)}(p+p'-p_1-p_2)M^{\alpha_1 \alpha_2 \beta_1 \beta_2}M^{\gamma_1\gamma_2 \delta_1 \delta_2}  \nonumber\\
&\qquad \times  \left[\partial_p^\mu f(x,p)\right]\left[\partial_\mu^x f(x,p')\right] h_{\beta_1 \gamma_1}(p_1,\s_1)  h_{\beta_2 \gamma_2} (p_2,\s_2) h_{\delta_2 \alpha_2}(p',\s')  \left[h(p,\s),h(p,\bar{\s})\right]_{\delta_1 \alpha_1}\;.\label{eq:coll_nonloc_3}
\end{align}
Note that, after performing the $\d\bar{S}(p)$-integration trivially, this term vanishes as a consequence of our assumptions that polarization effects enter at order $\mathcal{O}(\hbar)$ and that $\overline{M}=M$. 

Lastly, there are $u_1$-derivatives acting on the momentum-conserving delta function in the loss term, which can be rewritten as derivatives with respect to $p'$ and then act both on projectors and the Wigner function, giving 
\begin{align}
\mathfrak{C}^{\text{nonlocal}}_{\text{on-shell, IV}}(x,p,\s)
&=-\frac{i\hbar}{4m}\frac{m^4}{4} \int \d \Gamma_1 \, \d \Gamma_2 \, \d\Gamma' \, \d \bar{S}(p)\, (2\pi\hbar)^4\delta^{(4)}(p+p'-p_1-p_2)M^{\alpha_1 \alpha_2 \beta_1 \beta_2}M^{\gamma_1\gamma_2 \delta_1 \delta_2} \nonumber\\
&\times  h_{\beta_1 \gamma_1}(p_1,\s_1)  h_{\beta_2 \gamma_2} (p_2,\s_2) \left[\partial_\mu^x f(x,p)\right] \left[ h_{\delta_2 \alpha_2}(p',\s') \partial_{p'}^\mu f(x,p')+\left\{h(p',\s'),\gamma^\mu\right\}_{\delta_2 \alpha_2} f(x,p') \right] \nonumber \\
 & \times \left[h(p,\s),h(p,\bar{\s})\right]_{\delta_1 \alpha_1}\;.\label{eq:coll_nonloc_4}
\end{align}
Like Eq.\ \eqref{eq:coll_nonloc_3}, this contribution vanishes due to our assumptions.

\subsection{Summary}

Collecting both the local and the nonvanishing nonlocal contributions, we find 
\begin{eqnarray}
p\cdot \partial_x f(x,p,\s)
&=&  \frac14 \int \d \Gamma_1 \, \d \Gamma_2 \, \d \Gamma' \, \d \bar{S}(p)\, (2\pi\hbar)^4\delta^{(4)}\left( p  +p' -p_1-p_2  \right) \mathcal{W} \nonumber\\
&&\times\Big[f(x+\Delta_1-\Delta,p_1,\s_1) f(x+\Delta_2-\Delta,p_2,\s_2) - f(x,p,\bar{\s})f(x+\Delta'-\Delta,p',\s')\Big]\;,
\label{eq:coll_final}
\end{eqnarray}
where we defined the local transition rate
\begin{equation}
\mathcal{W}\coloneqq m^4 M^{\alpha_1 \alpha_2 \beta_1 \beta_2}M^{\gamma_1\gamma_2 \delta_1 \delta_2}  h_{\beta_1 \gamma_1}(p_1,\s_1)h_{\beta_2 \gamma_2} (p_2,\s_2)h_{\delta_2 \alpha_2}(p',\s')  \left\{h(p,\s),h(p,\bar{\s})\right\}_{\delta_1\alpha_1}\;,\label{eq:def_W}
\end{equation}
and the nonlocal shifts read
\begin{subequations}
\begin{eqnarray}
\Delta_{1}^\mu &\coloneqq&-\frac{i\hbar}{4m} \frac{m^4}{\mathcal{W}}  M^{\alpha_1 \alpha_2 \beta_1 \beta_2}M^{\gamma_1\gamma_2 \delta_1 \delta_2}h_{\beta_2 \gamma_2}(p_2,\s_2)   h_{\delta_2 \alpha_2}(p',\s') h_{\delta_1 \alpha_1}(p,\s) \left[h(p_1,\s_1),\gamma^\mu\right]_{\beta_1 \gamma_1}\;,\label{eq:def_Delta_1}\\
\Delta_{2}^\mu &\coloneqq&-\frac{i\hbar}{4m} \frac{m^4}{\mathcal{W}}  M^{\alpha_1 \alpha_2 \beta_1 \beta_2}M^{\gamma_1\gamma_2 \delta_1 \delta_2}h_{\beta_1\gamma_1}(p_1,\s_1)   h_{\delta_2 \alpha_2}(p',\s') h_{\delta_1 \alpha_1}(p,\s) \left[h(p_2,\s_2),\gamma^\mu\right]_{\beta_2 \gamma_2}\;,\label{eq:def_Delta_2}\\
\Delta'^\mu &\coloneqq&-\frac{i\hbar}{4m} \frac{m^4}{\mathcal{W}}  M^{\alpha_1 \alpha_2 \beta_1 \beta_2}M^{\gamma_1\gamma_2 \delta_1 \delta_2}h_{\beta_1\gamma_1}(p_1,\s_1) h_{\beta_2 \gamma_2}(p_2,\s_2)    h_{\delta_1 \alpha_1}(p,\s) \left[h(p',\s'),\gamma^\mu\right]_{\delta_2 \alpha_2}\;,\label{eq:def_Delta_prime}\\
\Delta^\mu &\coloneqq&-\frac{i\hbar}{4m} \frac{m^4}{\mathcal{W}}  M^{\alpha_1 \alpha_2 \beta_1 \beta_2}M^{\gamma_1\gamma_2 \delta_1 \delta_2}h_{\beta_1\gamma_1}(p_1,\s_1)h_{\beta_2 \gamma_2}(p_2,\s_2)   h_{\delta_2 \alpha_2}(p',\s')  \left[h(p,\s),\gamma^\mu\right]_{\delta_1 \alpha_1}\;.\label{eq:def_Delta}
\end{eqnarray}
\end{subequations}
At this point, two remarks are in order. Firstly, the factor $m^4$ in the local transition rate \eqref{eq:def_W} does not necessarily imply that $\mathcal{W}$ vanishes in the massless limit, since it cancels with appropriate inverse factors in the energy projectors $\Lambda^{+}(p)$. Indeed, considering the case where the distribution functions do not depend on spin, it is apparent that the spin-integrated transition rate becomes in the massless limit
\begin{equation}
 \int \d S_1 (p_1) \, \d S_2 (p_2) \, \d S' (p') \,\d \bar{S}(p)\, \mathcal{W}=\frac{1}{16}M^{\alpha_1 \alpha_2 \beta_1 \beta_2}M^{\gamma_1\gamma_2 \delta_1 \delta_2}\slashed{p}_{1, \beta_1 \gamma_1} \slashed{p}_{2,\beta_2\gamma_2}
 \slashed{p}'_{\delta_2\alpha_2}
 \left[(\mathds{1}+\gamma_5\slashed{\s})\slashed{p}\right]_{\delta_1\alpha_1}\;,
\end{equation}
which is (assuming that the vertices $M$ do not diverge) manifestly finite. Nevertheless, in order to properly assess the collision term in the massless limit, the calculation in the GLW approach should be repeated taking into account the different equations of motion for $W(x,p)$ in this case, which is beyond the scope of this work.

Secondly, it is reassuring to see how Eq.\ \eqref{eq:coll_final} agrees with the known expression for binary elastic scattering in the limit of local collisions and distribution functions independent of spin \cite{DeGroot:1980dk}. In this case, the integration
over $\d \bar{S}(p)$ in the first line of Eq.\ \eqref{eq:coll_final} will produce a factor of two, similar to the other spin-space integrals. Finally, there is
a factor of $1/2$ on the right-hand side of Eq.\ \eqref{eq:coll_final}, which can be interpreted as a symmetry factor due to the indistinguishable nature of the particles.

Upon inserting the explicit NJL-type interaction \eqref{eq:M_NJL}, we arrive at the main result of this work,
i.e., Eqs.\ (\ref{eq:def_Deltas_NJL}) with
Eq.\ (\ref{eq:W_NJL}). Note that, in order to arrive at the precise form of those equations, we switched the index pairs $\beta_1\leftrightarrow\beta_2$, $\gamma_1\leftrightarrow\gamma_2$, and used the symmetries of the vertices, i.e.,  $M^{\alpha_1\alpha_2\beta_1\beta_2}=-M^{\alpha_1\alpha_2\beta_2\beta_1}=-M^{\alpha_2\alpha_1\beta_1\beta_2}$.

\section{The nonlocal collision term in the KB approach}\label{sec:KB}

In this section, we
derive the collision term within the KB approach. We first discuss
the mass-shell constraint and
the Boltzmann-type equation in the semi-classical expansion, i.e., order by order in $\hbar$.
We then compute the various collision terms in $T$-matrix approximation.
The advantage of the KB approach
as compared to the GLW approach is that full quantum statistics is retained.

\subsection{Equations of motion in semi-classical expansion}
\label{sec:h-expand-kb}

We first derive a mass-shell constraint
and a Boltzmann-type equation for
the single-particle distribution function
$\mathfrak{f}(x,p,\ms)$ in extended phase space. Taking
Eqs.\ (\ref{eq:massshell-wigner}), (\ref{eq:boltzmann-wigner}) for 
$\Gamma_a = \dblone$ and $\Gamma_a = \gamma_5 \gamma^\mu$, multiplying the latter with
$\s^\mu$, and adding them, we obtain with
Eq.\ (\ref{eq:f})
\begin{subequations}
\label{eq:mass_boltz_KB}
\begin{align}
\left(p^{2} -\frac{\hbar^{2}}{4}\partial_x^{2}-m^{2}\right)\,
\mathfrak{f}(x,p,\ms) & = 
\frac{1}{2}\, \mathrm{Re}\mathrm{Tr}\left[(\dblone +  \gamma_5 \slashed{\ms}) (\slashed{K}+m)I_{\mathrm{coll}}\right]\;,
\label{eq:massshell-f} \\
\hbar\,  p\cdot\partial_x\, \mathfrak{f}(x,p,\ms) & = 
\frac{1}{2} \,\mathrm{Im}\mathrm{Tr}\left[(\dblone +  \gamma_5 \slashed{\ms})(\slashed{K}+m)I_{\mathrm{coll}}\right] \;.
\label{eq:boltzmann-f}
\end{align}
\end{subequations}
We now expand Eqs.~(\ref{eq:mass_boltz_KB}) 
up to second order in $\hbar$, i.e., we need
$\mathfrak{f}(x,p,\ms)$ up to second order in $\hbar$, cf.\ Eq.\ (\ref{eq:expansion_f}).
Furthermore, the collision term (\ref{eq:collision_term}) is already of order $\mathcal{O}(\hbar)$, i.e.,
\begin{equation}
I_{\textrm{coll}} = \hbar \,I_{\textrm{coll}}^{(1)} + \hbar^2 I_{\textrm{coll}}^{(2)} + \mathcal{O}(\hbar^3)\;.
\end{equation}
The Wigner functions $G^{\lessgtr}$ and the self-energies $\Sigma^\gtrless$ entering the collision term are 
therefore only required up to order $\mathcal{O}(\hbar)$,
\begin{subequations}
\begin{align}
G^{\lessgtr} (x,p) & = G^{\lessgtr (0)}(x,p) + \hbar \,G^{\lessgtr (1)}(x,p)\;, \\
\Sigma^{\gtrless} (x,p) & = \Sigma^{\gtrless (0)}(x,p) + \hbar \,\Sigma^{\gtrless (1)}(x,p)\;. 
\end{align}
\end{subequations}
Inserting this into Eq.~(\ref{eq:collision_term}), we obtain
\begin{subequations}
\begin{align}
I_{\textrm{coll}}^{(1)} & =  \frac{i}{2} \left( \Sigma^{<(0)} G^{>(0)} - \Sigma^{>(0)} G^{<(0)} \right)\;, \\
I_{\textrm{coll}}^{(2)} & = \Delta I_{\textrm{coll}}^{(1)}  + I_{\textrm{coll, PB}}^{(0)}\;,
\end{align}
\end{subequations}
where
\begin{subequations}
\begin{align}
\Delta I_{\textrm{coll}}^{(1)} & \coloneqq  \frac{i}{2} \left(  \Sigma^{<(1)} G^{>(0)} - \Sigma^{>(1)} G^{<(0)} 
+  \Sigma^{<(0)} G^{>(1)} - \Sigma^{>(0)} G^{<(1)} \right)\;, \\
I_{\textrm{coll, PB}}^{(0)} & \coloneqq  \frac{1}{4} \left( \left\{ \Sigma^{<(0)}, G^{>(0)} \right\}_\textrm{PB}
- \left\{ \Sigma^{>(0)}, G^{<(0)} \right\}_\textrm{PB} \right)\;.
\end{align}
\end{subequations}

\subsubsection{Zeroth order in $\hbar$}
\label{zeroth-order-h-expand}

At $\mathcal{O}(\hbar^{0})$, the collision term (\ref{eq:collision_term}) vanishes,
$I_{\mathrm{coll}}^{(0)} =0$, since it is at least of order $\mathcal{O}(\hbar)$.
Equation (\ref{eq:boltzmann-f}) is trivially fulfilled, while Eq.~(\ref{eq:massshell-f}) becomes
\begin{equation}
\mathcal{O}(\hbar)\;: \quad (p^2 - m^2)\, \mathfrak{f}^{(0)}(x,p) = 0\;.
\end{equation}
This confirms that $\mathfrak{f}^{(0)}(x,p) \sim \delta(p^2-m^2)$, i.e., it is on-shell,
cf.\ Eq.\ (\ref{eq:Gsmallerlarger_orderzero}).

\subsubsection{First order in $\hbar$}
\label{first-order-h-expand}

At $\mathcal{O}(\hbar)$, Eqs.~(\ref{eq:massshell-f}) and (\ref{eq:boltzmann-f}) become
\begin{subequations}
\begin{align}
(p^2 - m^2) \, \mathfrak{f}^{(1)}(x,p,\ms) & = - \frac{1}{4}\, \mathrm{Im} \mathrm{Tr} 
\left[ (\dblone + \gamma_5 \slashed{\ms}) (\slashed{p}+m) \left( \Sigma^{<(0)} G^{>(0)} - \Sigma^{>(0)} G^{<(0)} \right)
\right] \;, \label{eq:massshell_orderone} \\
p \cdot \partial_x \,\mathfrak{f}^{(0)}(x,p) & = \frac{1}{4} \,\mathrm{Re} \mathrm{Tr} 
\left[ (\dblone + \gamma_5 \slashed{\ms}) (\slashed{p}+m) \left( \Sigma^{<(0)} G^{>(0)} - \Sigma^{>(0)} G^{<(0)} \right)
\right] \;. \label{eq:Boltzmann_orderone}
\end{align}
\end{subequations}

\subsubsection{Second order in $\hbar$}
\label{second-order-h-expand}

At $\mathcal{O}(\hbar^2)$, we also need to take into account the $\mathcal{O}(\hbar)$ contribution
to the operator $K^\mu$, cf.~Eq.~(\ref{eq:k-op}), when computing the right-hand sides of Eqs.~(\ref{eq:massshell-f})
and (\ref{eq:boltzmann-f}). We thus obtain
\begin{subequations}
\begin{align}
(p^2 - m^2) \, \mathfrak{f}^{(2)}(x,p,\ms)  & - \frac{1}{4} \partial_x^2 \, \mathfrak{f}^{(0)}(x,p) 
= - \frac{1}{8}\, \mathrm{Re} \mathrm{Tr} 
\left[ (\dblone + \gamma_5 \slashed{\ms}) \slashed{\partial}_x \left( \Sigma^{<(0)} G^{>(0)} - \Sigma^{>(0)} G^{<(0)} \right)
\right] \nonumber \\
& - \frac{1}{4} \,\mathrm{Im}\mathrm{Tr} \left[ (\dblone + \gamma_5 \slashed{\ms}) (\slashed{p}+m)
\left( \Sigma^{<(1)} G^{>(0)} - \Sigma^{>(1)} G^{<(0)} +\Sigma^{<(0)} G^{>(1)} - \Sigma^{>(0)} G^{<(1)}  \right) \right]
\nonumber \\
& + \frac{1}{8}  \, \mathrm{Re} \mathrm{Tr} \left[ (\dblone + \gamma_5 \slashed{\ms}) (\slashed{p}+m)
\left( \left\{ \Sigma^{<(0)}, G^{>(0)} \right\}_\textrm{PB}
- \left\{ \Sigma^{>(0)}, G^{<(0)} \right\}_\textrm{PB} \right) \right]\;, \label{eq:massshell_ordertwo} \\
p \cdot \partial_x \, \mathfrak{f}^{(1)}(x,p,\ms) & = - \frac{1}{8} \,\mathrm{Im} \mathrm{Tr} 
\left[ (\dblone + \gamma_5 \slashed{\ms}) \slashed{\partial}_x \left( \Sigma^{<(0)} G^{>(0)} - \Sigma^{>(0)} G^{<(0)} \right)
\right] \nonumber \\
& + \frac{1}{4} \,\mathrm{Re}\mathrm{Tr} \left[ (\dblone + \gamma_5 \slashed{\ms}) (\slashed{p}+m)
\left( \Sigma^{<(1)} G^{>(0)} - \Sigma^{>(1)} G^{<(0)} +\Sigma^{<(0)} G^{>(1)} - \Sigma^{>(0)} G^{<(1)}  \right) \right]
\nonumber \\
& + \frac{1}{8} \, \mathrm{Im} \mathrm{Tr} \left[ (\dblone + \gamma_5 \slashed{\ms}) (\slashed{p}+m)
\left( \left\{ \Sigma^{<(0)}, G^{>(0)} \right\}_\textrm{PB}
- \left\{ \Sigma^{>(0)}, G^{<(0)} \right\}_\textrm{PB} \right) \right]\;. \label{eq:Boltzmann_ordertwo}
\end{align}
\end{subequations}

\subsection{Collision terms}

\subsubsection{Self-energies in
$T$-matrix approximation}

\begin{figure}
\caption{\label{fig:feynman}Feynman diagrams for (a,c) $\Sigma^{>}(x,p)$ and
(b,d) $\Sigma^{<}(x,p)$. Solid lines represent fermion propagators,
dashed lines represent the one-boson-exchange interaction of coupling strength $G_c$ or $G_d$, respectively. 
Vertices denote elements $\Gamma^{(c)}$, $\Gamma^{(d)}$ of the Clifford algebra corresponding to the 
interaction channels $(c)$ or $(d)$, respectively.}
\includegraphics[scale=0.5]{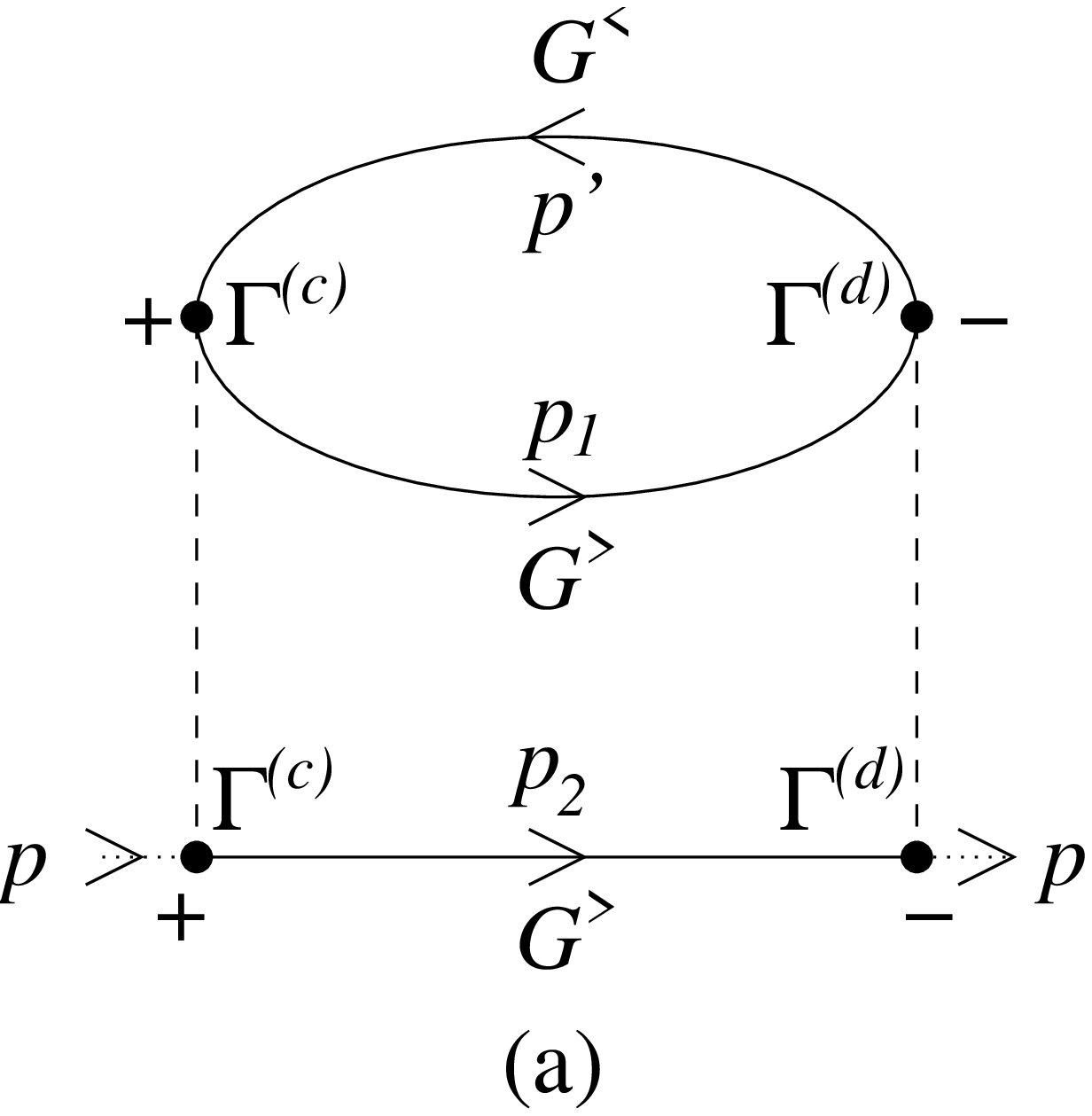} \hspace*{2cm}
\includegraphics[scale=0.5]{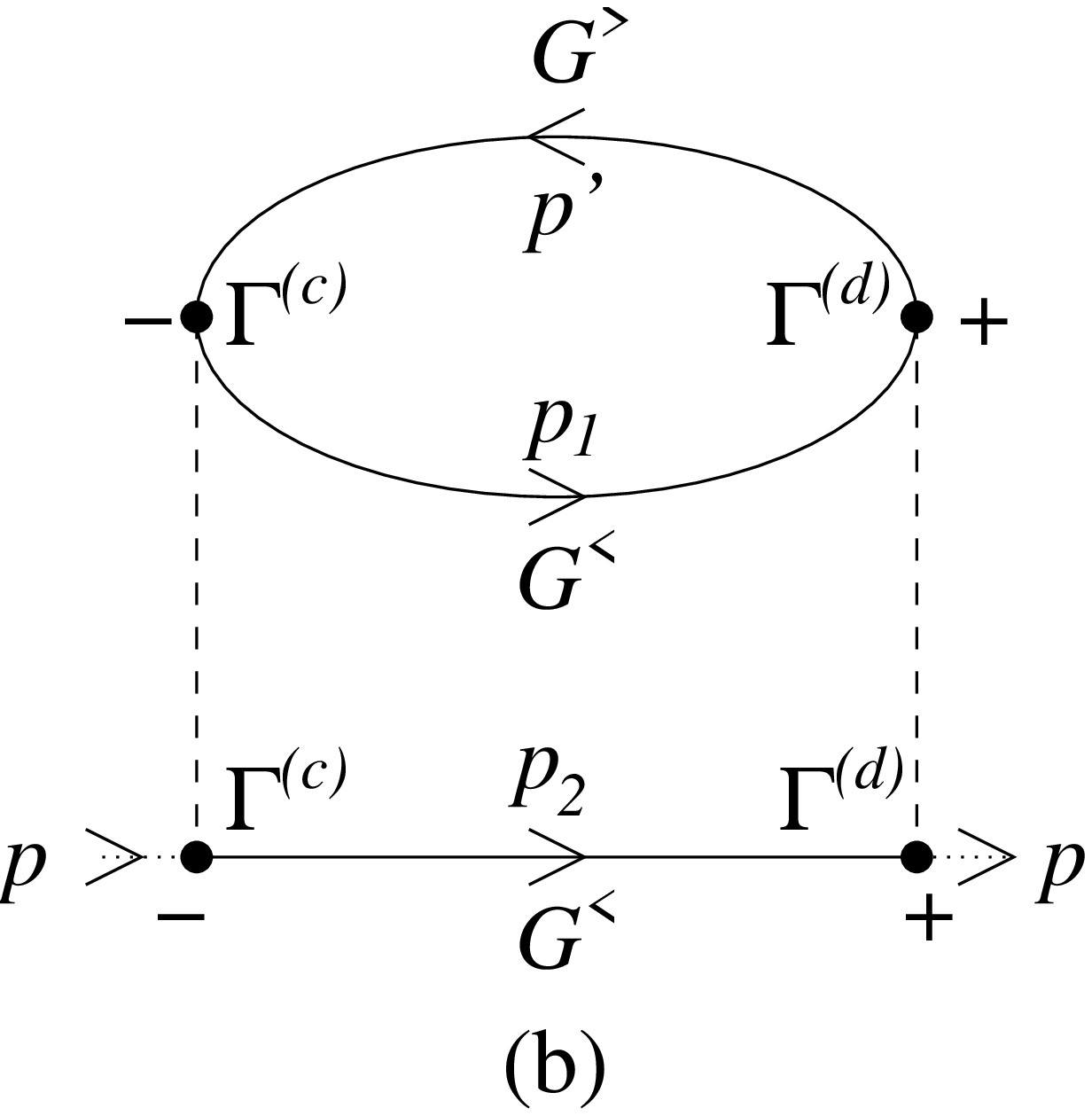} \\[0.3cm]
\includegraphics[scale=0.5]{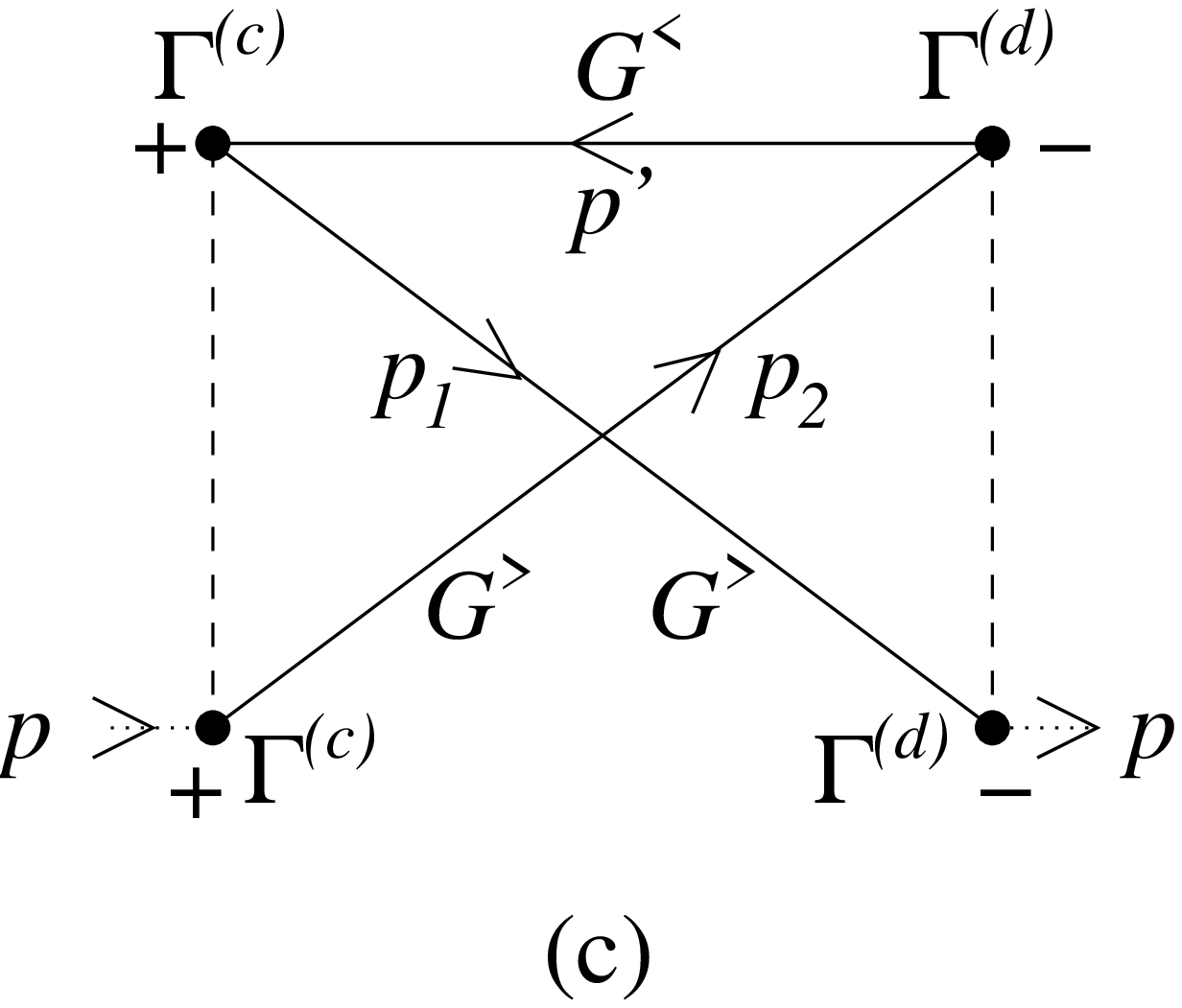} \hspace*{2cm}
\includegraphics[scale=0.5]{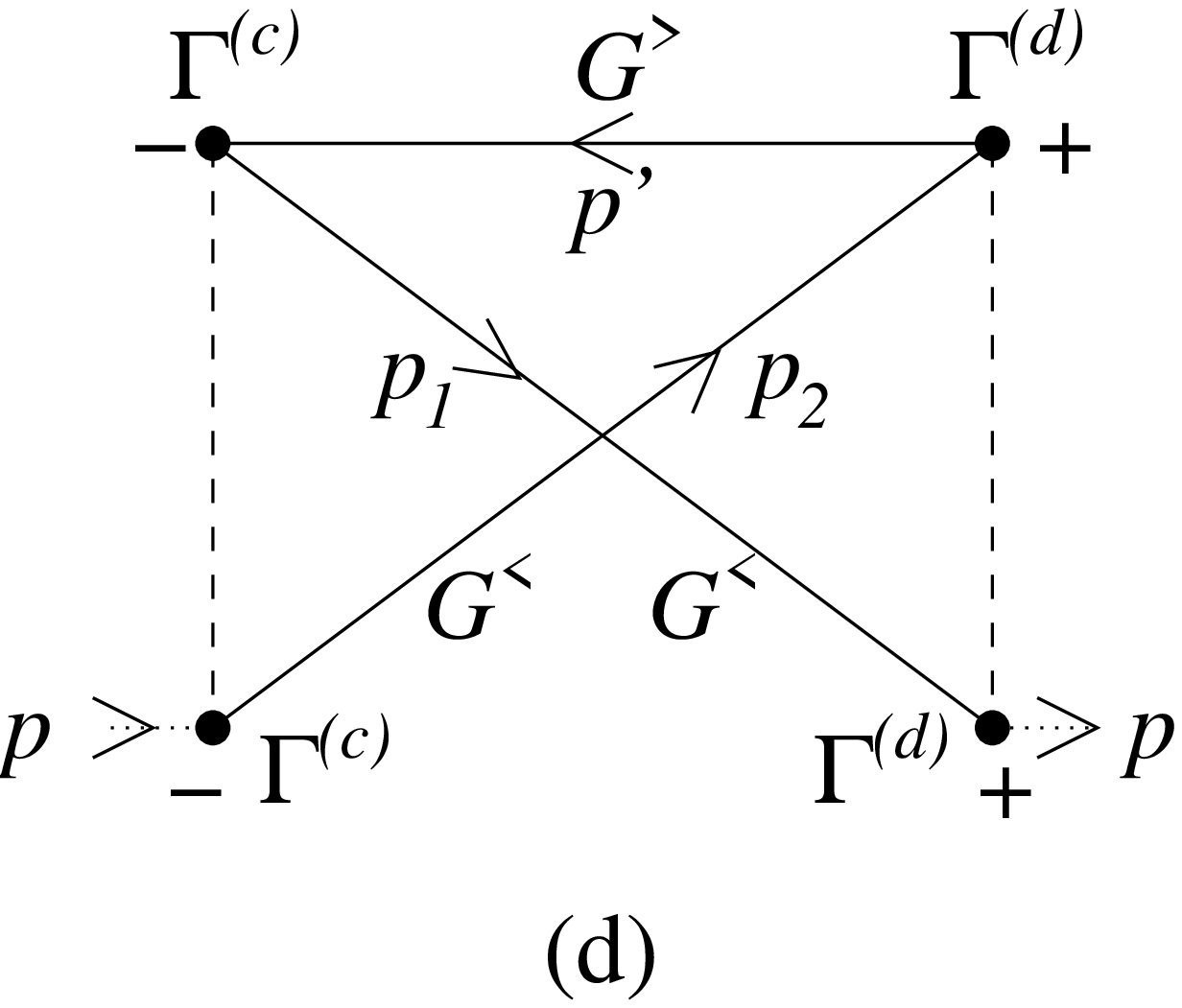}
\end{figure}

For binary elastic scattering, the self-energies $\Sigma^{\gtrless}(x,p)$ will
be taken in $T$-matrix approximation, where they are
given by the Feynman diagrams shown in Fig.\ \ref{fig:feynman},
\begin{align}
\Sigma^{\gtrless}(x,p) & =  4 \frac{G_{c}G_{d}}{\hbar^2} \int \frac{\d^4 p_1}{(2 \pi \hbar)^4}
\frac{\d^4 p_2}{(2 \pi \hbar)^4} \frac{\d^4 p'}{(2 \pi \hbar)^4} \,
(2\pi\hbar)^{4}\delta^{(4)}(p+p'-p_{1}-p_{2})\nonumber\\
&\times \left\{ \mathrm{Tr}\left[\Gamma^{(d)}G^{\gtrless}(x,p_{1})\Gamma^{(c)}G^{\lessgtr}(x,p')\right]
 \Gamma^{(d)}G^{\gtrless}(x,p_{2})\Gamma^{(c)} - \Gamma^{(d)}G^{\gtrless}(x,p_{1})
 \Gamma^{(c)}G^{\lessgtr}(x,p')\Gamma^{(d)}
 G^{\gtrless}(x,p_{2})\Gamma^{(c)} \right\}\;,\label{eq:self-en-1} 
\end{align}
The first term in the second line of Eq.~(\ref{eq:self-en-1}) 
corresponds to the ``direct diagrams'' of Figs.\ \ref{fig:feynman}(a) and (b),
while the last term corresponds to the ``exchange diagrams''  of Figs.\ \ref{fig:feynman}(c) and (d).
The coupling constants $G_{c}$ and $G_{d}$ are associated with the one-boson-exchange interactions
in channel $(c)$ or channel $(d)$, respectively. Each vertex carries a factor $\hbar^{-1}$, giving rise to
the factor $\hbar^{-2}$ (which was missed
in Eq.\ (131) of Ref.\ \cite{Sheng:2021kfc}).

\subsubsection{First order in $\hbar$}

To first order in $\hbar$, we need to compute the real and imaginary part of
\begin{equation} \label{eq:collision_orderone}
\mathcal{I}_0 \coloneqq \mathrm{Tr} 
\left[ (\dblone + \gamma_5 \slashed{\ms}) (\slashed{p}+m) \left( \Sigma^{<(0)} G^{>(0)} - \Sigma^{>(0)} G^{<(0)} \right)
\right]\;,
\end{equation}
cf.\ Eqs.\ (\ref{eq:massshell_orderone}), (\ref{eq:Boltzmann_orderone}).
The self-energies $\Sigma^{\lessgtr(0)}$ are given by Eq.\ (\ref{eq:self-en-1}), with all Wigner functions
taken at zeroth order in $\hbar$, i.e., by Eq.\ (\ref{eq:Gsmallerlarger_orderzero}). Inserting these expressions
into $\Sigma^{\lessgtr(0)}$ and the result into the trace (\ref{eq:collision_orderone}), we obtain
with the cyclicity property of the trace, with Eq.\ (\ref{eq:useful_id}), as well as using the idempotency of $\Lambda^+(p)$ the result
\begin{align}
\mathcal{I}_0 & = 8 m^4 \frac{G_c G_d}{\hbar^2} \, 4 \pi m \hbar\, \delta(p^2-m^2)  \int \d P_1 \, \d P_2 \,\d P' \, (2 \pi \hbar)^4 \delta^{(4)}(p+p' - p_1 - p_2) 
\nonumber \\
& \hspace*{4.5cm}  \times 
\mathcal{T}_0
\left[ f^{(0)}_1 f^{(0)}_2 \bar{f}^{\prime(0)} \bar{f}^{(0)} -  \bar{f}^{(0)}_1 \bar{f}^{(0)}_2 f^{\prime(0)} f^{(0)} \right]
\;, \label{eq:collisionterm_orderone}
\end{align}
where we introduced the abbreviations
\begin{equation}
f^{(0)}_1 \coloneqq f^{(0)}(x,p_1)\;, \quad f^{(0)}_2 \coloneqq f^{(0)}(x,p_2) \;, \quad f^{\prime(0)} \coloneqq f^{(0)}(x,p')
\;, \quad f^{(0)} \coloneqq f^{(0)}(x,p)\;,
\end{equation}
and similarly for $\bar{f}^{(0)}_1$, $\bar{f}^{(0)}_2$, $\bar{f}^{\prime(0)}$, and $\bar{f}^{(0)}$, respectively.
We also defined
\begin{align}
\mathcal{T}_0 & \coloneqq \mathrm{Tr} \left[ (\dblone +  \gamma_5 \slashed{\ms}) \Lambda^+(p)
\Gamma^{(d)}\Lambda^+(p_2) \Gamma^{(c)}\right]
\mathrm{Tr} \left[  \Gamma^{(d)} \Lambda^+(p_1) \Gamma^{(c)}  \Lambda^+(p')\right] \nonumber \\
& \; - \mathrm{Tr} \left[ (\dblone +  \gamma_5 \slashed{\ms}) \Lambda^+(p)
\Gamma^{(d)} \Lambda^+(p_1) \Gamma^{(c)}  \Lambda^+(p') \Gamma^{(d)} \Lambda^+(p_2) \Gamma^{(c)} \right]\;.  \label{eq:T0}
\end{align}
In App.\ \ref{app:B1} we prove
that, because of the symmetry of
the integrand in Eq.\ (\ref{eq:collisionterm_orderone}) under the
exchange $p_1^\mu \leftrightarrow p_2^\mu$, only the real
part of $\mathcal{T}_0$ contributes, wherefore in the following we will
set $\mathrm{Im} \mathcal{T}_0 =0$
under the integral.

Inserting Eq.\ (\ref{eq:collisionterm_orderone}) into Eqs.\ (\ref{eq:massshell_orderone}) and 
(\ref{eq:Boltzmann_orderone}), we obtain with Eq.\ (\ref{eq:frakf_onshell}) 
for the mass-shell constraint and the Boltzmann equation at order $\mathcal{O}(\hbar)$
\begin{subequations}
\begin{align}
(p^2 - m^2) \, \mathfrak{f}^{(1)}(x,p,\ms) 
 & = 0 \;, 
\label{eq:massshell_orderone_2} \\
p \cdot \partial_x \,f^{(0)}(x,p) & = 2m^4  \frac{ G_c G_d}{\hbar^2}   \int  \d P_1 \, \d P_2 \, \d P' \,(2 \pi \hbar)^4 \delta^{(4)}(p+p' - p_1 - p_2) 
\nonumber \\
& \hspace*{2.5cm}\times \mathrm{Re}\mathcal{T}_0 \,
\left[ f^{(0)}_1 f^{(0)}_2 \bar{f}^{\prime(0)} \bar{f}^{(0)} -  \bar{f}^{(0)}_1 \bar{f}^{(0)}_2 f^{\prime(0)} f^{(0)} \right] \;. 
\label{eq:Boltzmann_orderone_2}
\end{align}
\end{subequations}
The right-hand side of Eq.\ (\ref{eq:massshell_orderone_2})
vanishes because
$\mathrm{Im} \mathcal{T}_0 =0$.
This has the consequence that
$\mathfrak{f}^{(1)}$ is
on-shell, which is consistent
with Eq.\ (\ref{eq:f1}), see also Ref.~\cite{Wang:2020pej}.

In order to facilitate
comparison with the result
from the GLW approach, we extend
the integration on
the right-hand side of Eq.\ (\ref{eq:Boltzmann_orderone_2})
to extended phase space, 
$\d P_1 \d P_2 \d P' \rightarrow \d \Gamma_1 \d \Gamma_2 \d \Gamma' \d \bar{S}(p)$, using
the relations (\ref{eq:intspin}).
Since the integrand does not depend on any of the spin variables
$\ms^\mu_1$, $\ms^\mu_2$, $\ms^{\prime \mu}$, and $\bar{\ms}^\mu$, we may also extend
the definition of $\mathcal{T}_0$
under the integral,
\begin{equation} \label{eq:def_T_trace}
\mathcal{T}_0
\equiv\mathcal{T}
\coloneqq 32 \left\{ \mathrm{Tr} \left[ 
h \Gamma^{(d)} h_2 \Gamma^{(c)}
\bar{h} \right]
\mathrm{Tr} \left[  \Gamma^{(d)} h_1
\Gamma^{(c)}  h'\right] 
- \mathrm{Tr} \left[ h
\Gamma^{(d)} h_1 \Gamma^{(c)}  
h' \Gamma^{(d)} h_2 \Gamma^{(c)} 
\bar{h} \right]
\right\}\;,
\end{equation}
where we used Eq.\ (\ref{eq:def_h}).
With this, we obtain from
Eq.\ (\ref{eq:Boltzmann_orderone_2})
\begin{align}p \cdot \partial_x \,f^{(0)}(x,p) & = \frac{1}{4}  \int  \d \Gamma_1 \, \d \Gamma_2 \, \d \Gamma' \, \d \bar{S}(p)\, (2 \pi \hbar)^4 \delta^{(4)}(p+p' - p_1 - p_2) \, \mathcal{W} 
\left[ f^{(0)}_1 f^{(0)}_2 \bar{f}^{\prime(0)} \bar{f}^{(0)} -  \bar{f}^{(0)}_1 \bar{f}^{(0)}_2 f^{\prime(0)} f^{(0)} \right] \;, 
\label{eq:Boltzmann_orderone_3}
\end{align}
where
\begin{equation}
\label{eq:def_W_element}
\mathcal{W} \equiv
\frac{m^4}{2} \frac{G_c G_d}{\hbar^2}
\, \mathrm{Re} \mathcal{T}\;,
\end{equation}
cf.\ Eq.\ (\ref{eq:W_NJL}).

\subsubsection{Second order in \protect $\hbar$}

At order $\mathcal{O}(\hbar^2)$, we need to compute three different traces, cf.\ 
Eqs.\ (\ref{eq:massshell_ordertwo}) and (\ref{eq:Boltzmann_ordertwo}). The first one is
\begin{equation}
\mathcal{I}_1 \coloneqq \mathrm{Tr} 
\left[ (\dblone + \gamma_5 \slashed{\ms}) \slashed{\partial}_x \left( \Sigma^{<(0)} G^{>(0)} - \Sigma^{>(0)} G^{<(0)} \right)
\right]\;.
\end{equation}
The last term in parentheses under the trace is the same which already occurred in Eq.\ (\ref{eq:collision_orderone}).
Accounting for the additional $\gamma_\mu$ matrix in $\slashed{\partial}_x$ and the fact that
the partial derivative in this term acts on all distribution functions appearing
in the self-energies $\Sigma^{\lessgtr(0)}$ and in the Wigner functions $G^{\gtrless(0)}$, we readily obtain
\begin{align}
\mathcal{I}_1 &  = 4 m^3\frac{ G_c G_d}{\hbar^2} \, 4 \pi m \hbar\,\delta(p^2-m^2) \int \d P_1 \, \d P_2 \, \d P' \, (2 \pi \hbar)^4 \delta^{(4)}(p+p' - p_1 - p_2) 
\nonumber \\
& \times  \mathcal{T}^{(a)}_\mu \left[ \left( \partial_x^\mu f^{(0)}_1 \right) f^{(0)}_2 \bar{f}^{\prime (0)} \bar{f}^{(0)}
- \left( \partial_x^\mu \bar{f}^{(0)}_1 \right) \bar{f}^{(0)}_2 f^{\prime (0)} f^{(0)} 
 + f^{(0)}_1 \left( \partial_x^\mu f^{(0)}_2 \right)  \bar{f}^{\prime (0)} \bar{f}^{(0)}
- \bar{f}^{(0)}_1 \left( \partial_x^\mu \bar{f}^{(0)}_2 \right)  f^{\prime (0)} f^{(0)} \right. \nonumber \\
& \hspace*{1cm} + \left. f^{(0)}_1 f^{(0)}_2 \left( \partial_x^\mu \bar{f}^{\prime (0)} \right)   \bar{f}^{(0)}
- \bar{f}^{(0)}_1 \bar{f}^{(0)}_2 \left( \partial_x^\mu f^{\prime (0)} \right)   f^{(0)} 
 +  f^{(0)}_1 f^{(0)}_2 \bar{f}^{\prime (0)} \left( \partial_x^\mu \bar{f}^{(0)}  \right)  
- \bar{f}^{(0)}_1 \bar{f}^{(0)}_2 f^{\prime (0)} \left( \partial_x^\mu f^{(0)} \right)    \right] \;, \label{eq:I1}
\end{align}
where
\begin{align}
\mathcal{T}^{(a)}_\mu & \coloneqq \mathrm{Tr} \left[ (\dblone + \gamma_5 \slashed{\ms}) \Lambda^+(p) \gamma_\mu
\Gamma^{(d)} \Lambda^+(p_2) \Gamma^{(c)} \right]
\mathrm{Tr} \left[  \Gamma^{(d)} \Lambda^+(p_1) \Gamma^{(c)}  \Lambda^+(p')\right] \nonumber \\
& \; - \mathrm{Tr} \left[ (\dblone + \gamma_5 \slashed{\ms}) \Lambda^+(p) \gamma_\mu
\Gamma^{(d)} \Lambda^+(p_1) \Gamma^{(c)}  \Lambda^+(p') \Gamma^{(d)} \Lambda^+(p_2) 
\Gamma^{(c)} \right] \;.\label{eq:def_Ta}
\end{align}

The second trace we need to compute is
\begin{equation} \label{eq:trace2_ordertwo}
\mathcal{I}_2 \coloneqq \mathrm{Tr} \left[ (\dblone +  \gamma_5 \slashed{\ms}) (\slashed{p}+m)
\left( \Sigma^{<(1)} G^{>(0)} - \Sigma^{>(1)} G^{<(0)} +\Sigma^{<(0)} G^{>(1)} - \Sigma^{>(0)} G^{<(1)}  \right) \right]\;.
\end{equation}
Here, 
\begin{align}
\lefteqn{\Sigma^{\gtrless(1)}(x,p)  =  4 \frac{G_{c}G_{d}}{\hbar^2}\int  \frac{\d^4 p_1}{(2 \pi \hbar)^4}
\frac{\d^4 p_2}{(2 \pi \hbar)^4}\frac{\d^4 p'}{(2 \pi \hbar)^4}\, 
(2\pi\hbar)^{4}\delta^{(4)}(p+p'-p_{1}-p_{2})}\nonumber\\
&\times \left\{ \mathrm{Tr}\left[\Gamma^{(d)}G^{\gtrless(1)}(x,p_{1})\Gamma^{(c)}G^{\lessgtr(0)}(x,p')\right]
 \Gamma^{(d)}G^{\gtrless(0)}(x,p_{2})\Gamma^{(c)} - \Gamma^{(d)}G^{\gtrless(1)}(x,p_{1})
 \Gamma^{(c)}G^{\lessgtr(0)}(x,p')\Gamma^{(d)}
 G^{\gtrless(0)}(x,p_{2})\Gamma^{(c)}  \right. \nonumber \\
& +\, \mathrm{Tr}\left[\Gamma^{(d)}G^{\gtrless(0)}(x,p_{1})\Gamma^{(c)}G^{\lessgtr(1)}(x,p')\right]
 \Gamma^{(d)}G^{\gtrless(0)}(x,p_{2})\Gamma^{(c)} - \Gamma^{(d)}G^{\gtrless(0)}(x,p_{1})
 \Gamma^{(c)}G^{\lessgtr(1)}(x,p')\Gamma^{(d)}
 G^{\gtrless(0)}(x,p_{2})\Gamma^{(c)}   \nonumber \\
& + \left. \mathrm{Tr}\left[\Gamma^{(d)}G^{\gtrless(0)}(x,p_{1})\Gamma^{(c)}G^{\lessgtr(0)}(x,p')\right]
 \Gamma^{(d)}G^{\gtrless(1)}(x,p_{2})\Gamma^{(c)} - \Gamma^{(d)}G^{\gtrless(0)}(x,p_{1})
 \Gamma^{(c)}G^{\lessgtr(0)}(x,p')\Gamma^{(d)}
 G^{\gtrless(1)}(x,p_{2})\Gamma^{(c)}   
 \right\}\;.\label{eq:self-en-orderone} 
\end{align}
According to Eq.\ (\ref{eq:Gonesplit}), each of the Wigner functions $G^{\gtrless(1)}$ contains
two terms, a quasi-classical contribution and a gradient contribution. Since
$G^{\gtrless(1)}$ appears linearly in all terms in the trace (\ref{eq:trace2_ordertwo}), the latter also
splits into two parts, 
\begin{equation} \label{eq:I2}
\mathcal{I}_{2} \equiv \mathcal{I}_{2 \textrm{qc}} + \mathcal{I}_{2 \nabla}\;.
\end{equation} 
The first part, $\mathcal{I}_{2 \textrm{qc}}$,  contains 
the quasi-classical parts $G^{\gtrless(1)}_{\textrm{qc}}$, and
the second part, $\mathcal{I}_{2 \nabla}$, contains 
the gradient parts $G^{\gtrless(1)}_{\nabla}$. 

Let us first focus on $\mathcal{I}_{2 \textrm{qc}}$.
Inserting $G^{\lessgtr(1)}_{\textrm{qc}}$ from Eqs.\ (\ref{eq:Gsmallerqp1}) and (\ref{eq:Glargerqp1})
as well as $G^{\lessgtr(0)}$ from Eq.\ (\ref{eq:Gsmallerlarger_orderzero}), we obtain
\begin{align}
\mathcal{I}_{2 \textrm{qc}} & = \frac{m^4}{2} \, \frac{ G_c G_d}{\hbar^2} \, 4 \pi m \hbar\,\delta(p^2-m^2) 
\int \d\Gamma_1 \, \d\Gamma_2\, \d\Gamma' \d\bar{S}(p)\,
(2 \pi \hbar)^4 \delta^{(4)}(p + p'-p_1 - p_2)\nonumber \\
& \times \left\{ \mathcal{T}_1 \left[ f^{(1)}_1 f^{(0)}_2 \bar{f}^{\prime(0)} \bar{f}^{(0)} 
-  \bar{f}^{(1)}_1 \bar{f}^{(0)}_2 f^{\prime(0)} f^{(0)} \right]  
+  \mathcal{T}_2 \left[ f^{(0)}_1 f^{(1)}_2 \bar{f}^{\prime(0)} \bar{f}^{(0)} 
-  \bar{f}^{(0)}_1 \bar{f}^{(1)}_2 f^{\prime(0)} f^{(0)} \right] \right.  \nonumber \\
& \hspace*{0.1cm}+ \left. \mathcal{T}' \left[ f^{(0)}_1 f^{(0)}_2 \bar{f}^{\prime(1)} \bar{f}^{(0)} 
-  \bar{f}^{(0)}_1 \bar{f}^{(0)}_2 f^{\prime(1)} f^{(0)} \right]   
 + \bar{\mathcal{T}} \left[ f^{(0)}_1 f^{(0)}_2 \bar{f}^{\prime(0)} \bar{f}^{(1)} 
-  \bar{f}^{(0)}_1 \bar{f}^{(0)}_2 f^{\prime(0)} f^{(1)} \right] \right\}   \;,  \label{eq:I2qp}
\end{align}
where we defined
\begin{subequations}
\label{eq:def_Ts}
\begin{align}
\mathcal{T}_1 &\coloneqq \mathrm{Tr} \left[ (\dblone + \gamma_5 \slashed{\ms}) \Lambda^+(p)
\Gamma^{(d)} \Lambda^+(p_2) \Gamma^{(c)} \right]
\mathrm{Tr} \left[  \Gamma^{(d)} (\dblone + \gamma_5 \slashed{\ms}_1)\Lambda^+(p_1) 
\Gamma^{(c)}  \Lambda^+(p')\right] \nonumber \\
& \;- \mathrm{Tr} \left[ (\dblone + \gamma_5 \slashed{\ms}) \Lambda^+(p)
\Gamma^{(d)} (\dblone + \gamma_5 \slashed{\ms}_1)\Lambda^+(p_1) \Gamma^{(c)}  
\Lambda^+(p') \Gamma^{(d)} \Lambda^+(p_2) \Gamma^{(c)} \right] \;,\label{T_1} \\
\mathcal{T}_2 &\coloneqq \mathrm{Tr} \left[ (\dblone + \gamma_5 \slashed{\ms}) \Lambda^+(p)
\Gamma^{(d)} (\dblone + \gamma_5 \slashed{\ms}_2)\Lambda^+(p_2) \Gamma^{(c)} \right]
\mathrm{Tr} \left[  \Gamma^{(d)} \Lambda^+(p_1) 
\Gamma^{(c)}  \Lambda^+(p')\right] \nonumber \\
& \;- \mathrm{Tr} \left[ (\dblone + \gamma_5 \slashed{\ms}) \Lambda^+(p)
\Gamma^{(d)} \Lambda^+(p_1) \Gamma^{(c)}  
\Lambda^+(p') \Gamma^{(d)} (\dblone + \gamma_5 \slashed{\ms}_2)\Lambda^+(p_2) \Gamma^{(c)} \right] \;, \\
\mathcal{T}' &\coloneqq \mathrm{Tr} \left[ (\dblone + \gamma_5 \slashed{\ms}) \Lambda^+(p)
\Gamma^{(d)} \Lambda^+(p_2) \Gamma^{(c)} \right]
\mathrm{Tr} \left[  \Gamma^{(d)} \Lambda^+(p_1) 
\Gamma^{(c)}  (\dblone + \gamma_5 \slashed{\ms}')\Lambda^+(p')\right] \nonumber \\
&\; - \mathrm{Tr} \left[ (\dblone + \gamma_5 \slashed{\ms}) \Lambda^+(p)
\Gamma^{(d)} \Lambda^+(p_1) \Gamma^{(c)}  (\dblone + \gamma_5 \slashed{\ms}')
\Lambda^+(p') \Gamma^{(d)} \Lambda^+(p_2) \Gamma^{(c)} \right] \;, \\
\bar{\mathcal{T}} &\coloneqq \mathrm{Tr} \left[ (\dblone + \gamma_5 \slashed{\ms}) \Lambda^+(p)
\Gamma^{(d)} \Lambda^+(p_2) \Gamma^{(c)} (\dblone + \gamma_5 \Dslash{\bar{\ms}})\right]
\mathrm{Tr} \left[  \Gamma^{(d)} \Lambda^+(p_1) 
\Gamma^{(c)}  \Lambda^+(p')\right] \nonumber \\
& \;- \mathrm{Tr} \left[ (\dblone + \gamma_5 \slashed{\ms}) \Lambda^+(p)
\Gamma^{(d)} \Lambda^+(p_1) \Gamma^{(c)}  
\Lambda^+(p') \Gamma^{(d)} \Lambda^+(p_2) \Gamma^{(c)} (\dblone + \gamma_5 \Dslash{\bar{\ms}})\right] \;,\label{T_bar}
\end{align}
\end{subequations}
and introduced the abbreviations
\begin{equation}
f^{(1)}_1 \coloneqq f^{(1)}(x,p_1,\ms_1)\;, \quad f^{(1)}_2 \coloneqq f^{(1)}(x,p_2,\ms_2) \;, \quad 
f^{\prime(1)} \coloneqq f^{(1)}(x,p',\ms')\;, \quad f^{(1)} \coloneqq f^{(1)}(x,p,\bar{\ms})\;,
\end{equation}
and similarly for $\bar{f}^{(1)}_1$, $\bar{f}^{(1)}_2$, $\bar{f}^{\prime(1)}$, and $\bar{f}^{(1)}$, respectively.
Note that, in Eq.\ (\ref{eq:I2qp}), we extended the phase-space integration from $\d P_1\, \d P_2\, \d P'$ to
$\d\Gamma_1 \,\d\Gamma_2 \,\d\Gamma' \,\d\bar{S}(p)/16$. Because of Eq.\ (\ref{eq:intspin}), this merely inserts 
a factor of one in all terms which do not depend on the respective spin vector. Because of this, we
may also extend the definition
of the quantities 
(\ref{eq:def_Ts}) so that, under
the extended phase-space integration,
all become identical,
\begin{equation}
 \mathcal{T}_1  \equiv
\mathcal{T}_2 \equiv \mathcal{T}'
\equiv \bar{\mathcal{T}} 
\equiv \mathcal{T} 
\;, 
\end{equation}
where we used Eq.\ (\ref{eq:def_T_trace}).
Consequently,
\begin{align}
\mathcal{I}_{2 \textrm{qc}} & = \frac{m^4}{2} \, \frac{ G_c G_d}{\hbar^2} \, 4 \pi m \hbar\, \delta(p^2-m^2) 
\int \d\Gamma_1 \, \d\Gamma_2\, \d\Gamma' \d\bar{S}(p)\,
(2 \pi \hbar)^4 \delta^{(4)}(p + p'-p_1 - p_2) \nonumber \\
& \times \mathcal{T}  \left[ f^{(1)}_1 f^{(0)}_2 \bar{f}^{\prime(0)} \bar{f}^{(0)} 
-  \bar{f}^{(1)}_1 \bar{f}^{(0)}_2 f^{\prime(0)} f^{(0)} + f^{(0)}_1 f^{(1)}_2 \bar{f}^{\prime(0)} \bar{f}^{(0)} 
-  \bar{f}^{(0)}_1 \bar{f}^{(1)}_2 f^{\prime(0)} f^{(0)}  \right.  \nonumber \\
& \hspace*{0.4cm}+ \left.  f^{(0)}_1 f^{(0)}_2 \bar{f}^{\prime(1)} \bar{f}^{(0)} 
-  \bar{f}^{(0)}_1 \bar{f}^{(0)}_2 f^{\prime(1)} f^{(0)} + f^{(0)}_1 f^{(0)}_2 \bar{f}^{\prime(0)} \bar{f}^{(1)} 
-  \bar{f}^{(0)}_1 \bar{f}^{(0)}_2 f^{\prime(0)} f^{(1)} \right]   \;. \label{eq:I2qp_2}
\end{align}

We now consider the gradient part $\mathcal{I}_{2 \nabla}$. Inserting 
$G^{\lessgtr(1)}_{\nabla}$ from Eqs.\ (\ref{eq:Gsmallernabla1}) and (\ref{eq:Glargernabla1})
as well as $G^{\lessgtr(0)}$ from Eq.\ (\ref{eq:Gsmallerlarger_orderzero}), we obtain
\begin{align}
\mathcal{I}_{2 \nabla} & = 2 m^2 \,  \frac{ G_c G_d}{\hbar^2} \, 4 \pi m \hbar\,\delta(p^2-m^2) 
\int \d P_1 \, \d P_2\, \d P' \, (2 \pi \hbar)^4 \delta^{(4)}(p + p'-p_1 - p_2)\nonumber \\
& \times \left\{ \mathcal{T}_{1,\mu \nu} \, p_1^\nu \left[ \left( \partial_x^\mu f^{(0)}_1 \right) 
f^{(0)}_2 \bar{f}^{\prime(0)} \bar{f}^{(0)} 
-  \left( \partial_x^\mu \bar{f}^{(0)}_1 \right)  \bar{f}^{(0)}_2 f^{\prime(0)} f^{(0)} \right]  \right. \nonumber \\
& \hspace*{0.3cm} +  \mathcal{T}_{2,\mu \nu} \, p_2^\nu \left[ f^{(0)}_1 \left( \partial_x^\mu f^{(0)}_2 \right)  
\bar{f}^{\prime(0)} \bar{f}^{(0)} -  \bar{f}^{(0)}_1 \left( \partial_x^\mu \bar{f}^{(0)}_2 \right)  f^{\prime(0)} f^{(0)} \right] 
\nonumber \\
& \hspace*{0.3cm}+ \mathcal{T}^{\prime}_{\mu \nu} \, p^{\prime \nu} \left[ f^{(0)}_1 f^{(0)}_2 
\left( \partial_x^\mu \bar{f}^{\prime (0)} \right)  \bar{f}^{(0)} 
-  \bar{f}^{(0)}_1 \bar{f}^{(0)}_2 \left( \partial_x^\mu f^{\prime(0)} \right)  f^{(0)} \right]   \nonumber \\
&\hspace*{0.3cm} +\left. \mathcal{T}_{\mu \nu} \, p^\nu \left[ f^{(0)}_1 f^{(0)}_2 
\bar{f}^{\prime(0)} \left( \partial_x^\mu \bar{f}^{(0)} \right)  
-  \bar{f}^{(0)}_1 \bar{f}^{(0)}_2 f^{\prime(0)} \left( \partial_x^\mu f^{(0)} \right)  \right] \right\}   \;, 
\label{eq:I2nabla}
\end{align}
where we defined
\begin{subequations}\label{eq:def_T_tens}
\begin{align}
\mathcal{T}_{1,\mu \nu} &\coloneqq \mathrm{Tr} \left[ (\dblone + \gamma_5 \slashed{\ms}) \Lambda^+(p)
\Gamma^{(d)} \Lambda^+(p_2) \Gamma^{(c)} \right]
\mathrm{Tr} \left[  \Gamma^{(d)} \sigma_{\mu \nu} \Gamma^{(c)}  \Lambda^+(p')\right] \nonumber \\
& \; - \mathrm{Tr} \left[ (\dblone + \gamma_5 \slashed{\ms}) \Lambda^+(p)
\Gamma^{(d)} \sigma_{\mu \nu} \Gamma^{(c)}  
\Lambda^+(p') \Gamma^{(d)} \Lambda^+(p_2) \Gamma^{(c)} \right] \;, \label{def_T1_tens} \\
\mathcal{T}_{2,\mu \nu} &\coloneqq \mathrm{Tr} \left[ (\dblone + \gamma_5 \slashed{\ms}) \Lambda^+(p)
\Gamma^{(d)} \sigma_{\mu \nu} \Gamma^{(c)} \right]
\mathrm{Tr} \left[  \Gamma^{(d)} \Lambda^+(p_1) 
\Gamma^{(c)}  \Lambda^+(p')\right] \nonumber \\
& \; - \mathrm{Tr} \left[ (\dblone +  \gamma_5 \slashed{\ms}) \Lambda^+(p)
\Gamma^{(d)} \Lambda^+(p_1) \Gamma^{(c)}  
\Lambda^+(p') \Gamma^{(d)} \sigma_{\mu \nu} \Gamma^{(c)} \right] \;,\label{def_T2_tens}  \\
\mathcal{T}_{\mu \nu}^{\prime} &\coloneqq \mathrm{Tr} \left[ (\dblone + \gamma_5 \slashed{\ms}) \Lambda^+(p)
\Gamma^{(d)} \Lambda^+(p_2) \Gamma^{(c)} \right]
\mathrm{Tr} \left[  \Gamma^{(d)} \Lambda^+(p_1) 
\Gamma^{(c)}  \sigma_{\mu \nu}\right] \nonumber \\
& \; - \mathrm{Tr} \left[ (\dblone + \gamma_5 \slashed{\ms}) \Lambda^+(p)
\Gamma^{(d)} \Lambda^+(p_1) \Gamma^{(c)}  \sigma_{\mu \nu} 
\Gamma^{(d)} \Lambda^+(p_2) \Gamma^{(c)} \right] \;, \label{def_T'_tens} \\
\mathcal{T}_{\mu \nu} &\coloneqq \mathrm{Tr} \left[ (\dblone + \gamma_5 \slashed{\ms}) \Lambda^+(p)
\Gamma^{(d)} \Lambda^+(p_2) \Gamma^{(c)} \sigma_{\mu \nu}\right]
\mathrm{Tr} \left[  \Gamma^{(d)} \Lambda^+(p_1) 
\Gamma^{(c)}  \Lambda^+(p')\right] \nonumber \\
& \; - \mathrm{Tr} \left[ (\dblone + \gamma_5 \slashed{\ms}) \Lambda^+(p)
\Gamma^{(d)} \Lambda^+(p_1) \Gamma^{(c)}  
\Lambda^+(p') \Gamma^{(d)} \Lambda^+(p_2) \Gamma^{(c)} \sigma_{\mu \nu}\right] \;.
\end{align}
\end{subequations}

 The third trace we need to compute is
\begin{equation} \label{eq:I3}
\mathcal{I}_3 \coloneqq \mathrm{Tr} \left[ (\dblone + \gamma_5 \slashed{\ms}) (\slashed{p}+m)
\left( \left\{ \Sigma^{<(0)}, G^{>(0)} \right\}_\textrm{PB}
- \left\{ \Sigma^{>(0)}, G^{<(0)} \right\}_\textrm{PB} \right) \right]\;.
\end{equation}
For the Poisson-bracket terms, we need
\begin{subequations}
\begin{align}
\partial_x^\mu G^{<(0)} & = 4 \pi m \hbar\, \delta(p^2-m^2) \Lambda^+(p) \partial_x^\mu f^{(0)}\;, \quad
\partial_p^\mu G^{<(0)}  = 4 \pi m \hbar\, \delta(p^2-m^2) \left[ \frac{\gamma^\mu}{2m} + \Lambda^+(p) \partial_p^\mu \right] f^{(0)}\;, 
\label{eq:gradientsGsmaller}\\
\partial_x^\mu G^{>(0)}  & = 4 \pi m \hbar\, \delta(p^2-m^2) \Lambda^+(p) \partial_x^\mu \bar{f}^{(0)}\;, \quad
\partial_p^\mu G^{>(0)}  = 4 \pi m \hbar\, \delta(p^2-m^2) \left[ \frac{\gamma^\mu}{2m} + \Lambda^+(p) \partial_p^\mu \right] 
\bar{f}^{(0)}\;, \label{eq:gradientsGlarger}
\end{align}
\end{subequations}
where we neglected off-shell terms $\sim \delta'(p^{2}-m^2)$
in the equalities on the right-hand side. We also need
\begin{subequations}
\begin{align}
\partial_x^\mu \Sigma^{<(0)}(x,p) & =  4 m^3 \frac{ G_c G_d}{\hbar^2}
\int \d P_1 \, \d P_2\, \d P' \, (2\pi\hbar)^{4}\delta^{(4)}(p+p'-p_{1}-p_{2})\nonumber\\
&\times \left\{ \mathrm{Tr}\left[\Gamma^{(d)}\Lambda^+(p_{1})\Gamma^{(c)}\Lambda^+(p')\right]
 \Gamma^{(d)}\Lambda^+(p_2)\Gamma^{(c)} - \Gamma^{(d)}\Lambda^+(p_{1})
 \Gamma^{(c)}\Lambda^+(p')\Gamma^{(d)}
 \Lambda^+(p_2) \Gamma^{(c)} \right\} \nonumber \\
 & \times \left[ \left(  \partial_x^\mu f^{(0)}_1 \right) f^{(0)}_2 \bar{f}^{\prime(0)} 
+ f^{(0)}_1 \left( \partial_x^\mu  f^{(0)}_2\right)  \bar{f}^{\prime(0)}  
+  f^{(0)}_1 f^{(0)}_2\left( \partial_x^\mu  \bar{f}^{\prime(0)}   \right) \right]   \;, \\
\partial_x^\mu \Sigma^{>(0)}(x,p) & =  4 m^3 \frac{ G_c G_d}{\hbar^2}
\int \d P_1 \, \d P_2\, \d P' \, (2\pi\hbar)^{4}\delta^{(4)}(p+p'-p_{1}-p_{2})\nonumber\\
&\times \left\{ \mathrm{Tr}\left[\Gamma^{(d)}\Lambda^+(p_{1})\Gamma^{(c)}\Lambda^+(p')\right]
 \Gamma^{(d)}\Lambda^+(p_2)\Gamma^{(c)} - \Gamma^{(d)}\Lambda^+(p_{1})
 \Gamma^{(c)}\Lambda^+(p')\Gamma^{(d)}
 \Lambda^+(p_2) \Gamma^{(c)} \right\} \nonumber \\
 & \times \left[ \left(  \partial_x^\mu \bar{f}^{(0)}_1 \right) \bar{f}^{(0)}_2 f^{\prime(0)} 
+ \bar{f}^{(0)}_1 \left( \partial_x^\mu  \bar{f}^{(0)}_2\right)  f^{\prime(0)}  
+  \bar{f}^{(0)}_1 \bar{f}^{(0)}_2\left( \partial_x^\mu  f^{\prime(0)}   \right) \right]   \;, \\
\partial_p^\mu \Sigma^{<(0)}(x,p) & =  - 4 m^3 \frac{ G_c G_d}{\hbar^2}
\int \d P_1 \, \d P_2\, \d P' \, (2\pi\hbar)^{4}\delta^{(4)}(p+p'-p_{1}-p_{2})\nonumber\\
&\times \left( \mathrm{Tr}\left\{\Gamma^{(d)}\Lambda^+(p_{1})\Gamma^{(c)}
\left[ \frac{\gamma^\mu}{2m} + \Lambda^+(p') \partial_{p'}^\mu \right] \right\}
 \Gamma^{(d)}\Lambda^+(p_2)\Gamma^{(c)} \right. \nonumber \\
 & \hspace*{0.3cm} - \left. \Gamma^{(d)}\Lambda^+(p_{1})
 \Gamma^{(c)}\left[ \frac{\gamma^\mu}{2m} + \Lambda^+(p') \partial_{p'}^\mu \right]\Gamma^{(d)}
 \Lambda^+(p_2) \Gamma^{(c)} \right)   f^{(0)}_1 f^{(0)}_2\bar{f}^{\prime(0)}   \;, \label{eq:gradientspSigmasmaller} \\
\partial_p^\mu \Sigma^{>(0)}(x,p) & =  - 4 m^3 \frac{ G_c G_d}{\hbar^2}
\int \d P_1 \, \d P_2\, \d P' \, (2\pi\hbar)^{4}\delta^{(4)}(p+p'-p_{1}-p_{2})\nonumber\\
&\times \left( \mathrm{Tr}\left\{\Gamma^{(d)}\Lambda^+(p_{1})\Gamma^{(c)}
\left[ \frac{\gamma^\mu}{2m} + \Lambda^+(p') \partial_{p'}^\mu \right] \right\}
 \Gamma^{(d)}\Lambda^+(p_2)\Gamma^{(c)} \right. \nonumber \\
 & \hspace*{0.3cm} - \left. \Gamma^{(d)}\Lambda^+(p_{1})
 \Gamma^{(c)}\left[ \frac{\gamma^\mu}{2m} + \Lambda^+(p') \partial_{p'}^\mu \right]\Gamma^{(d)}
 \Lambda^+(p_2) \Gamma^{(c)} \right)  \bar{f}^{(0)}_1 \bar{f}^{(0)}_2f^{\prime(0)}   \;.
 \label{eq:gradientspSigmalarger}
\end{align}
\end{subequations}
In Eqs.\ (\ref{eq:gradientspSigmasmaller}) and (\ref{eq:gradientspSigmalarger}), we used the fact that,
because of the energy-momentum conserving delta function, $\partial^\mu_p \equiv \partial^\mu_{p'}$, and then integrated by parts. In the course of the latter, we have neglected off-shell terms $\sim \delta'(p^{\prime 2}-m^2)$. Inserting Eqs.\ (\ref{eq:gradientsGsmaller})
-- (\ref{eq:gradientspSigmalarger}) into Eq.\ (\ref{eq:I3}), we arrive at
\begin{align}
\mathcal{I}_3 & = 4 m^3  \frac{ G_c G_d}{\hbar^2} \, 4 \pi m \hbar\, \delta(p^2-m^2) \int \d P_1 \, \d P_2\, \d P' \, (2\pi\hbar)^{4}\delta^{(4)}(p+p'-p_{1}-p_{2})
\nonumber\\
& \times \left\{ 2 m \, \mathcal{T}_0 \left[ \left( \partial_x^\mu f^{(0)}_1 \right) f^{(0)}_2 \bar{f}^{\prime(0)} 
\left( \partial_\mu^p \bar{f}^{(0)} \right) -  \left( \partial_x^\mu \bar{f}^{(0)}_1 \right)  \bar{f}^{(0)}_2 f^{\prime(0)} 
\left( \partial_\mu^p f^{(0)} \right) \right. \right. \nonumber \\
& \hspace*{1.3cm} + f^{(0)}_1  \left( \partial_x^\mu f^{(0)}_2\right)  \bar{f}^{\prime(0)} 
\left( \partial_\mu^p \bar{f}^{(0)} \right)
-  \bar{f}^{(0)}_1\left( \partial_x^\mu \bar{f}^{(0)}_2 \right)   f^{\prime(0)} \left(\partial_\mu^p f^{(0)} \right) \nonumber \\
& \hspace*{1.3cm} + f^{(0)}_1 f^{(0)}_2 \left( \partial_x^\mu\bar{f}^{\prime(0)} \right)  
\left( \partial_\mu^p \bar{f}^{(0)} \right)
-  \bar{f}^{(0)}_1 \bar{f}^{(0)}_2\left( \partial_x^\mu  f^{\prime(0)} \right)   \left(\partial_\mu^p f^{(0)} \right) \nonumber \\
& \hspace*{1.3cm} + \left. f^{(0)}_1 f^{(0)}_2 \left( \partial_{p'}^\mu\bar{f}^{\prime(0)} \right)  
\left( \partial_\mu^x \bar{f}^{(0)} \right)
-  \bar{f}^{(0)}_1 \bar{f}^{(0)}_2\left( \partial_{p'}^\mu  f^{\prime(0)} \right)  \left( \partial_\mu^x f^{(0)} \right)
\right]\nonumber \\
& \hspace*{0.3cm} + \mathcal{T}^{(b)}_\mu \left[ \left( \partial_x^\mu f^{(0)}_1 \right) f^{(0)}_2 
\bar{f}^{\prime (0)} \bar{f}^{(0)} - \left( \partial_x^\mu \bar{f}^{(0)}_1 \right) \bar{f}^{(0)}_2 f^{\prime (0)} f^{(0)} 
\right. \nonumber \\
&  \hspace*{1.3cm} + f^{(0)}_1 \left( \partial_x^\mu f^{(0)}_2 \right)  \bar{f}^{\prime (0)} \bar{f}^{(0)}
- \bar{f}^{(0)}_1 \left( \partial_x^\mu \bar{f}^{(0)}_2 \right)  f^{\prime (0)} f^{(0)}  \nonumber \\
& \hspace*{1.3cm} + \left. f^{(0)}_1 f^{(0)}_2 \left( \partial_x^\mu \bar{f}^{\prime (0)} \right)   \bar{f}^{(0)}
- \bar{f}^{(0)}_1 \bar{f}^{(0)}_2 \left( \partial_x^\mu f^{\prime (0)} \right)   f^{(0)} \right] \nonumber \\
& \hspace*{0.3cm}+ \left. \mathcal{T}^{(c)}_\mu \left[  f^{(0)}_1  f^{(0)}_2 \bar{f}^{\prime (0)}
 \left(\partial_{x}^{\mu} \bar{f}^{(0)}\right)
-  \bar{f}^{(0)}_1 \bar{f}^{(0)}_2 f^{\prime (0)}\left( \partial_{x}^{\mu}f^{(0)} \right)\right] \right\}\;, \label{eq:I3final}
\end{align}
where we have used Eq.\ (\ref{eq:T0}) and defined
\begin{subequations}
\label{eq:def_Tmubc}
\begin{align}
\mathcal{T}^{(b)}_\mu & \coloneqq \mathrm{Tr} \left[ (\dblone + \gamma_5 \slashed{\ms}) \Lambda^+(p) 
\Gamma^{(d)} \Lambda^+(p_2) \Gamma^{(c)} \gamma_\mu\right]
\mathrm{Tr} \left[  \Gamma^{(d)} \Lambda^+(p_1) \Gamma^{(c)}  \Lambda^+(p')\right] \nonumber \\
& \; - \mathrm{Tr} \left[ (\dblone + \gamma_5 \slashed{\ms}) \Lambda^+(p) 
\Gamma^{(d)} \Lambda^+(p_1) \Gamma^{(c)}  \Lambda^+(p') \Gamma^{(d)} \Lambda^+(p_2) 
\Gamma^{(c)} \gamma_\mu \right] \;, \label{eq:def_Tb}\\
\mathcal{T}^{(c)}_\mu & \coloneqq \mathrm{Tr} \left[ (\dblone + \gamma_5 \slashed{\ms}) \Lambda^+(p) 
\Gamma^{(d)} \Lambda^+(p_2) \Gamma^{(c)} \right]
\mathrm{Tr} \left[  \Gamma^{(d)} \Lambda^+(p_1) \Gamma^{(c)}  \gamma_\mu \right] \nonumber \\
& \; - \mathrm{Tr} \left[ (\dblone + \gamma_5 \slashed{\ms}) \Lambda^+(p) 
\Gamma^{(d)} \Lambda^+(p_1) \Gamma^{(c)}  \gamma_\mu\Gamma^{(d)} \Lambda^+(p_2) 
\Gamma^{(c)} \right] \;.
\label{eq:def_Tmuc}
\end{align}
\end{subequations}
In App.\ \ref{app:B2} we prove that,
under an integral of
the same type as in Eq.\ (\ref{eq:I3final}), 
$\mathcal{T}_\mu^{(b)}
\equiv \mathcal{T}_\mu^{(a)*}$,
while in App.\ \ref{app:B3} we
prove that under the same type of integral the imaginary part of $\mathcal{T}_\mu^{(c) }$ vanishes.

Inserting Eqs.\ (\ref{eq:I1}), (\ref{eq:I2}) with (\ref{eq:I2qp_2}) and (\ref{eq:I2nabla}), and (\ref{eq:I3final}) into
Eqs.\ (\ref{eq:massshell_ordertwo}) and (\ref{eq:Boltzmann_ordertwo}),
and using the identities
of App.\ \ref{app:T}, we obtain 
\begin{subequations}
\begin{align}
 (p^2 - m^2) & \,\mathfrak{f}^{(2)}(x,p,\ms)  - \pi m \hbar \, \delta(p^2-m^2)\partial_x^2 \, f^{(0)}(x,p) 
 \nonumber \\
& = - \frac{m^3}{16} \frac{G_c G_d}{\hbar^2} \,4 \pi m \hbar\, \delta(p^2-m^2) \int \d\Gamma_1 \, \d\Gamma_2\, \d\Gamma' 
\d\bar{S}(p) \, (2\pi\hbar)^{4}\delta^{(4)}(p+p'-p_{1}-p_{2})
\nonumber\\
& \times \left\{  \left[2m \, \mathrm{Im} \mathcal{T} \,  f^{(1)}_1 + \mathrm{Re} X_{1,\mu}  \left( \partial_x^\mu f^{(0)}_1 \right) 
\right] f^{(0)}_2 
\bar{f}^{\prime (0)} \bar{f}^{(0)} 
- \left[2m \, \mathrm{Im} \mathcal{T} \,  \bar{f}^{(1)}_1 + \mathrm{Re} X_{1,\mu}  \left( \partial_x^\mu \bar{f}^{(0)}_1 \right) 
\right] \bar{f}^{(0)}_2 
f^{\prime (0)} f^{(0)} \right. \nonumber \\
& \hspace*{0.25cm} + f^{(0)}_1  \left[2m \, \mathrm{Im} \mathcal{T} \,  f^{(1)}_2 + \mathrm{Re} X_{2,\mu}  \left( \partial_x^\mu f^{(0)}_2 \right) 
\right] 
\bar{f}^{\prime (0)} \bar{f}^{(0)} 
- \bar{f}^{(0)}_1 \left[2m \, \mathrm{Im} \mathcal{T} \,  \bar{f}^{(1)}_2 + \mathrm{Re} X_{2,\mu}  \left( \partial_x^\mu \bar{f}^{(0)}_2 \right) 
\right] 
f^{\prime (0)} f^{(0)} \nonumber \\
& \hspace*{0.25cm}+ f^{(0)}_1
f^{(0)}_2\left[2m \, \mathrm{Im} \mathcal{T} \,  \bar{f}^{\prime (1)} + \mathrm{Re} X_{\mu}'  \left( \partial_x^\mu \bar{f}^{\prime(0)} \right) 
\right]  \bar{f}^{(0)} - \bar{f}^{(0)}_1 
\bar{f}^{(0)}_2 \left[2m \, \mathrm{Im} \mathcal{T} \,  f^{\prime (1)} + \mathrm{Re} X_{\mu}'  \left( \partial_x^\mu f^{\prime (0)} \right) 
\right] f^{(0)} 
\nonumber \\
& \hspace*{0.25cm}+ f^{(0)}_1
f^{(0)}_2 \bar{f}^{\prime (0)}\left[2m \, \mathrm{Im} \mathcal{T} \,  \bar{f}^{(1)} + \mathrm{Re} X_{\mu}  \left( \partial_x^\mu \bar{f}^{(0)} \right) 
\right]  - \bar{f}^{(0)}_1 
\bar{f}^{(0)}_2 f^{\prime (0)} \left[2m \, \mathrm{Im} \mathcal{T} \,  f^{(1)} + \mathrm{Re} X_{\mu}  \left( \partial_x^\mu f^{(0)} \right) 
\right]  
\nonumber \\
& \hspace*{0.25cm}- m \, \mathrm{Re} \mathcal{T}_0 \left[ \left( \partial_x^\mu f^{(0)}_1 \right) f^{(0)}_2 \bar{f}^{\prime(0)} 
\left( \partial_\mu^p \bar{f}^{(0)} \right) -  \left( \partial_x^\mu \bar{f}^{(0)}_1 \right)  \bar{f}^{(0)}_2 f^{\prime(0)} 
\left( \partial_\mu^p f^{(0)} \right) \right.  \nonumber \\
& \hspace*{1.5cm} + f^{(0)}_1  \left( \partial_x^\mu f^{(0)}_2\right)  \bar{f}^{\prime(0)} 
\left( \partial_\mu^p \bar{f}^{(0)} \right)
-  \bar{f}^{(0)}_1\left( \partial_x^\mu \bar{f}^{(0)}_2 \right)   f^{\prime(0)} \left(\partial_\mu^p f^{(0)} \right) \nonumber \\
& \hspace*{1.5cm} + f^{(0)}_1 f^{(0)}_2 \left( \partial_x^\mu\bar{f}^{\prime(0)} \right)  
\left( \partial_\mu^p \bar{f}^{(0)} \right)
-  \bar{f}^{(0)}_1 \bar{f}^{(0)}_2\left( \partial_x^\mu  f^{\prime(0)} \right)   \left(\partial_\mu^p f^{(0)} \right) \nonumber \\
& \hspace*{1.5cm} + \left. \left. f^{(0)}_1 f^{(0)}_2 \left( \partial_{p'}^\mu\bar{f}^{\prime(0)} \right)  
\left( \partial_\mu^x \bar{f}^{(0)} \right)
-  \bar{f}^{(0)}_1 \bar{f}^{(0)}_2\left( \partial_{p'}^\mu  f^{\prime(0)} \right)  \left( \partial_\mu^x f^{(0)} \right)
\right] \right\}\;, \label{eq:massshell_ordertwo_2}
\end{align} 
\begin{align}
p \cdot \partial_x & \, f^{(1)}(x,p,\ms)  = \frac{m^3}{16} 
\frac{G_c G_d}{\hbar^2} 
\int \d\Gamma_1 \, \d\Gamma_2\, \d\Gamma' 
\d\bar{S}(p) \, (2\pi\hbar)^{4}\delta^{(4)}(p+p'-p_{1}-p_{2})
\nonumber\\
& \times \left\{  \left[2m \, \mathrm{Re} \mathcal{T} \,  f^{(1)}_1 - \mathrm{Im} X_{1,\mu}  \left( \partial_x^\mu f^{(0)}_1 \right) 
\right] f^{(0)}_2 
\bar{f}^{\prime (0)} \bar{f}^{(0)} 
- \left[2m \, \mathrm{Re} \mathcal{T} \,  \bar{f}^{(1)}_1 - \mathrm{Im} X_{1,\mu}  \left( \partial_x^\mu \bar{f}^{(0)}_1 \right) 
\right] \bar{f}^{(0)}_2 
f^{\prime (0)} f^{(0)} \right. \nonumber \\
& \hspace*{0.25cm} + f^{(0)}_1  \left[2m \, \mathrm{Re} \mathcal{T} \,  f^{(1)}_2 - \mathrm{Im} X_{2,\mu}  \left( \partial_x^\mu f^{(0)}_2 \right) 
\right] 
\bar{f}^{\prime (0)} \bar{f}^{(0)} 
- \bar{f}^{(0)}_1 \left[2m \, \mathrm{Re} \mathcal{T} \,  \bar{f}^{(1)}_2 - \mathrm{Im} X_{2,\mu}  \left( \partial_x^\mu \bar{f}^{(0)}_2 \right) 
\right] 
f^{\prime (0)} f^{(0)} \nonumber \\
& \hspace*{0.25cm}+ f^{(0)}_1
f^{(0)}_2 \left[2m \, \mathrm{Re} \mathcal{T} \,  \bar{f}^{\prime (1)} - \mathrm{Im} X_{\mu}'  \left( \partial_x^\mu \bar{f}^{\prime(0)} \right) 
\right]  \bar{f}^{(0)} - \bar{f}^{(0)}_1 
\bar{f}^{(0)}_2 \left[2m \, \mathrm{Re} \mathcal{T} \,  f^{\prime (1)} - \mathrm{Im} X_{\mu}'  \left( \partial_x^\mu f^{\prime (0)} \right) 
\right]  f^{(0)} 
\nonumber \\
& \hspace*{0.25cm} \left. + f^{(0)}_1
f^{(0)}_2 \bar{f}^{\prime (0)} \left[2m \, \mathrm{Re} \mathcal{T} \,  \bar{f}^{(1)} - \mathrm{Im} X_{\mu}  \left( \partial_x^\mu \bar{f}^{(0)} \right) 
\right] -\bar{f}^{(0)}_1 
\bar{f}^{(0)}_2 f^{\prime (0)} \left[2m \, \mathrm{Re} \mathcal{T} \,  f^{(1)} - \mathrm{Im} X_{\mu}  \left( \partial_x^\mu f^{(0)} \right) 
\right]  
 \right\}\;, \label{eq:Boltzmann_ordertwo_2}
\end{align}
\end{subequations}
where we have used
Eq.\ (\ref{eq:f1}), employed $\mathrm{Im} \mathcal{T}_0 \equiv 0$ under the $\d P_1\, \d P_2$-integral, and defined
\begin{subequations}
\label{eq:def_X}
\begin{align}
X_{1,\mu} & \coloneqq \frac{1}{2} \left[ \mathcal{T}_\mu^{(a)} - \mathcal{T}_\mu^{(a)*} \right] - \frac{i}{2m}
\mathcal{T}_{1,\mu \nu} p_1^\nu
= i \left[
\mathrm{Im} \mathcal{T}^{(a)}_\mu - \frac{1}{2m} \mathcal{T}_{1,\mu \nu} p_1^\nu \right]\;, \label{eq:X1}\\
X_{2,\mu} & \coloneqq \frac{1}{2} \left[ \mathcal{T}_\mu^{(a)} - \mathcal{T}_\mu^{(a)*} \right] - \frac{i}{2m}
\mathcal{T}_{2,\mu \nu} p_2^\nu = i \left[
\mathrm{Im} \mathcal{T}^{(a)}_\mu - \frac{1}{2m} \mathcal{T}_{2,\mu \nu} p_2^\nu \right]\;, 
\label{eq:X2}\\
X_{\mu}^\prime & \coloneqq \frac{1}{2} \left[ \mathcal{T}_\mu^{(a)} - \mathcal{T}_\mu^{(a)*} \right] - \frac{i}{2m}
\mathcal{T}_{\mu \nu}^\prime p^{\prime \nu}= i \left[
\mathrm{Im} \mathcal{T}^{(a)}_\mu - \frac{1}{2m} \mathcal{T}_{\mu \nu}^\prime p^{\prime \nu} \right]\;, \label{eq:Xprime}\\
X_{\mu} & \coloneqq \frac{1}{2} \left[ \mathcal{T}_\mu^{(a)} - \mathcal{T}_\mu^{(c)} \right] - \frac{i}{2m}
\mathcal{T}_{\mu \nu} p^\nu
\equiv \mathrm{Re} \mathcal{T}_\mu^{(a)}
- \frac{1}{2} \mathcal{T}_\mu^{(c)} - \frac{p_\mu}{2m} \mathcal{T}_0\;.\label{eq:X}
\end{align}
\end{subequations}
In Eqs.\ (\ref{eq:X1}) -- (\ref{eq:X}), we used
the fact that $\mathcal{T}_\mu^{(b)} \equiv\mathcal{T}_\mu^{(a)*}$
under
the $\d \Gamma_1 \d \Gamma_2$ integral. 
For the second equality
in Eq.\ (\ref{eq:X}) we
employed Eq.\ (\ref{eq:Tmunupnu}).
Note that $X_\mu$ is
purely real under
the $\d \Gamma_1 \d \Gamma_2$ integral.

In the following, we focus on the Boltzmann equation (\ref{eq:Boltzmann_ordertwo_2}) and define
\begin{subequations}\label{eq:Delta_all}
\begin{align}
\bar{\Delta}_{1,\mu} & \coloneqq - \frac{\mathrm{Im} X_{1,\mu} }{
2m\, \mathrm{Re} \mathcal{T}} 
\equiv \frac{1}{2m\, \mathrm{Re} \mathcal{T}}\, \left[\frac{1}{2m}
\, \mathrm{Re} \mathcal{T}_{1,\mu \nu} p_1^\nu - \mathrm{Im} \mathcal{T}_\mu^{(a)}
 \right]\equiv \frac{1}{\hbar}\left(\Delta_{1,\mu}-\Delta_\mu\right) \;, \label{eq:Delta1}  \\
\bar{\Delta}_{2,\mu} & \coloneqq - \frac{\mathrm{Im} X_{2,\mu}}{
2m\, \mathrm{Re} \mathcal{T}} 
\equiv  \frac{1}{2m\, \mathrm{Re} \mathcal{T}}\, \left[
\frac{1}{2m}
\, \mathrm{Re} \mathcal{T}_{2,\mu \nu} p_2^\nu  - \mathrm{Im} \mathcal{T}_\mu^{(a)}
\right]\equiv \frac{1}{\hbar}\left(\Delta_{2,\mu}-\Delta_\mu\right) \;, \label{eq:Delta2}  \\
\bar{\Delta}_{\mu}^\prime & \coloneqq - \frac{\mathrm{Im} X_{\mu}^\prime }{
2m\, \mathrm{Re} \mathcal{T}} 
\equiv  \frac{1}{2m\, \mathrm{Re} \mathcal{T}}\, \left[ \frac{1}{2m}
\, \mathrm{Re} \mathcal{T}_{\mu \nu}^\prime p^{\prime \nu} -\mathrm{Im} \mathcal{T}_\mu^{(a)}
 \right] \;\,\equiv \frac{1}{\hbar}\left(\Delta'_{\mu}-\Delta_\mu\right) \;, \label{eq:Deltaprime}  \\
\bar{\Delta}_{\mu} & \coloneqq - \frac{\mathrm{Im} X_{\mu} }{
2m\, \mathrm{Re} \mathcal{T}} 
\equiv 0
\;. \label{eq:Delta} 
\end{align}
\end{subequations}
All identities right after
the definitions hold under the integral. 
The final identities,
which relate the 
barred $\Delta$'s on the left-hand sides to the space-time shifts \eqref{eq:def_Deltas_NJL} 
on the right-hand sides are
proven in App.\ \ref{app:B6},
using identities from
App.\ \ref{app:B4}.

Then, with Eq.\ (\ref{eq:def_W_element})
the Boltzmann equation 
(\ref{eq:Boltzmann_ordertwo_2}) reads
\begin{align}
p \cdot \partial_x  \, f^{(1)}(x,p,\ms) & =  \frac{1}{4}  
\int \d\Gamma_1 \, \d\Gamma_2\, \d\Gamma' 
\d\bar{S}(p) \, (2\pi\hbar)^{4}\delta^{(4)}(p+p'-p_{1}-p_{2}) \, \mathcal{W}
\nonumber\\
& \times \left\{  \left[ f^{(1)}_1 + 
\bar{\Delta}_{1,\mu}  \left( \partial_x^\mu f^{(0)}_1 \right) 
\right] f^{(0)}_2 
\bar{f}^{\prime (0)} \bar{f}^{(0)} 
- \left[\bar{f}^{(1)}_1 + \bar{\Delta}_{1, \mu}  \left( \partial_x^\mu \bar{f}^{(0)}_1 \right) 
\right] \bar{f}^{(0)}_2 
f^{\prime (0)} f^{(0)} \right. \nonumber \\
& \hspace*{0.25cm} + f^{(0)}_1 \left[ f^{(1)}_2 + 
\bar{\Delta}_{2,\mu}  \left( \partial_x^\mu f^{(0)}_2 \right) 
\right]  
\bar{f}^{\prime (0)} \bar{f}^{(0)} 
-  \bar{f}^{(0)}_1 \left[ \bar{f}^{(1)}_2 + 
\bar{\Delta}_{2,\mu}  \left( \partial_x^\mu \bar{f}^{(0)}_2 \right) 
\right] 
f^{\prime (0)} f^{(0)} \nonumber \\
& \hspace*{0.25cm}+ f^{(0)}_1
f^{(0)}_2  \left[ \bar{f}^{\prime (1)} + 
\bar{\Delta}_{\mu}^\prime \left( \partial_x^\mu \bar{f}^{\prime (0)} \right) 
\right] \bar{f}^{(0)} - \bar{f}^{(0)}_1 
\bar{f}^{(0)}_2 \left[ f^{\prime (1)} + 
\bar{\Delta}_{\mu}^\prime \left( \partial_x^\mu f^{\prime (0)} \right) 
\right]  f^{(0)} 
\nonumber \\
& \hspace*{0.25cm} \left. + f^{(0)}_1
f^{(0)}_2 \bar{f}^{\prime (0)} \left[ \bar{f}^{(1)} + 
\bar{\Delta}_{\mu} \left( \partial_x^\mu \bar{f}^{(0)} \right) 
\right]  -\bar{f}^{(0)}_1 
\bar{f}^{(0)}_2 f^{\prime (0)} \left[ f^{(1)} + 
\bar{\Delta}_{\mu} \left( \partial_x^\mu f^{(0)} \right) 
\right] 
 \right\}\;. \label{eq:Boltzmann_ordertwo_4}
\end{align}

\subsection{Summary}

Expanding $f(x+ \hbar \bar{\Delta},p, \ms)$ in powers of $\hbar$ around
$f^{(0)}(x,p)$, 
\begin{equation}
f(x + \hbar \bar{\Delta},p,\ms) =
f^{(0)}(x,p) + \hbar f^{(1)}(x,p,\ms)
+ \hbar \bar{\Delta}_\mu \partial^\mu_x
f^{(0)}(x,p) + \mathcal{O}(\hbar^2)\;,
\end{equation}
and similarly for $f_1$, $f_2$, $f'$, $\bar{f}$, $\bar{f}_1$, $\bar{f}_2$,
and $\bar{f}'$, we 
now combine the results
(\ref{eq:Boltzmann_orderone_3}),
(\ref{eq:Delta_all}), and (\ref{eq:Boltzmann_ordertwo_4}) to
write the complete Boltzmann equation for $f(x,p,\ms)$ up to first order
in $\hbar$ as,
\begin{align}
p \cdot \partial_x  \, f(x,p,\ms) & =  \frac{1}{4}  
\int \d\Gamma_1 \, \d\Gamma_2\, \d\Gamma' 
\d\bar{S}(p) \, (2\pi\hbar)^{4}\delta^{(4)}(p+p'-p_{1}-p_{2}) \, \mathcal{W}
\nonumber\\
& \times  \left[ 
f (x+ \Delta_1-\Delta,p_1,\ms_1) 
f (x+\Delta_2-\Delta,p_2,\ms_2)
\bar{f} (x+\Delta'-\Delta,p',\ms') \bar{f} (x,p, \bar{\ms}) \right. \nonumber \\
& \left. \;  - \bar{f} (x+\Delta_1-\Delta,p_1,\ms_1) 
\bar{f} (x+\Delta_2-\Delta,p_2,\ms_2)
f (x+\Delta'-\Delta,p',\ms') 
f (x,p, \bar{\ms}) \right] + \mathcal{O}(\hbar^2) \;. \label{eq:Boltzmann_final_KB}
\end{align}
This result agrees
with that of the GLW approach,
Eq.\ (\ref{eq:finalcollisionterm}), in the limit of
Boltzmann statistics,
where $\bar{f}$, $\bar{f}'$, $\bar{f}_1$, $\bar{f}_2 \rightarrow 1$, and generalizes it to the
case of quantum statistics.

\section{Conclusions and Outlook}
\label{sec:summary}

In this work, we have revisited the nonlocal collision term derived in Refs.\ \cite{Weickgenannt:2020aaf,Weickgenannt:2021cuo}. In those works, the nonlocality of the collision term manifested itself by certain space-time shifts (given by Eq.\ (\ref{eq:deltanon}) for a particle with momentum $p^\mu$ and spin vector $\s^\mu$) of the collision partners. However, the explicit dependence of these space-time shifts on the frame vector $\hat{t}^\mu = (1, \mathbf{0})$ violates Lorentz covariance.
In this work, we restored Lorentz covariance by carefully recomputing the collision term in the GLW approach \cite{DeGroot:1980dk}, and confirmed the result by a calculation within the KB \cite{Mrowczynski:1992hq} approach. This results in the more complicated, but manifestly Lorentz-covariant expressions (\ref{eq:def_Deltas_NJL}) for the space-time shifts. 

In future work, one should repeat the derivation of spin hydrodynamics along the lines of Ref.\ \cite{Weickgenannt:2022qvh,Weickgenannt:2022zxs} with the Lorentz-covariant space-time shifts (\ref{eq:def_Deltas_NJL}). For such a calculation one cannot use the space-time shifts (\ref{eq:def_Deltas_NJL}) directly, since the trace
$\mathcal{T}$
appears in the denominator, which depends on the spin variables, cf.\ Eq.\ (\ref{eq:def_T_trace}).
However, in order to perform the integrations over spin space, the latter should appear in the numerator, where one can apply the relations (\ref{eq:intspin}).
One therefore needs to resort
to the form of the nonlocal collision term as given, e.g., in Eq.\ (\ref{eq:Boltzmann_ordertwo_2}).
Here, both $\mathcal{T}$ and
the first-order distribution functions are linear in the spin variables and thus appear under the integral in a form where the relations (\ref{eq:intspin}) are 
applicable. In App.\ \ref{app:traces} we give the trace terms required for such a calculation for scalar boson exchange as a simple example.

\begin{acknowledgments}
The authors thank V.\ Ambru\cb{s}, U.\ Heinz, X.-L.\ Sheng, and Q.\ Wang for enlightening
discussions. This work is supported by the Deutsche Forschungsgemeinschaft
(DFG, German Research Foundation) through the Collaborative Research
Center CRC-TR 211 ``Strong-interaction matter under extreme conditions''
- project number 315477589 - TRR 211 and by the State of Hesse within the Research Cluster
ELEMENTS (Project ID 500/10.006).
D.W.\ acknowledges support by the Studienstiftung des deutschen Volkes 
(German Academic Scholarship Foundation).
\end{acknowledgments}

\appendix

\section{Prefactor in Eqs.\ (\ref{eq:def_onshell_terms}) and (\ref{eq:frakf_onshell})}
\label{EMTensor}

In this appendix, we convince ourselves of the 
correctness of the prefactor in 
Eqs.\ (\ref{eq:def_onshell_terms}) and (\ref{eq:frakf_onshell}).
To this end, consider the expression for the canonical
energy-momentum tensor in terms of the vector 
component of the Wigner function,
\begin{equation}
T^{\mu \nu} = \int \frac{\d^4p}{(2 \pi \hbar)^4}\, 
p^\nu \mathcal{V}^\mu\;.
\end{equation}
At order $\mathcal{O}(\hbar^0)$,
$\mathcal{V}^{(0)\mu} = p^\mu \mathcal{F}^{(0)}/m$,
cf.\ Eq.\ (\ref{eq:orderzero}), 
and we obtain with Eqs.\ (\ref{eq:f_frakf})
and (\ref{eq:frakf_onshell})
\begin{equation}
T^{\mu \nu} = 2 \int \frac{\d^4p}{(2 \pi \hbar)^4}\, 
p^\nu \frac{p^\mu}{m} \, 4 \pi m\hbar \, 
\delta(p^2 - m^2) \, f^{(0)}(x,p)
= 2 \int \d P
\, p^\mu p^\nu \, f^{(0)}(x,p)\;.
\end{equation}
This is the standard expression for the
energy-momentum tensor for a non-interacting
gas in kinetic theory. In equilibrium, $f^{(0)}$ is
the Fermi-Dirac distribution function,
$f^{(0)}_{\text{eq}} = [e^{(u \cdot p - \mu)/T}+1]^{-1}$. 
The overall factor 2 counts the
two spin degrees of freedom. Thus, Eqs.\ \eqref{eq:def_W_onshell} and 
(\ref{eq:frakf_onshell}) are correct.

\section{Properties of trace
terms}
\label{app:T}

In this appendix, we
prove a set of identities for
the traces $\mathcal{T}_0$,
$\mathcal{T}_\mu^{(a)}$,
$\mathcal{T}_\mu^{(b)}$,
$\mathcal{T}_\mu^{(c)}$,
$\mathcal{T}_{1,\mu \nu}p_1^\nu$,
$\mathcal{T}_{2,\mu \nu}p_2^\nu$,
$\mathcal{T}_{\mu \nu}'p^{'\nu}$,
and 
$\mathcal{T}_{\mu \nu}p^\nu$,
respectively. These identities
make use of the symmetry properties of the  $\d P_1 \d P_2$
integrals under which these various traces appear.

\subsection{$\mathcal{T}_0$
is real}
\label{app:B1}

We prove that, under
an integral of the form
\begin{equation}
\label{eq:integral}
\int \d P_1 \, \d P_2\,
\mathcal{T}_0\, f(p_1,p_2)\;,
\end{equation}
where $f(p_1,p_2)
= f(p_2,p_1)$, 
the imaginary part of 
$\mathcal{T}_0$ can be set to zero.
To this end, we compute
\begin{align}
\mathcal{T}_0^* & = \mathrm{Tr} \left[ (\dblone + \gamma_5\slashed{\s}
) \Lambda^+(p)
\Gamma^{(d)} \Lambda^+(p_2) \Gamma^{(c)} \right]^*
\mathrm{Tr} \left[  \Gamma^{(d)} \Lambda^+(p_1) \Gamma^{(c)} \Lambda^+(p')\right]^*  \nonumber \\
& - \mathrm{Tr} \left[ 
(\dblone + \gamma_5 \slashed{\s})
\Lambda^+(p)
\Gamma^{(d)} \Lambda^+(p_1) \Gamma^{(c)}  \Lambda^+(p') \Gamma^{(d)} \Lambda^+(p_2) \Gamma^{(c)} \right]^* \nonumber \\
& = \mathrm{Tr} \left[  \Gamma^{(c)\dagger} \Lambda^{+}(p_2)^\dagger 
\Gamma^{(d)\dagger } \Lambda^{+}(p)^\dagger
(\dblone + \gamma_5 \slashed{\s})^\dagger  \right]
\mathrm{Tr} \left[ \Lambda^{+}(p')^\dagger \Gamma^{(c)\dagger} \Lambda^{+}(p_1)^\dagger \Gamma^{(d)\dagger}\right] \nonumber \\
&- \mathrm{Tr} \left[ 
\Gamma^{(c)\dagger } \Lambda^{+}(p_2)^\dagger \Gamma^{(d)\dagger} \Lambda^{+}(p')^\dagger \Gamma^{(c)\dagger} \Lambda^{+}(p_1)^\dagger \Gamma^{(d)\dagger} \Lambda^{+}(p)^\dagger
(\dblone + \gamma_5 \slashed{\s})^\dagger \right] \;.
\end{align}
Using $\Lambda(p)^\dagger = \gamma^0 \Lambda^+(p) \gamma^0$ (and similarly for
$\Lambda^+(p_1)$, $\Lambda^+(p_2)$,
and $\Lambda^+(p^\prime)$), 
$(\dblone + \gamma_5 \slashed{\ms})^\dagger= 
\gamma^0(\dblone + \gamma_5 \slashed{\ms}) \gamma^0$,  as well as
$\Gamma^{(c)\dagger} = \gamma^0 \Gamma^{(c)} \gamma^0$ we obtain
\begin{align}
\mathcal{T}_0^* & = \mathrm{Tr} \left[  \Gamma^{(c)} \Lambda^{+}(p_2)
\Gamma^{(d)} \Lambda^{+}(p)
(\dblone + \gamma_5 \slashed{\s}) \right]
\mathrm{Tr} \left[ \Lambda^{+}(p') \Gamma^{(c)} \Lambda^{+}(p_1) \Gamma^{(d)}\right] \nonumber \\
&- \mathrm{Tr} \left[ 
\Gamma^{(c)} \Lambda^{+}(p_2) \Gamma^{(d)} \Lambda^{+}(p') \Gamma^{(c)} \Lambda^{+}(p_1) \Gamma^{(d)} \Lambda^{+}(p)
(\dblone + \gamma_5 \slashed{\s})\right]  \;.
\end{align}
We now exchange 
the summation indices
$c \leftrightarrow d$
in both terms, which is
possible, since the prefactor
$\sim G_c G_d$
of the collision integral is also symmetric under this exchange. Under an
integral of the form (\ref{eq:integral}), we are also
allowed to
exchange the integration 
variables $p_1^\mu
\leftrightarrow p_2^\mu$ in
the second term, because 
$f(p_1,p_2)$ is symmetric under this exchange. Finally, using Eq.\ (\ref{eq:useful_id})
and the cyclic 
property of the trace, we
arrive at
\begin{equation}
\mathcal{T}_0^* \equiv \mathcal{T}_0\;,\label{eq:T0_real}
\end{equation}
which proves that the imaginary part of $\mathcal{T}_0$ vanishes under the integral
(\ref{eq:integral}).

\subsection{ $\mathcal{T}_\mu^{(b)}$ is complex conjugate of
$\mathcal{T}_\mu^{(a)}$}
\label{app:B2}

We prove that, under an
integral of the form
\begin{equation}
\label{eq:integral2}
\int \d P_1 \, \d P_2\,
\mathcal{T}_\mu^{(b)}\, f(p_1,p_2)\;,
\end{equation}
where $f(p_1,p_2)
= f(p_2,p_1)$, we may set
$\mathcal{T}_\mu^{(b)}\equiv \mathcal{T}_\mu^{(a)*}$.
From Eq.\ (\ref{eq:def_Ta}) we compute
\begin{align}
\mathcal{T}^{(a)*}_\mu & = \mathrm{Tr} \left[ (\dblone + \gamma_5 \slashed{\ms}) \Lambda^+(p) \gamma_\mu
\Gamma^{(d)} \Lambda^+(p_2) \Gamma^{(c)} \right]^*
\mathrm{Tr} \left[  \Gamma^{(d)} \Lambda^+(p_1) \Gamma^{(c)}  \Lambda^+(p')\right]^* \nonumber \\
& \; - \mathrm{Tr} \left[ (\dblone + \gamma_5 \slashed{\ms}) \Lambda^+(p) \gamma_\mu
\Gamma^{(d)} \Lambda^+(p_1) \Gamma^{(c)}  \Lambda^+(p') \Gamma^{(d)} \Lambda^+(p_2) 
\Gamma^{(c)} \right]^* \nonumber\\
& = \mathrm{Tr} \left[  \Gamma^{(c)\dagger} \Lambda^{+}(p_2)^\dagger 
\Gamma^{(d)\dagger } \gamma_\mu^\dagger \Lambda^{+}(p)^\dagger
(\dblone + \gamma_5 \slashed{\s})^\dagger  \right]
\mathrm{Tr} \left[ \Lambda^{+}(p')^\dagger \Gamma^{(c)\dagger} \Lambda^{+}(p_1)^\dagger \Gamma^{(d)\dagger}\right] \nonumber \\
&- \mathrm{Tr} \left[ 
\Gamma^{(c)\dagger } \Lambda^{+}(p_2)^\dagger \Gamma^{(d)\dagger} \Lambda^{+}(p')^\dagger \Gamma^{(c)\dagger} \Lambda^{+}(p_1)^\dagger \Gamma^{(d)\dagger} 
\gamma_\mu^\dagger \Lambda^{+}(p)^\dagger
(\dblone + \gamma_5 \slashed{\s})^\dagger \right] \;.
\end{align}
Performing similar steps
as in App.\ \ref{app:B1},
because $f(p_1,p_2)$ in 
the integral
(\ref{eq:integral2}) is symmetric under $p_1^\mu \leftrightarrow p_2^\mu$, we then show that
\begin{align}
\mathcal{T}_\mu^{(a)*} & = \mathrm{Tr} \left[ 
(\dblone + \gamma_5 \slashed{\s}) \Lambda^{+}(p) \Gamma^{(d)} \Lambda^{+}(p_2)
\Gamma^{(c)}  \gamma_\mu \right]
\mathrm{Tr} \left[  \Gamma^{(d)} \Lambda^{+}(p_1) \Gamma^{(c)}
\Lambda^{+}(p')\right] \nonumber \\
&- \mathrm{Tr} \left[ 
(\dblone + \gamma_5 \slashed{\s})
\Lambda^{+}(p)
\Gamma^{(d)} \Lambda^{+}(p_1) \Gamma^{(c)} \Lambda^{+}(p') \Gamma^{(d)} \Lambda^{+}(p_2) \Gamma^{(c)} \gamma_\mu\right] 
\equiv \mathcal{T}_\mu^{(b)}\;,\label{eq:Ta_cc}
\end{align}
cf.\ Eq.\ (\ref{eq:def_Tb}).

\subsection{$\mathcal{T}_\mu^{(c)}$ is real}
\label{app:B3}

We show that,
under an integral
\begin{equation}
\label{eq:integral3}
\int \d P_1 \, \d P_2\,
\mathcal{T}_\mu^{(c)}\, f(p_1,p_2)\;,
\end{equation}
where $f(p_1,p_2)
= f(p_2,p_1)$, 
the imaginary part of $\mathcal{T}_\mu^{(c)}$ vanishes. To this end, we compute
\begin{align}
 \mathcal{T}_\mu^{(c)*} & =\mathrm{Tr} \left[ (\dblone + \gamma_5 \slashed{\ms}) \Lambda^+(p) 
\Gamma^{(d)} \Lambda^+(p_2) \Gamma^{(c)} \right]^*
\mathrm{Tr} \left[  \Gamma^{(d)} \Lambda^+(p_1) \Gamma^{(c)}  \gamma_\mu \right]^* \nonumber \\
& \; - \mathrm{Tr} \left[ (\dblone + \gamma_5 \slashed{\ms}) \Lambda^+(p) 
\Gamma^{(d)} \Lambda^+(p_1) \Gamma^{(c)}  \gamma_\mu\Gamma^{(d)} \Lambda^+(p_2) 
\Gamma^{(c)} \right]^* \nonumber \\
& = \mathrm{Tr} \left[  \Gamma^{(c)\dagger} \Lambda^{+}(p_2)^\dagger 
\Gamma^{(d)\dagger }  \Lambda^{+}(p)^\dagger
(\dblone + \gamma_5 \slashed{\s})^\dagger  \right]
\mathrm{Tr} \left[ \gamma_\mu^\dagger \Gamma^{(c)\dagger} \Lambda^{+}(p_1)^\dagger \Gamma^{(d)\dagger}\right] \nonumber \\
&- \mathrm{Tr} \left[ 
\Gamma^{(c)\dagger } \Lambda^{+}(p_2)^\dagger \Gamma^{(d)\dagger} \gamma_\mu^\dagger  \Gamma^{(c)\dagger} \Lambda^{+}(p_1)^\dagger \Gamma^{(d)\dagger} 
\Lambda^{+}(p)^\dagger
(\dblone + \gamma_5 \slashed{\s})^\dagger \right] \;.
\end{align}
Employing similar steps as 
in App.\ \ref{app:B1}, we obtain
\begin{align}
\mathcal{T}_\mu^{(c)*} & = \mathrm{Tr} \left[ 
(\dblone + \gamma_5 \slashed{\s}) \Lambda^{+}(p) \Gamma^{(d)} \Lambda^{+}(p_2)
\Gamma^{(c)}  \right]
\mathrm{Tr} \left[  \Gamma^{(d)} \Lambda^{+}(p_1) \Gamma^{(c)}
\gamma_\mu \right] \nonumber \\
&- \mathrm{Tr} \left[ 
(\dblone + \gamma_5 \slashed{\s})
\Lambda^{+}(p)
\Gamma^{(d)} \Lambda^{+}(p_1) \Gamma^{(c)} \gamma_\mu  \Gamma^{(d)} \Lambda^{+}(p_2) \Gamma^{(c)} \right] 
\equiv \mathcal{T}_\mu^{(c)}\;.\label{eq:Tc_cc}
\end{align}

\subsection{Identities for $\mathcal{T}_{1,\mu\nu}p_1^\nu$, $\mathcal{T}_{2,\mu\nu}p_2^\nu$, $\mathcal{T}'_{\mu\nu}p'^\nu$,
and $\mathcal{T}_{\mu\nu}p^\nu$}
\label{app:B4}

We first show that, under an
integral of the form
\begin{equation}
\label{eq:integral4a}
\int \d P_1 \, \d P_2\, 
\mathcal{T}^\prime_{\mu \nu}
 p^{\prime\nu} \, f(p_1,p_2)\;,
\end{equation}
where $f(p_1,p_2) = f(p_2,p_1)$,
$\mathcal{T}^\prime_{\mu \nu} p^{\prime \nu}$
assumes the form
\begin{align}
    \mathcal{T}_{\mu \nu}^{\prime}p'^\nu &\equiv 2m\,\mathrm{Im}\bigg\{\mathrm{Tr} \left[ (\dblone + \gamma_5 \slashed{\ms}) \Lambda^+(p)
\Gamma^{(d)} \Lambda^+(p_2) \Gamma^{(c)} \right]
\mathrm{Tr} \left[  \Gamma^{(d)} \Lambda^+(p_1) 
\Gamma^{(c)}  \Lambda^+(p')\gamma_\mu  \right] \nonumber \\
& \hspace*{1.3cm}\! - \mathrm{Tr} \left[ (\dblone + \gamma_5 \slashed{\ms}) \Lambda^+(p)
\Gamma^{(d)} \Lambda^+(p_1) \Gamma^{(c)}   \Lambda^+(p') \gamma_\mu
\Gamma^{(d)} \Lambda^+(p_2) \Gamma^{(c)} \right]\bigg\} \;. \label{eq:rel_T'}
\end{align}
To this end, note that
$\sigma_{\mu \nu} p^{\prime\nu}
= \frac{i}{2} [ \gamma_\mu, \slashed{p}'] = i m [ \gamma_\mu,
\Lambda^+(p')]$.
Therefore, Eq.\ (\ref{def_T'_tens}) (multiplied by $p^{\prime \nu}$) can be written as
\begin{align}
\mathcal{T}_{\mu \nu}^{\prime} p^{\prime \nu}  &= im \left\{ \mathrm{Tr} \left[ (\dblone + \gamma_5 \slashed{\ms}) \Lambda^+(p)
\Gamma^{(d)} \Lambda^+(p_2) \Gamma^{(c)} \right]
\mathrm{Tr} \left[  \Gamma^{(d)} \Lambda^+(p_1) 
\Gamma^{(c)}  \gamma_\mu
\Lambda^+(p') \right] \right. \nonumber \\
& \hspace*{0.75cm} - \mathrm{Tr} \left[ (\dblone + \gamma_5 \slashed{\ms}) \Lambda^+(p)
\Gamma^{(d)} \Lambda^+(p_1) \Gamma^{(c)}  \gamma_{\mu} \Lambda^+(p') 
\Gamma^{(d)} \Lambda^+(p_2) \Gamma^{(c)} \right] \nonumber \\
& \hspace*{0.75cm}  - \mathrm{Tr} \left[ (\dblone + \gamma_5 \slashed{\ms}) \Lambda^+(p)
\Gamma^{(d)} \Lambda^+(p_2) \Gamma^{(c)} \right]
\mathrm{Tr} \left[  \Gamma^{(d)} \Lambda^+(p_1) 
\Gamma^{(c)} \Lambda^+(p')  \gamma_\mu
\right] \nonumber \\
& \hspace*{0.75cm}   + \left. \! \mathrm{Tr} \left[ (\dblone + \gamma_5 \slashed{\ms}) \Lambda^+(p)
\Gamma^{(d)} \Lambda^+(p_1) \Gamma^{(c)}   \Lambda^+(p') \gamma_{\mu}
\Gamma^{(d)} \Lambda^+(p_2) \Gamma^{(c)} \right] \right\}\;.
\label{eq:Tprimemunupprimenu}
\end{align}
Using similar steps as
in App.\ \ref{app:B1}, we
exploit the symmetry under
the integral (\ref{eq:integral4a})
to show that the terms in the
first two lines in Eq.\ (\ref{eq:Tprimemunupprimenu}) 
are the complex conjugates
of the terms in the last two lines. This proves
Eq.\ (\ref{eq:rel_T'}).

Next, we show that, under an
integral of the form
\begin{equation}
\label{eq:integral4b}
\int \d P_1\, \d P_2 \left[\mathcal{T}_{1,\mu \nu}p_1^\nu \, f(p_1,p_2)
+ \mathcal{T}_{2,\mu \nu}p_2^\nu \, g(p_1,p_2) \right]\;,
\end{equation}
with real-valued
functions $f(p_1,p_2)$, $g(p_1,p_2)$ which fulfill $f(p_1,p_2) = g(p_2,p_1)$,
we may use the following
expressions for $\mathcal{T}_{1,\mu \nu}p_1^\nu $ and
$\mathcal{T}_{2,\mu \nu}p_2^\nu$,
\begin{subequations}
\begin{align}
\mathcal{T}_{1,\mu \nu}p_{1}^{\nu} &\equiv 2m\,\mathrm{Im}\bigg\{\mathrm{Tr} \left[ (\dblone + \gamma_5 \slashed{\ms}) \Lambda^+(p)
\Gamma^{(d)} \Lambda^+(p_2) \Gamma^{(c)} \right]
\mathrm{Tr} \left[  \Gamma^{(d)}\Lambda^+(p_1) \gamma_\mu \Gamma^{(c)}  \Lambda^+(p')\right] \nonumber \\
& \hspace*{1.3cm}\! - \mathrm{Tr} \left[ (\dblone + \gamma_5 \slashed{\ms}) \Lambda^+(p)
\Gamma^{(d)}  \Lambda^+(p_1) \gamma_\mu \Gamma^{(c)}  
\Lambda^+(p') \Gamma^{(d)} \Lambda^+(p_2) \Gamma^{(c)} \right]\bigg\} \;, \label{eq:rel_T1} \\
\mathcal{T}_{2,\mu \nu}p_{2}^{\nu} &\equiv 2m\,\mathrm{Im}\bigg\{\mathrm{Tr} \left[ (\dblone + \gamma_5 \slashed{\ms}) \Lambda^+(p)
\Gamma^{(d)}  \Lambda^+(p_2) \gamma_\mu\Gamma^{(c)} \right]
\mathrm{Tr} \left[  \Gamma^{(d)} \Lambda^+(p_1) 
\Gamma^{(c)}  \Lambda^+(p')\right] \nonumber \\
& \hspace*{1.3cm}\! - \mathrm{Tr} \left[ (\dblone +  \gamma_5 \slashed{\ms}) \Lambda^+(p)
\Gamma^{(d)} \Lambda^+(p_1) \Gamma^{(c)}  
\Lambda^+(p') \Gamma^{(d)} \Lambda^+(p_2)\gamma_\mu \Gamma^{(c)} \right] \bigg\}\;.\label{eq:rel_T2}
\end{align}
\end{subequations}
The proof follows similar steps
as that of Eq.\ (\ref{eq:rel_T'}).
We first write $\sigma_{\mu \nu} p_{1,2}^\nu = \frac{i}{2} [ \gamma_\mu, \slashed{p}_{1,2}]
= i m [ \gamma_\mu , \Lambda^+(p_{1,2})]$
and insert this into Eq.\
(\ref{def_T1_tens}) (multiplied by
$p_1^\nu$) and Eq.\
(\ref{def_T2_tens}) (multiplied by $p_2^\nu$), respectively,
\begin{subequations}
\begin{align}
\mathcal{T}_{1,\mu \nu} p_1^\nu  &= im \left\{ \mathrm{Tr} \left[ (\dblone + \gamma_5 \slashed{\ms}) \Lambda^+(p)
\Gamma^{(d)} \Lambda^+(p_2) \Gamma^{(c)} \right]
\mathrm{Tr} \left[  \Gamma^{(d)}  \gamma_\mu \Lambda^+(p_1) 
\Gamma^{(c)} 
\Lambda^+(p') \right] \right. \nonumber \\
& \hspace*{0.75cm} - \mathrm{Tr} \left[ (\dblone + \gamma_5 \slashed{\ms}) \Lambda^+(p)
\Gamma^{(d)} \gamma_{\mu} \Lambda^+(p_1) \Gamma^{(c)}  \Lambda^+(p') 
\Gamma^{(d)} \Lambda^+(p_2) \Gamma^{(c)} \right] \nonumber \\
& \hspace*{0.75cm}  - \mathrm{Tr} \left[ (\dblone + \gamma_5 \slashed{\ms}) \Lambda^+(p)
\Gamma^{(d)} \Lambda^+(p_2) \Gamma^{(c)} \right]
\mathrm{Tr} \left[  \Gamma^{(d)} \Lambda^+(p_1) \gamma_\mu
\Gamma^{(c)} \Lambda^+(p')  
\right] \nonumber \\
& \hspace*{0.75cm}   + \left. \! \mathrm{Tr} \left[ (\dblone + \gamma_5 \slashed{\ms}) \Lambda^+(p)
\Gamma^{(d)} \Lambda^+(p_1) \gamma_{\mu}\Gamma^{(c)}   \Lambda^+(p') 
\Gamma^{(d)} \Lambda^+(p_2) \Gamma^{(c)} \right] \right\}\;,
\label{eq:T1munup1nu} \\
\mathcal{T}_{2,\mu \nu} p_2^\nu  &= im \left\{ \mathrm{Tr} \left[ (\dblone + \gamma_5 \slashed{\ms}) \Lambda^+(p)
\Gamma^{(d)} \gamma_\mu \Lambda^+(p_2) \Gamma^{(c)} \right]
\mathrm{Tr} \left[  \Gamma^{(d)}  \Lambda^+(p_1) 
\Gamma^{(c)} 
\Lambda^+(p') \right] \right. \nonumber \\
& \hspace*{0.75cm} - \mathrm{Tr} \left[ (\dblone + \gamma_5 \slashed{\ms}) \Lambda^+(p)
\Gamma^{(d)} \Lambda^+(p_1) \Gamma^{(c)}  \Lambda^+(p') 
\Gamma^{(d)} \gamma_{\mu} \Lambda^+(p_2) \Gamma^{(c)} \right] \nonumber \\
& \hspace*{0.75cm}  - \mathrm{Tr} \left[ (\dblone + \gamma_5 \slashed{\ms}) \Lambda^+(p)
\Gamma^{(d)} \Lambda^+(p_2) \gamma_\mu\Gamma^{(c)} \right]
\mathrm{Tr} \left[  \Gamma^{(d)} \Lambda^+(p_1) 
\Gamma^{(c)} \Lambda^+(p')  
\right] \nonumber \\
& \hspace*{0.75cm}   + \left. \! \mathrm{Tr} \left[ (\dblone + \gamma_5 \slashed{\ms}) \Lambda^+(p)
\Gamma^{(d)} \Lambda^+(p_1) \Gamma^{(c)}   \Lambda^+(p') 
\Gamma^{(d)} \Lambda^+(p_2) \gamma_{\mu}\Gamma^{(c)} \right] \right\}\;,
\label{eq:T2munup2nu}
\end{align}
\end{subequations}
Multiplying
$\mathcal{T}_{1,\mu \nu}p_1^\nu$
with $f(p_1,p_2)$ and 
$\mathcal{T}_{2,\mu \nu}p_2^\nu$
with $g(p_1,p_2)$, and taking the sum, we then prove that, under the integral (\ref{eq:integral4b}) and using $f(p_1,p_2) = g(p_2,p_1)$, the
complex conjugate of terms resulting from the last
two lines in Eqs.\ (\ref{eq:T1munup1nu}) and
(\ref{eq:T2munup2nu}) is
\begin{align}
& \left\{ \mathrm{Tr} \left[ (\dblone + \gamma_5 \slashed{\ms}) \Lambda^+(p)
\Gamma^{(d)} \Lambda^+(p_2) \Gamma^{(c)} \right]
\mathrm{Tr} \left[  \Gamma^{(d)}  \Lambda^+(p_1) \gamma_\mu 
\Gamma^{(c)} 
\Lambda^+(p') \right] \right. \nonumber \\
& - \left. \! \mathrm{Tr} \left[ (\dblone + \gamma_5 \slashed{\ms}) \Lambda^+(p)
\Gamma^{(d)}  \Lambda^+(p_1) \gamma_{\mu}\Gamma^{(c)}  \Lambda^+(p') 
\Gamma^{(d)} \Lambda^+(p_2) \Gamma^{(c)} \right] \right\}^*
f(p_1,p_2) \nonumber \\
 + & \left\{ \mathrm{Tr} \left[ (\dblone + \gamma_5 \slashed{\ms}) \Lambda^+(p)
\Gamma^{(d)}  \Lambda^+(p_2) \gamma_\mu \Gamma^{(c)} \right]
\mathrm{Tr} \left[  \Gamma^{(d)}  \Lambda^+(p_1) 
\Gamma^{(c)} 
\Lambda^+(p') \right] \right. \nonumber \\
& - \left.\! \mathrm{Tr} \left[ (\dblone + \gamma_5 \slashed{\ms}) \Lambda^+(p)
\Gamma^{(d)} \Lambda^+(p_1) \Gamma^{(c)}  \Lambda^+(p') 
\Gamma^{(d)} \Lambda^+(p_2) \gamma_{\mu} \Gamma^{(c)} \right] 
\right\}^* g(p_1,p_2) \nonumber \\
= & \quad \, \mathrm{Tr} \left[ (\dblone + \gamma_5 \slashed{\ms}) \Lambda^+(p)
\Gamma^{(d)} \Lambda^+(p_2) \Gamma^{(c)} \right]
\mathrm{Tr} \left[  \Gamma^{(d)}  \gamma_\mu \Lambda^+(p_1) 
\Gamma^{(c)} 
\Lambda^+(p') \right] f(p_1,p_2) \nonumber \\
& - \mathrm{Tr} \left[ (\dblone + \gamma_5 \slashed{\ms}) \Lambda^+(p)
\Gamma^{(d)} \Lambda^+(p_1) \Gamma^{(c)}  \Lambda^+(p') 
\Gamma^{(d)} \gamma_{\mu}\Lambda^+(p_2)  \Gamma^{(c)} \right] g(p_1,p_2) \nonumber \\
 & + \mathrm{Tr} \left[ (\dblone + \gamma_5 \slashed{\ms}) \Lambda^+(p)
\Gamma^{(d)} \gamma_\mu  \Lambda^+(p_2) \Gamma^{(c)} \right]
\mathrm{Tr} \left[  \Gamma^{(d)}  \Lambda^+(p_1) 
\Gamma^{(c)} 
\Lambda^+(p') \right] g(p_1,p_2) \nonumber \\
& - \mathrm{Tr} \left[ (\dblone + \gamma_5 \slashed{\ms}) \Lambda^+(p)
\Gamma^{(d)} \gamma_{\mu} \Lambda^+(p_1)  \Gamma^{(c)}  \Lambda^+(p') 
\Gamma^{(d)}  \Lambda^+(p_2) \Gamma^{(c)} \right] 
f(p_1,p_2)\;.
\end{align}
This shows that, under
the integral (\ref{eq:integral4b})
and in the combination
$\mathcal{T}_{1,\mu \nu}p_1^\nu \, f(p_1,p_2) +
\mathcal{T}_{2,\mu \nu}p_2^\nu\, g(p_1,p_2)$,
the terms in the first and
second lines of Eqs.\ (\ref{eq:T1munup1nu}),
(\ref{eq:T2munup2nu}) are (up to a minus sign) the complex conjugates
of the terms in
the third and fourth lines.
This completes the proof of
Eqs.\ (\ref{eq:rel_T1}), (\ref{eq:rel_T2}).

Finally, we compute
\begin{align}
\frac{i}{2m}\, \mathcal{T}_{\mu \nu} p^\nu & = \frac{1}{2m}
\mathrm{Tr} \left[ (\dblone + \gamma_5 \slashed{\ms}) \Lambda^+(p)
\Gamma^{(d)} \Lambda^+(p_2) \Gamma^{(c)} i \sigma_{\mu \nu}p^\nu\right]
\mathrm{Tr} \left[  \Gamma^{(d)} \Lambda^+(p_1) 
\Gamma^{(c)}  \Lambda^+(p')\right] \nonumber \\
& \; - \frac{1}{2m} \mathrm{Tr} \left[ (\dblone + \gamma_5 \slashed{\ms}) \Lambda^+(p)
\Gamma^{(d)} \Lambda^+(p_1) \Gamma^{(c)}  
\Lambda^+(p') \Gamma^{(d)} \Lambda^+(p_2) \Gamma^{(c)} i \sigma_{\mu \nu}p^\nu\right] \;.
\end{align}
Using $i \sigma_{\mu \nu}p^\nu= - \gamma_\mu \slashed{p} + p_\mu$,
the cyclicity of the trace,
the relation
$\slashed{p} (\dblone + \gamma_5\slashed{\ms}) =(\dblone + \gamma_5\slashed{\ms}) \slashed{p}$, which holds since $p \cdot \ms =0$,
and the identity $\slashed{p} \Lambda^+(p)
= m \Lambda^+(p)$, which is valid for
on-shell particles, we
arrive with Eqs.\ (\ref{eq:T0}) and
(\ref{eq:def_Tb}) at
\begin{equation}
\label{eq:Tmunupnu}
\frac{i}{2m}\, \mathcal{T}_{\mu \nu} p^\nu = \frac{p_\mu}{2m}\, \mathcal{T}_0 - \frac{1}{2} \mathcal{T}_\mu^{(b)}
\equiv \frac{p_\mu}{2m}\, \mathcal{T}_0 - \frac{1}{2} \mathcal{T}_\mu^{(a)*} \;.
\end{equation}
The last identity uses the
result of App.\ \ref{app:B2},
which is possible since the term on the left-hand side
appears under an integral of
the type (\ref{eq:integral2}),
cf., e.g., Eq.\ (\ref{eq:Boltzmann_ordertwo_2}).

\subsection{Relation between traces and space-time shifts}
\label{app:B6}

Consider the
traces in Eqs.\ \eqref{eq:T0}, \eqref{eq:def_Ta}, \eqref{eq:def_T_tens}, and
\eqref{eq:def_Tmubc}.
Under the $\d \Gamma_1 \, \d \Gamma_2 \, \d \Gamma'$ integral,  we can replace
the energy projectors by the
quantities defined in Eq.\ (\ref{eq:def_h}) as, e.g., $\Lambda^+(p_1)\equiv 2 h(p_1,\s_1)$, since the additional terms vanish because of
Eq.\ (\ref{eq:intspin}). 
Employing this for $\Lambda^+(p_1),\,\Lambda^+(p_2)$, and $\Lambda^+(p')$, we find
with Eqs.\ \eqref{eq:T0_real}, \eqref{eq:Ta_cc}, \eqref{eq:Tc_cc}, 
\eqref{eq:rel_T'}, \eqref{eq:rel_T1}, \eqref{eq:rel_T2}, and \eqref{eq:Tmunupnu},
\begin{subequations}\label{eq:rel_T_Delta}
\begin{eqnarray}
    \mathrm{Im}\mathcal{T}_\mu^{(a)} \equiv - \mathrm{Im}\mathcal{T}_\mu^{(b)}&\equiv& 4m\, \frac{\hbar^2}{G_cG_d}\, \frac{\mathcal{W}}{m^4}\, 
    \frac{\Delta_{\mu}}{\hbar}  \;,\\
    \mathrm{Im}\mathcal{T}_\mu^{(c)}&\equiv& 0\;,\\
    \frac{1}{2m}\mathrm{Re}\mathcal{T}_{1,\mu\nu}p_{1}^{\nu}&\equiv&  4m\, \frac{\hbar^2}{G_cG_d}\, \frac{\mathcal{W}}{m^4}\, \frac{\Delta_{1,\mu}}{\hbar} \;,\\
    \frac{1}{2m}\mathrm{Re}\mathcal{T}_{2,\mu\nu}p_{2}^{\nu}&\equiv& 4m\, \frac{\hbar^2}{G_cG_d}\, \frac{\mathcal{W}}{m^4}\, \frac{\Delta_{2,\mu}}{\hbar} \;,\\
    \frac{1}{2m}\mathrm{Re}\mathcal{T}'_{\mu\nu}p^{\prime \nu}&\equiv&  4m\, \frac{\hbar^2}{G_cG_d}\, \frac{\mathcal{W}}{m^4}\,
    \frac{\Delta^\prime_\mu}{\hbar}\;,\\
    \frac{1}{2m}\mathrm{Re}\mathcal{T}_{\mu\nu}p^{\nu}&\equiv&  2m\, \frac{\hbar^2}{G_cG_d}\, \frac{\mathcal{W}}{m^4}\,
    \frac{\Delta_\mu}{\hbar} \;,
\end{eqnarray}
\end{subequations}
with the space-time shifts
(\ref{eq:def_Deltas_NJL}) and
$\mathcal{W}$ from Eq.\ \eqref{eq:W_NJL}. Using
Eq.\ (\ref{eq:def_W_element}), we
immediately prove Eqs.\ (\ref{eq:Delta_all}).

\section{Traces for scalar boson exchange}
\label{app:traces}

In this appendix, we evaluate the
various traces occurring in the
collision term for scalar boson exchange, 
$\Gamma^{(c)}\equiv \dblone$. 
Using the Mandelstam variables
\begin{equation}
s \coloneqq (p + p')^2 = (p_1 + p_2)^2\;,\quad
t \coloneqq (p-p_1)^2 = (p'-p_2)^2\; \quad
u \coloneqq (p-p_2)^2 = (p'-p_1)^2\;,
\end{equation}
with $s+t+u=4m^2$, Eq.\ (\ref{eq:T0}) becomes
\begin{align}
\mathcal{T}_0 & = \frac{(s+t)(2s+t)}{8m^4} - \frac{s-t}{2 m^2} + \frac{i}{m^3} \epsilon_{\mu \nu \alpha \beta}
\ms^\mu p^\nu p_1^\alpha p_2^\beta\;.
\end{align}
We denote the momentum $p^\mu$ 
in the C.M.\ frame as $p^\mu_\star \coloneqq (E_\star, \mathbf{p}_\star)$, with
$E_\star \coloneqq \sqrt{p_\star^2 + m^2}$, where $p_\star \coloneqq |\mathbf{p}_\star|$.
It then follows that
$p^{\prime \mu}_\star \equiv
(E_\star, - \mathbf{p}_\star)$.
Furthermore, we define $p_{1 \star}^\mu
\coloneqq (E_{\star}, \mathbf{p}_{1 \star})$, such that $p_{2,\star}^\mu \equiv (E_\star, - \mathbf{p}_{1\star})$.
Here we have used that, in the
C.M.\ frame, $|\mathbf{p}_\star|
= |\mathbf{p}_{1\star}|
\equiv p_\star$, such
that the on-shell energies of all
particles are equal.
Introducing the scattering angle $\Theta = \measuredangle(\mathbf{p}_\star, \mathbf{p}_{1\star})$,
we have $s = 4(m^2 + p_\star^2)$, $t= - 2p_\star^2(1- \cos\Theta)$, $u= - 2p_\star^2(1+ \cos\Theta)$, and therefore
\begin{equation}
\mathrm{Re} \mathcal{T}_0 = 2 \left[ 1 + \frac{p_\star^2}{m^2} (1 + 2 \cos \Theta)
+ \frac{p_\star^4}{m^4}( 1 + \cos\Theta) (3 + \cos \Theta) \right]\;.
\end{equation}
The low-energy limit is therefore $\lim_{p_\star \rightarrow 0} \mathrm{Re}\mathcal{T}_0 = 2$, and the
limit  of small scattering angles is $\lim_{\Theta \rightarrow 0} \mathrm{Re}\mathcal{T}_0 = 2 \left( 1+ 3\frac{p_\star^2}{m^2}
+ \frac{5p_\star^4}{4m^4}\right)$.

Similarly, we evaluate the traces \eqref{eq:def_Ts}. This is
simplified by taking the difference
of these traces and 
$\mathcal{T}_0$. Using
energy-momentum conservation multiple times to eliminate the
dependence on $p^{\prime \mu}$, we find
\begin{subequations}
\label{eq:T_i_minus_T_0}
\begin{align}
    \mathcal{T}_1-\mathcal{T}_0 & =  \s\cdot \s_1 \left[ \frac{s-t}{2m^2} + \frac{t(s+t)}{8m^4} \right] -  \frac{\s \cdot p_1}{m}  \left( \frac{u}{4m^2}  \, \frac{\s_1 \cdot p}{m} + \frac{\s_1 \cdot p_2}{m} \right) - \frac{ \s \cdot p_2 \, \s_1 \cdot p }{m^2}  \nonumber \\
    & + \frac{i}{m^3}\epsilon_{\mu\nu\alpha\beta}\s_{1}^\mu p^\nu p_{1}^\alpha
    \, p_{2}^\beta \;,\label{eq:T1_scalar}\\
    \mathcal{T}_2-\mathcal{T}_0 & = \s\cdot \s_2 \left[ \frac{s-t}{2m^2} - \frac{(2s+t)(s+t)}{8m^4} \right] +  \frac{\s \cdot p_2}{m}  \left( \frac{2s+t}{4m^2}  \, \frac{\s_2 \cdot p}{m} - \frac{\s_2 \cdot p_1}{m} \right) - \frac{ \s \cdot p_1 \, \s_2 \cdot p }{m^2}  \nonumber \\
    & + \frac{i}{m^3}\epsilon_{\mu\nu\alpha\beta}\s_{2}^\mu p^\nu p_{1}^\alpha
    \, p_{2}^\beta \;,\label{eq:T2_scalar}\\
    \mathcal{T}' - \mathcal{T}_0 & = 
    \s\cdot \s' \left[ \frac{s-t}{2m^2} + \frac{t(s+t)}{8m^4} \right] -  \frac{\s \cdot p_1}{m} \left( 
    \frac{\s' \cdot p}{m}
    + \frac{u}{4m^2} \, \frac{\s' \cdot p_2}{m} \right)
    - \frac{\s \cdot p_2}{m} \left(  \frac{\s' \cdot p}{m}
    + \frac{t}{4m^2} \, \frac{\s' \cdot p_1}{m} \right)  \nonumber \\
    & + \frac{i}{m^3}\epsilon_{\mu\nu\alpha\beta}\s^{\prime\mu} p^\nu p_{1}^\alpha
    \, p_{2}^\beta \;,\\
    \bar{\mathcal{T}}-\mathcal{T}_0 & = \s\cdot \bar{\s} \left[ \frac{s-t}{2m^2} - \frac{(2s+t)(s+t)}{8m^4} \right]
    + \frac{\s \cdot p_1\, \bar{\s} \cdot p_2}{m^2} -
    \frac{\s \cdot p_2\, \bar{\s} \cdot p_1}{m^2} \nonumber \\
    & + \frac{i}{m^3} \epsilon_{\mu\nu\alpha \beta}\bar{\s}^{\mu}p^\nu p_{1}^{\alpha}\, p_{2}^{\beta}\;.\label{eq:Tbar_scalar}
\end{align}
\end{subequations}
We note that the last two terms
in the first line of Eq.\ (\ref{eq:Tbar_scalar}) are antisymmetric under the exchange of
$p_1 \leftrightarrow p_2$ and thus vanish under a $\d P_1\, \d P_2$  integral where the remainder of the integrand is symmetric under this exchange.

In the low-energy limit, i.e.,
where the C.M.\ momentum
$\mathbf{p}_\star \rightarrow 0$, one can
show that all four-products between
spin vectors and momenta
vanish. This is most easily seen in the C.M.\ frame and using the orthogonality of the spin four-vector with the four-momentum. Also,
in this limit all momenta just have a time component, e.g., $p^\mu \rightarrow (m, \mathbf{0})$, and similarly for the other momenta.
Then, all imaginary parts in Eqs.\ (\ref{eq:T_i_minus_T_0}) vanish. This then yields
\begin{subequations}
\begin{eqnarray}
    \lim_{p_\star \rightarrow 0} \left(\mathcal{T}_1-\mathcal{T}_0\right)& = & 2 \,\s\cdot \s_1 \;,\\
    \lim_{p_\star \rightarrow 0} \left(\mathcal{T}_2-\mathcal{T}_0\right) & = & - 2\, \s\cdot \s_2  \;,\\
    \lim_{p_\star \rightarrow 0} \left(\mathcal{T}'-\mathcal{T}_0\right) & = &  2 \,\s\cdot \s'  \;,\\
    \lim_{p_\star \rightarrow 0} \left(\bar{\mathcal{T}}-\mathcal{T}_0\right)& = &-2 \, \s\cdot \bar{\s}  \;.
\end{eqnarray}
\end{subequations}

Similarly, we compute from
Eqs.\ (\ref{eq:def_T_tens})
\begin{subequations}
\begin{align}
\mathcal{T}_{1, \mu \nu} p_1^\nu & =
\epsilon_{\mu \nu \alpha \beta} \left[ \frac{\s \cdot p_1}{m}\,
\frac{p^\nu p_1^\alpha p_2^\beta}{2m^2} + 
\left( 1+\frac{t}{4m^2}\right) \frac{\s^\nu p_1^\alpha p_2^\beta}{m} 
+ \frac{t-u}{4m^2}\, \frac{\s^\nu p^\alpha p_1^\beta}{m}  \right]
\nonumber \\
& + i \left[
\left( 1 - \frac{3s}{4m^2}\right) p_\mu + 
\frac{s-t}{4m^2} \, p_{1,\mu}
+ \left( 1- \frac{3t}{4m^2}\right)
 p_{2, \mu} \right]\;, \\ 
\mathcal{T}_{2, \mu \nu} p_2^\nu & =
- 
\epsilon_{\mu \nu \alpha \beta} \left[ \frac{\s \cdot p_2}{m} \,
\frac{p^\nu p_1^\alpha p_2^\beta}{2m^2}  + 
\left( 1+ \frac{u}{4m^2}\right) \, \frac{\s^\nu p_1^\alpha p_2^\beta}{m}
+ \frac{t-u- 4(s+t)}{4m^2}
\, \frac{\s^\nu p^\alpha p_2^\beta}{m}  \right]
\nonumber \\
& - i \left[
\left( 1 - \frac{3s}{4m^2}\right) p_\mu 
+ \left( 1 - \frac{3u}{4m^2} \right)
 p_{1, \mu} + 
\frac{s-u}{4m^2} \, p_{2,\mu}
\right]\;, \\ 
\mathcal{T}_{\mu \nu}^\prime p^{\prime \nu} & = -
 \epsilon_{\mu \nu \alpha \beta} \left[ \frac{\s \cdot (p_1-  p_2)}{m} \,
\frac{p^\nu p_1^\alpha p_2^\beta}{2m^2}  + 
\frac{t-u}{4m^2}\, 
\frac{\s^\nu p_1^\alpha p_2^\beta}{m} 
+ \left( 1+ \frac{t}{4m^2} \right)
\, \frac{\s^\nu p^\alpha p_1^\beta }{m}
+ \left( 1 + \frac{u}{4m^2}\right)
\, \frac{\s^\nu p^\alpha p_2^\beta}{m}  \right]
\nonumber \\
& - i \left(
\frac{u-t}{4m^2}\, p_\mu 
+ \frac{s-t}{4m^2}
\, p_{1, \mu} + 
\frac{u-s}{4m^2} \, p_{2,\mu}
\right)\;, 
\end{align}
\end{subequations}
and from Eq.\ (\ref{eq:def_Ta})
\begin{align}
\mathcal{T}_\mu^{(a)} & = 
\left( \frac{s + 3t}{4m^2} \, \frac{p_\mu}{m} - \frac{s+t}{4m^2}\,  \frac{p_{1, \mu} }{m} +
\frac{s}{2m^2}\, \frac{p_{2,\mu}}{m} \right)  - i \, \epsilon_{\mu \nu \alpha \beta} \, \s^\nu 
\left(
 \frac{p_1^\alpha
p_2^\beta}{m^2}  - 
\frac{u}{4m^2}\, \frac{p^\alpha p_1^\beta}{m^2}  + \frac{s+t-u}{4m^2} \,
\frac{p^\alpha p_2^\beta}{m^2} 
\right)\;,
\end{align}
as well as from Eq.\ (\ref{eq:def_Tmuc})
\begin{align}
\mathcal{T}_\mu^{(c)}
& = - \left[
\frac{u+t}{4m^2}\,  \frac{p_\mu}{m}  - 
\frac{3(s+t)}{4m^2}\, \frac{p_{1,\mu}}{m}
+ \frac{s+u}{4m^2}\, \frac{p_{2,\mu} }{m}
\right]
+ \frac{i}{2} \,
\epsilon_{\mu \nu \alpha \beta}\, \s^\nu \left( \frac{p_1^\alpha p_2^\beta }{m^2} -
\frac{p^\alpha p_1^\beta}{m^2} 
+ \frac{p^\alpha p_2^\beta}{m^2} \right) \;.
\end{align}
Note that some terms in these expression can be further simplified using the symmetry under the $\d P_1 \, \d P_2$ integral.

From these results 
we finally compute
the space-time shifts using Eqs.\ (\ref{eq:Delta_all}),
\begin{subequations}
\begin{align}
\Delta_{1,\mu} & = 
\frac{\hbar}{2m \mathrm{Re}\mathcal{T}}\,
\frac{1}{2m}\, \mathrm{Re}
\mathcal{T}_{1,\mu \nu}p_1^\nu \nonumber \\
& = 
\frac{\hbar}{2m \mathrm{Re}\mathcal{T}}\,
 \epsilon_{\mu \nu \alpha \beta} \left[  \frac{\s \cdot p_1}{m} \,
\frac{p^\nu p_1^\alpha p_2^\beta}{4m^3} + 
\left(1 + \frac{t}{4m^2}\right) \frac{\s^\nu p_1^\alpha p_2^\beta }{2m^2}
+ \frac{t-u}{4m^2}\, \frac{\s^\nu p^\alpha p_1^\beta}{2m^2}  \right]\;, \\
\Delta_{2,\mu} & = 
\frac{\hbar}{2m \mathrm{Re}\mathcal{T}}\,
\frac{1}{2m}\, \mathrm{Re}
\mathcal{T}_{2,\mu \nu}p_2^\nu \nonumber \\
& = -
\frac{\hbar}{2m \mathrm{Re}\mathcal{T}}\,
\epsilon_{\mu \nu \alpha \beta} \left[ \frac{\s \cdot p_2}{m}\,
\frac{p^\nu p_1^\alpha p_2^\beta}{4m^3}  + 
\left( 1+\frac{u}{4m^2}\right) \frac{\s^\nu p_1^\alpha p_2^\beta }{2m^2}
+ \frac{t-u-4(s+t)}{4m^2}
\, \frac{\s^\nu p^\alpha p_2^\beta}{2m^2}  \right] \;,
\\
\Delta^\prime_\mu & = 
\frac{\hbar}{2m \mathrm{Re} \mathcal{T}} \,
\mathrm{Re}
\mathcal{T}_{\mu \nu}^\prime p^{\prime \nu}  \nonumber \\
& =-
\frac{\hbar}{2m \mathrm{Re} \mathcal{T}} \,
\epsilon_{\mu \nu \alpha \beta} \left[ \frac{\s \cdot (p_1 - p_2)}{m} \,
\frac{p^\nu p_1^\alpha p_2^\beta}{4m^3}  + 
\frac{t-u}{4m^2}\, \frac{\s^\nu p_1^\alpha p_2^\beta}{2m^2} 
+ \left(1+ \frac{t}{4m^2}\right)
 \frac{\s^\nu p^\alpha p_1^\beta}{2m^2}
+ \left(1+ \frac{u}{4m^2}\right)
 \frac{\s^\nu p^\alpha p_2^\beta}{2m^2}  \right] \;, \\
\Delta_\mu & =
\frac{\hbar}{2m \mathrm{Re}\mathcal{T}}\,
\mathrm{Im} \mathcal{T}_\mu^{(a)} \nonumber \\
& = - \frac{\hbar}{2m \mathrm{Re}\mathcal{T}}\,
\epsilon_{\mu \nu \alpha \beta} \, \s^\nu\left( \frac{p_1^\alpha p_2^\beta}{m^2}- \frac{u}{4m^2}\,
\frac{p^\alpha p_1^\beta}{m^2} + \frac{s + t - u}{4m^2}\, \frac{p^\alpha p_2^\beta}{m^2}
\right) \;.
\end{align}
\end{subequations}

In the low-energy limit, the spacetime shifts vanish.

\bibliographystyle{apsrev}
\bibliography{biblio_paper_long}

\end{document}